\documentclass[12pt]{article}
\usepackage{amsmath,amsfonts, feynmp, epsf}
\usepackage{amssymb}
\usepackage{graphicx}
\usepackage{grffile}
\input epsf

\makeatletter\@addtoreset{equation}{section}\makeatother

\def\bff {\mathbf{f}}
\def\bh {\mathbf{h}}

\def\bv {\mathbf{v}}
\def\bP {\mathbf{P}}

\def\bG {\mathbf{G}}

\def\be{\begin{equation}}
\def\ee{\end{equation}}
\def\bea{\begin{eqnarray}}
\def\eea{\end{eqnarray}}
\def\ie{\begin{equation}\begin{aligned}}
\def\fe{\end{aligned}\end{equation}}

\newcommand{\m}{\mu}
\newcommand{\n}{\nu}

\newcommand{\A}{{\alpha}}
\newcommand{\B}{{\beta}}

\makeatletter\@addtoreset{equation}{section}\makeatother

\newcommand{\vev}[1]{{\left< {#1} \right>}}
\newcommand{\bra}[1]{{\left< {#1} \right|}}
\newcommand{\ket}[1]{{\left| {#1} \right>}}

\newcommand{\cC}{{\mathcal C}}

\newcommand{\cG}{{\mathcal G}}
\newcommand{\cJ}{{\mathcal J}}

\newcommand{\cN}{{\mathcal N}}
\newcommand{\cO}{{\mathcal O}}

\newcommand{\cQ}{{\mathcal Q}}

\newcommand{\cM}{{\mathcal M}}

\newcommand{\tV}{{\widetilde V}}

\renewcommand{\title}[1]{\vbox{\center\LARGE{#1}}\vspace{5mm}}
\renewcommand{\author}[1]{\vbox{\center#1}\vspace{5mm}}
\newcommand{\address}[1]{\vbox{\center\em#1}}
\newcommand{\email}[1]{\vbox{\center\tt#1}\vspace{5mm}}

\begin{document}

%\unitlength = .8mm
\begin{titlepage}
\begin{center}
\hfill \\
\hfill \\
\vskip 1cm

\title{Correlators in $W_N$ Minimal Model Revisited}

\author{Chi-Ming Chang$^{a}$ and
Xi Yin$^{b}$}

\address{
Jefferson Physical Laboratory, Harvard University,\\
Cambridge, MA 02138 USA}

\email{$^a$cmchang@physics.harvard.edu,
$^b$xiyin@fas.harvard.edu}

\end{center}

\abstract{
In this paper, we study a class of sphere and torus correlation functions in the $W_N$ minimal model. In particular, we show that a large class of exact sphere three-point functions of $W_N$ primaries, derived using affine Toda theory, exhibit large $N$ factorization. This allows us to identify some fundamental particles and their bound states in the holographic dual, including light states. We also derive the torus two-point function of basic primaries, by directly constructing the relevant conformal blocks. The result can then be analytically continued to give the Lorentzian thermal two-point functions.
}

\vfill

\end{titlepage}

\eject \tableofcontents

\section{Introduction}

The AdS/CFT correspondence \cite{Maldacena:1997re} is one of the most important insights that came out of the study of string theory. While it is often said that both strings and the holographic dimension emerge from the large $N$ and strong 't Hooft coupling limit of a gauge theory, there are really two separate dualities in play here. Firstly, a large $N$ CFT, regardless of whether the 't Hooft coupling is weak or strong, is holographically dual to some theory of gravity together with higher spin fields in $AdS$, whose coupling is controlled by $1/N$ \cite{Klebanov:2002ja}. It often happens that, then, as a 't Hooft coupling parameter varies from weak to strong, the bulk theory interpolates between a higher spin gauge theory and a string theory (where the $AdS$ radius becomes finite or large in string units). The duality as two separate stories: holography from large $N$, and the emergence of strings out of bound states of higher spin fields, has become particularly evident in \cite{Giombi:2011kc}.

The holographic dualities between higher spin gauge theories in AdS and vector model CFTs \cite{Klebanov:2002ja, Sezgin:2002rt, Gaberdiel:2010pz, Giombi:2011kc} are a nice class of examples in that they avoid the complication of the second story mentioned above.\footnote{See \cite{Giombi:2009wh, Giombi:2010vg, Koch:2010cy, Douglas:2010rc, Giombi:2011ya, Maldacena:2011jn} for recent nontrivial checks and progress toward deriving the duality with vector models.} Both sides of the duality can be studied order by order in the $1/N$ expansion. The AdS$_3$/CFT$_2$ version of this duality, proposed by Gaberdiel and Gopakumar \cite{Gaberdiel:2010pz}, relates a higher spin gauge theory coupled to scalar matter fields in AdS$_3$ \cite{Vasiliev:1999ba} and the $W_N$ minimal model in two dimensions \cite{Bouwknegt:1992wg}.\footnote{For works leading up to this duality, and explorations on its consequences, see \cite{Henneaux:2010xg, Campoleoni:2010zq, Gaberdiel:2011wb, Kiritsis:2010xc, Castro:2010ce, GGHR, Ahn:2011pv, Chang:2011mz, Papadodimas:2011pf, Gutperle:2011kf, Kraus:2011ds, Castro:2011iw}.} While it was proposed in \cite{Gaberdiel:2010pz} that the bulk theory is Vasiliev's system in $AdS_3$, it was pointed out in \cite{Chang:2011mz} and in \cite{Papadodimas:2011pf} that Vasiliev's system should be dual only perturbatively in $1/N$ to a subsector of the $W_N$ minimal model, while the full non-perturbative duality requires adding new perturbative states in the bulk.\footnote{See \cite{Castro:2011iw} however for intriguing candidates for some new bulk states in higher spin gauge theories in $AdS_3$.}

One of the key observations of \cite{Gaberdiel:2010pz} is that the $W_{N,k}$ minimal model has a 't Hooft-like limit, where $N$ is taken to be large while the ``'t Hooft coupling" $\lambda={N\over k+N}$ is held finite. The basic evidence is that the spectrum of operators organize into that of ``basic primaries", which are dual to elementary particles in the bulk, and the composite operators which are dual to bound states of elementary particles. It was not obvious, however, that the correlation functions obey large $N$ factorization, as for single trace operators in large $N$ gauge theories. This will be demonstrated in the current paper. In particular, we will understand which operators are the fundamental particles, and which ones are their bound states, by extracting such information from the $1/N$ expansion of exact correlation functions in the $W_N$ minimal model. 

Our main findings are summarized as follows.

1. We derive all sphere three point functions of primaries in the $W_N$ minimal model of the following form: one of the primaries is labelled by a pair of $SU(N)$ representations $(\Lambda_+,\Lambda_-)$, both of which are symmetric products of the fundamental (or anti-fundamental) representation ${\bf f}$ (or ${\bf \bar f}$), and the other two primaries are completely general.\footnote{The technique used in this paper allows us to go beyond this set using four point functions, but we will not present those results here.} We see the explicit large $N$ factorization in these three point functions. For example, denote by $\phi$ the primary $({\bf f},0)$ (on both left and right moving sector). The large $N$ factorization leads to the identification
\ie
&(A,0) \sim {1\over \sqrt{2}}\phi^2, \\
&(S,0) \sim {1\over \sqrt{2} \Delta_{(\bff,0)}} (\phi \partial\bar\partial\phi - \partial\phi \bar\partial\phi),
\fe
where $A$ and $S$ are the anti-symmetric and symmetric tensor product representation of ${\bf f}$, and $ \Delta_{(\bff,0)}=1+\lambda$ is the scaling dimension of $\phi$ at large $N$. This large $N$ factorization is a simple check of the duality, in verifying that $(A,0)$ and $(S,0)$ are indeed bound states of two elementary scalar particles in the bulk, and behave as two free particles in the infinite $N$ limit.

A less obvious example concerns the ``light" primary $({\bf f},{\bf f})$, which we denote by $\omega$. Its scaling dimension $ \Delta_{(\bff,\bff)}$ vanishes in the infinite $N$ limit, and is given by $\Delta_{(\bff,\bff)}= \lambda^2/N$ at order $1/N$. Two candidates for the lowest bound state of two $\omega$'s are $(A,A)$ and $(S,S)$, both of which have scaling dimension $2\Delta_{(\bff,\bff)}$ at order $1/N$. We will find that
\ie
{(A,A)+(S,S)\over \sqrt{2}} \sim {1\over \sqrt{2}}\omega^2
\fe
is the bound state of two $\omega$'s, while ${1\over \sqrt{2}}((A,A)-(S,S))$ is a new elementary light particle in the bulk. This shows that the elementary light particles in the bulk also interact weakly in the large $N$ limit.

A word of caution is that even in the infinite $N$ limit, the space of states is {\it not} the freely generated Fock space of single particle primary states and their descendants. As observed in \cite{Papadodimas:2011pf}, for instance, the level $(1,1)$ descendant of $\omega$, namely ${1\over \Delta_{(\bff,\bff)}}\partial\bar\partial\omega$, should be identified with the the two-particle state (or ``double trace operator") $\phi\,\widetilde\phi$, where $\widetilde\phi$ is the other basic primary $(0,\bff)$. We will see that this identification is consistent with the large $N$ factorization of composite operators made out of $\omega$, $\phi$, and $\widetilde\phi$. This suggests that the Hilbert space at infinite $N$ is a quotient of the freely generated Fock space, with identifications such as ${1\over \Delta_{(\bff,\bff)}}\partial\bar\partial\omega\sim \phi\widetilde\phi$. This peculiar feature is closed tied to the presence of light states. The large $N$ factorization in the $W_N$ minimal model holds only up to such identifications.

2. We compute the sphere four-point function of $({\bf f},0)$, $({\bf \bar f},0)$, with a general primary $(\Lambda_+,\Lambda_-)$ and its charge conjugate, which generalizes the four-point functions considered in \cite{Papadodimas:2011pf}. This result is not new and is in fact contained in \cite{Fateev:2007ab}. In \cite{Fateev:2007ab}, the sphere four-point function was obtained by solving the differential equation due to a null state, which we will review. The method gives the answer for general $N$, but is not easy to generalize to correlators on a Riemann surface of nonzero genus. We will then consider an alternative method, using contour integrals of screening charges. This second method requires knowing which contours correspond to which conformal blocks; they will be analyzed in detail through the investigation of monodromies. While this approach appears rather cumbersome due to the complexity of the contour integral, it allows for a straightforward generalization to the computation of torus two-point functions.\footnote{Our method is a direct generalization of \cite{Jayaraman:1988ex}, where the torus two-point function in the Virasoro minimal model was derived.}

3. We derive a contour integral expression for the torus two-point function of the basic primaries $({\bf f},0)$ and $({\bf \bar f},0)$. Since the result is exact, it can be analytically continued to Lorentzian signature, yielding the Lorentzian thermal two-point function on the circle. The latter is a useful probe of the dual bulk geometry. In a theory of ordinary gravity in $AdS_3$, at temperatures above the Hawking-Page transition, the dominant phase is the BTZ black hole. The thermal two-point function on the boundary should see the thermalization of the black hole reflected in an exponential decay behavior of the correlator, for a very long time before Poincar\'e recurrence kicks in.\footnote{In the $W_N$ minimal model, all scaling dimensions are integer multiples of ${1\over N(N+k)(N+k+1)}\sim {\lambda^2\over N^3}$, and hence Poincar\'e recurrence must already occur at no later than time scale $N^3$. In fact, we will see that the Poincare recurrence in the two-point function under consideration occurs at an even shorter time $N(k+N)$. But if the BTZ black hole dominates the bulk in some temperature of order 1, we should expect to see thermalization at time scale of order 1 (and $\ll N^2$).} While the BTZ black hole clearly exists in any higher spin gravity theory in $AdS_3$, it is unclear whether the BTZ black hole will be the dominant phase at any temperature at all, as there can be competing higher spin black hole solutions (see \cite{Gutperle:2011kf, Kraus:2011ds, Castro:2011iw}). Nonetheless, the question of whether thermalization occurs at the level of two-point functions can be answered definitively using the exact torus two-point function. So far, it appears to be difficult to extract the large $N$ behavior from our exact contour integral expression, which we leave to future work. In the $N=2$ case, i.e. Virasoro minimal model, where the contour integral involved is a relatively simple one, we computed numerically certain thermal two-point functions at integer values of times, as a demonstration in principle.

In section 2, we will summarize the definitions and convention for $W_N$ minimal model which will be used throughout this paper. Section 3 describes the strategy of the computation, namely using the Coulomb gas formalism. In section 4, 5, 6 we present a class of sphere three, four-point, and torus two-point functions, make various checks of the result, and discuss the implications. We conclude in section 7.

\section{Definitions and conventions for the $W_N$ minimal model}

The $W_N$ minimal model can be realized as the coset model
\ie
{SU(N)_k\oplus SU(N)_1\over SU(N)_{k+1}}.
\fe
A priori, through the coset construction, the $W_N$ primaries are labeled by a triple of representations of $SU(N)$ current algebra $(\rho, \mu; \nu)$ (at level $k,1$, and $k+1$ respectively.) By a slight abuse of notation, we will also denote by $\rho, \mu,\nu$ the corresponding highest weight vectors. The three representations are subject to the constraint that $\rho+\mu-\nu$ lies in the root lattice of $SU(N)$. Each representation is subject to the condition that the sum of $N-1$ Dynkin labels is less than or equal to the affine level. This condition determines $\mu$ uniquely, given $\rho$ and $\nu$. We will therefore label the primaries by the pair of the representations $(\rho;\nu) \equiv (\Lambda_+, \Lambda_-)$ from now on.

Let $\A_i$, $i=1,\cdots,N-1$, be the simple roots of $SU(N)$. They have inner product $\A_i\cdot \A_j = A_{ij}$, where $A_{ij}$ is the Cartan matrix. In particular, $\A_i^2=2$. Let $\omega^i$, $i=1,\cdots, N-1$, be the fundamental weights. They obey $\omega^i\cdot \A_j = \delta^i_j$. We write $F^{ij} = \omega^i\cdot \omega^j=(A^{-1})^{ij}$. The highest weight $\lambda$ of some representation $\Lambda$ takes the form
\ie
\lambda = \sum_{i=1}^{N-1} \lambda_i \omega^i,
\fe
where $(\lambda_1,\cdots,\lambda_{N-1})\in \mathbb{Z}_{\geq 0}^{N-1}$ are the Dynkin labels.

The Weyl vector is
\ie
\rho = \sum_{i=1}^{N-1} \omega^i, %= {1\over 2} \sum_{\A\in \Delta_+} \A,
\fe
i.e. it has Dynkin label $(1,1,\cdots,1)$. 

Given a root $\A$, the simple Weyl reflection with respect to $\A$ acts on a weight $\lambda$ by
\ie
s_\A (\lambda) = \lambda - (\A\cdot\lambda) \A.
\fe
A general Weyl group element $w$ can be written as $w=s_{\A_1}\cdots s_{\A_m}$. We will use the notation $w(\lambda)$ for the Weyl reflection of $\lambda$ by $w$. The {\it shifted} Weyl reflection $w\cdot\lambda$ is defined by
\ie
w\cdot\lambda = w(\lambda+\rho)-\rho.
\fe

Now let us discuss the $W_N$ character of a primary $(\Lambda_+, \Lambda_-)$. Throughout this paper, we use the notation $p=k+N$ and $p'=k+N+1$. The central charge is
\ie
c = N-1 - {N(N^2-1)\over pp'}.
\fe
Note that $\rho^2 = {1\over 12}N(N^2-1)$.
The conformal dimension of the primary is
\ie\label{confdim}
h_{(\Lambda_+,\Lambda_-)} = {1\over 2pp'} \left( |p'\Lambda_+ - p\Lambda_- +\rho|^2 - \rho^2 \right).
\fe
The character of $(\Lambda_+,\Lambda_-)$ can be written as a sum over {\it affine} Weyl group elements,
\ie
\chi^N_{(\Lambda_+,\Lambda_-)}(\tau) = {1\over \eta(\tau)^{N-1}} \sum_{\hat w\in \widehat W}
\epsilon(\hat w) q^{{1\over 2pp'}|p'\hat w(\Lambda_++\rho) - p (\Lambda_-+\rho)|^2},
\fe
where $\widehat W$ is given by the semi-direct product of $W$ with translations by $p$ times the root lattice, namely an element $\hat w\in \widehat W$ acts on a weight vector $\lambda$ by
\ie
\hat w(\lambda) = w(\lambda) + p n^i \A_i,~~~~w\in W, ~n_i\in\mathbb{Z}.
\fe
$\epsilon(\hat w) = \epsilon(w)$ is the signature of $\hat w$.

Let us illustrate this formula with the $N=2$ example, i.e. Virasoro minimal model. Write $\Lambda_+ = (r-1) \omega^1$, $1\leq r\leq p-1=k+1$, and $\Lambda_- = (s-1) \omega^1$, $1\leq s\leq p=k+2$. The Weyl group $\mathbb{Z}_2$ contains the reflection $w(\lambda) = -\lambda$. We have $\hat w(\Lambda_++\rho) = -r\omega^1+ p n \A_1 = (-r+2pn)\omega^1$. So
\ie
h_{r,s} = {(p'r-ps)^2-1\over 4pp'},
\fe
and
\ie
\chi_{r,s}(\tau) &= {1\over \eta(\tau)} \sum_{n\in\mathbb{Z}} \left[ q^{{1\over 4pp'}(p'(r + 2pn)-ps)^2} - q^{{1\over 4pp'}(p'(-r + 2pn)-ps)^2} \right]
\\
&=  {q^{{1\over 4pp'} (p'r-ps)^2}\over \eta(\tau)} \sum_{n\in\mathbb{Z}} \left[ q^{n(pp'n+p'r-ps)} - q^{(pn-r)(p'n-s)} \right]
\fe
The term corresponding to $(w,n=0)$ comes from the null state at level $rs$.

\section{Coulomb gas formalism}

The idea of Coulomb gas formalism is to represent operators in the $W_N$ minimal model by vertex operators constructed out of $N-1$ free bosons. This allows for the construction of all $W_N$ currents as well as the primaries of the correct scaling dimensions. However, the free boson correlators by themselves do not obey the correct fusion rules of the $W_N$ minimal model. To obtain the correct correlation functions, suitable screening operators must be inserted, and integrated along contours in a conformally invariant manner. More precisely, one obtains in this way the $W_N$ conformal blocks. One then needs to sums up the conformal blocks with coefficients determined by monodromies, etc. This strategy is explained below.

\subsection{Rewriting free boson characters}

Let us begin with the following character of $N-1$ free bosons, twisted by an $SU(N)$ weight vector $\lambda$,
\ie
\widetilde K^N_\lambda(\tau) &=  {1\over \eta(\tau)^{N-1}} \sum_{\A\in \Lambda_{root}} q^{{1\over 2pp'} |\lambda +pp' \A|^2} 
\\
&= {1\over \eta(\tau)^{N-1}} \sum_{(n^1,\cdots,n^{N-1})\in \mathbb{Z}^{N-1}} q^{{1\over 2pp'} |\lambda +pp' n^j \A_j|^2}.
\fe
Define the lattice
\ie
\Gamma_{x} = \sqrt{x} \Lambda_{root},
\fe
and its dual lattice
\ie
\Gamma_{x}^* = {1\over\sqrt{x}} \Lambda_{weight}.
\fe
We may then write
\ie
K^N_u (\tau) =  {1\over \eta(\tau)^{N-1}} \sum_{n \in \Gamma_{pp'}} q^{{1\over 2} (u+n)^2} 
\fe
for $u\in \Gamma_{pp'}^*$. In fact, $u$ may be defined in the quotient of lattices,
\ie
u \in \Gamma_{pp'}^*/\Gamma_{pp'}.
\fe
Note that the number of elements in $\Gamma_{pp'}^*/\Gamma_{pp'}$ is
\ie
\det(pp' A_{ij}) = N (pp')^{N-1}.
\fe 
It is useful to consider the decomposition 
\ie
u = \lambda + \lambda',~~~~ \lambda \in \Gamma_{p\over p'}^*/\Gamma_{pp'},~\lambda' \in \Gamma^*_{p'\over p}/\Gamma_{pp'}.
\fe
This decomposition is well defined up to the identification
\ie\label{lltid}
(\lambda, \lambda')\sim (\lambda + t, \lambda'-t),~~~t \in \Gamma_{1\over pp'}^*/\Gamma_{pp'} = (\Gamma_{p\over p'}^* \cap \Gamma^*_{p'\over p})/\Gamma_{pp'}.
\fe
Consider the action of a simple Weyl reflection on $v\in \Gamma_x^*$,
\ie
w_\A(v) = v - (\A\cdot v)\A,
\fe
where $\A$ is a root. Since $(\A\cdot v)\A\in x^{-{1\over 2}} \Lambda_{root} = \Gamma_{1\over x}$,
the Weyl action is trivial on $\Gamma_x^*/\Gamma_{1\over x}$. In particular, the Weyl action on $u$ is trivial on $\Gamma_{1\over pp'}^*/\Gamma_{pp'}$, and is well defined on $\lambda$ and $\lambda'$ separately. Therefore, one can define the {\it double} Weyl action by $W\times W$ on $\lambda$ and $\lambda'$ independently. This will be important in describing $W_N$ primaries.

Now consider $N-1$ free bosons compactified on the Narain lattice $\Gamma^{N-1,N-1}$, which is even, self-dual, of signature $(N-1,N-1)$, defined as\footnote{To see that $\Gamma^{N-1,N-1}$ is even, note that
\ie
(v,\bar v)\cdot (v,\bar v) = v^2 - \bar v^2 = v^2 - (v+n)^2 = -2v\cdot n - n^2,
\fe
where $n\in \Gamma_{pp'}$, and the RHS is an even integer. To see that it is self-dual, take a basis $(v^i, v^i)$ and $(v_i,0)$, $i=1,\cdots,N-1$,  where $v_i\in \Gamma_{pp'}$ and $v^i\in \Gamma_{pp'}^*$ are dual basis for the respective lattices. This basis is unimodular.}
\ie\label{gammann}
\Gamma^{N-1,N-1} = \{ (v,\bar v)|v,\bar v\in \Gamma_{pp'}^*, v-\bar v\in \Gamma_{pp'} \}.
\fe
The free boson partition function can be decomposed in terms of the characters as
\ie
Z^{bos}_{\Gamma^{N-1,N-1}}(\tau,\bar\tau) = \sum_{u \in \Gamma_{pp'}^*/\Gamma_{pp'}} |K^N_u(\tau )|^2.
\fe

\subsection{$W_N$ characters and partition function}

Consider a $W_N$ primary $(\Lambda_+, \Lambda_-)$. Using the decomposition $u=\lambda+\lambda'$ described in the previous subsection, we may rewrite the $W_N$ character
\ie
\chi^N_{(\Lambda_+,\Lambda_-)}(\tau) = {1\over \eta(\tau)^{N-1}} \sum_{\hat w\in \widehat W}
\epsilon(\hat w) q^{{1\over 2pp'}|p'\hat w(\Lambda_++\rho) - p (\Lambda_-+\rho)|^2}
\fe
in the form $\chi^N_{\lambda+\lambda'}(\tau)$, where
\ie
\lambda = \sqrt{p'\over p} (\Lambda_++\rho) \in \Gamma_{p\over p'}^*,~~~~\lambda' = -\sqrt{p\over p'} (\Lambda_-+\rho) \in \Gamma_{p'\over p}^*.
\fe
In other words, we write
\ie\label{WNcharacter}
\chi^N_{\lambda+\lambda'}(\tau) &= {1\over \eta(\tau)^{N-1}} \sum_{w\in W, n\in \Gamma_{pp'}}
\epsilon( w) q^{{1\over 2}| w(\lambda) + \lambda'+n|^2}
\\
& = \sum_{w\in W} \epsilon(w) K^N_{w(\lambda)+\lambda'}(\tau).
\fe
The rationale for the alternating sum in the above formula is the following. The dimension of the free boson vertex operator $e^{i (u-Q)\cdot X}$ corresponding to the character $K^N_u$, with linear dilaton (as will be described in the next subsection), is
\ie\label{dimm}
h_u={1\over 2}u^2-{1\over 2}Q^2.
\fe
Let $w$ be a simple Weyl reflection, by a root $\A_w$. A simple computation shows that
\ie
h_{w(\lambda)+\lambda'}&=h_{\lambda+\lambda'}+(\A_w\cdot\lambda)(-\A_w\cdot\lambda').
\fe
If we restrict $\lambda$ and $-\lambda'$ to sit in the identity Weyl chamber of $\Gamma_{p\over p'}^*$ and $\Gamma_{p'\over p}^*$, then $(\A_w\cdot\lambda)(-\A_w\cdot\lambda')$ is always a nonnegative integer. It is possible to subtract off the character $K^N_{w(\lambda)+\lambda'}$ to make the theory ``smaller". The alternating sum in (\ref{WNcharacter}) does this in a Weyl invariant manner\footnote{For $w$ not a simple Weyl reflection, one can show that $h_{w(\lambda)+\lambda'}-h_{\lambda+\lambda'}$ is still a nonnegative integer, when $\lambda$ and $-\lambda'$ sit in the identity Weyl chamber of $\Gamma_{p\over p'}^*$ and $\Gamma_{p'\over p}^*$.}, and gives the character $\chi^N_{\lambda+\lambda'}(\tau)$ of the $W_N$ minimal model.

Note that $\chi^N_{\lambda+\lambda'}(\tau)$ vanishes identically whenever $(\lambda,\lambda')$ is fixed by the action of a subgroup of the double Weyl group $W\times W$. The set of inequivalent characters are thus parameterized by
\ie
E = (\Gamma_{pp'}^*/\Gamma_{pp'} - \{{\rm fixed ~points}\})/W\times W.
\fe
This is also the set of inequivalent $W_N$ primaries.
The partition function of the $W_N$ minimal model is given by the diagonal modular invariant
\ie
Z^N_{p,p'}(\tau,\bar\tau) &= \sum_{(\Lambda_+, \Lambda_-)} |\chi^N_{(\Lambda_+,\Lambda_-)}(\tau)|^2
\\
&={1\over N(N!)^2}\sum_{\lambda\in \Gamma_{p\over p'}^*/\Gamma_{pp'}, ~\lambda'\in \Gamma_{p'\over p}^*/\Gamma_{pp'}} |\chi_{\lambda+\lambda'}^N(\tau)|^2
\\
&={1\over (N!)^2}\sum_{u\in \Gamma_{pp'}^*/\Gamma_{pp'}} |\chi_u^N(\tau)|^2,
\fe
where the first sum is only over inequivalent $(\Lambda_+,\Lambda_-)$ under shifted Weyl reflections. The decomposition $u=\lambda+\lambda'$ is understood in going between the last two lines ($\lambda, \lambda'$ are defined up to a shift by $t\in \Gamma^*_{1\over pp'}/\Gamma_{pp'}$).

Let us illustrate again with the $N=2$ example. In this case, $\Gamma_{pp'} = \sqrt{2pp'}\,\mathbb{Z}$, $\Gamma_{pp'}^* ={1\over \sqrt{2pp'}}\mathbb{Z}$. We have
\ie
\lambda \in \sqrt{p'\over 2p}\,\mathbb{Z},~~~\lambda' \in \sqrt{p\over 2p'}\,\mathbb{Z},~~~ t\in \sqrt{pp'\over 2}\,\mathbb{Z},
\fe
and
\ie
{\Gamma_{pp'}^*\over \Gamma_{pp'}} \simeq {\mathbb{Z}_{2p}\times \mathbb{Z}_{2p'}\over \mathbb{Z}_2} 
\fe
$W\simeq \mathbb{Z}_2$ acts on $\Gamma_x$ by reflection. The set of inequivalent characters is
\ie
E \simeq {\mathbb{Z}_p^\times \times \mathbb{Z}_{p'}^\times\over\mathbb{Z}_2},
\fe
where the $\mathbb{Z}_2$ identification on ${\mathbb{Z}_p^\times \times \mathbb{Z}_{p'}^\times}$ is
\ie
(r,s) \to (r+p,s+p')\sim (p-r,p'-s).
\fe

\bigskip

Returning to the general $W_N$ characters, the modular transformation on $\chi^N_u(\tau)$ takes the form
\ie
\chi^N_u (-1/\tau) = \sum_{\tilde u\in \Gamma_{pp'}^*/\Gamma_{pp'}} \tilde S_{u,\tilde u} \chi^N_{\tilde u}(\tau),
~~~~\tilde S_{u,\tilde u} = {1\over \sqrt{N(pp')^{N-1}}} e^{-2\pi i u\cdot \tilde u}.
\fe
The RHS is not yet written as a sum over independent characters. After doing so, we have
\ie
\chi^N_u (-1/\tau) = \sum_{\tilde u\in \left(\Gamma_{pp'}^*/\Gamma_{pp'}-{\rm fixed}\right)/W\times W} S_{u,\tilde u} \chi^N_{\tilde u}(\tau),
\fe
where
\ie
S_{u,\tilde u} = \sum_{(w,w')\in W\times W} \epsilon(w)\epsilon(w') S_{u,w(\tilde\lambda) + w'(\tilde\lambda')}.
\fe

\subsection{Coulomb gas representation of vertex operators and screening charge}

We have seen that the partition function of the $W_N$ minimal model may be obtained from that of the free bosons on the lattice $\Gamma^{N-1,N-1}$ by twisting by $\epsilon(w)$ in a sum over action by Weyl group elements $w\in W$. The free boson vertex operators corresponding to lattice vectors of $\Gamma^{N-1,N-1}$ take the form
\ie
e^{iv \cdot X + i\B \cdot X_L},
\fe
where $v\in \Gamma^*_{pp'}$, and $\B \in \Gamma_{pp'}$. The lowest weight states appearing in the characters $|K^N_u|^2$ are of the form $e^{iv\cdot X}$.

Given a $W_N$ primary labeled by $(\Lambda_+, \Lambda_-)$, we associate it with the free boson vertex operator $e^{i v\cdot X}$, with the identification
\ie\label{vlam}
v = \sqrt{p'\over p}\Lambda_+ - \sqrt{p\over p'}\Lambda_-.
\fe
In order to match the conformal dimensions, we need to turn on a linear dilaton background charge
$Q=2v_0 \rho$, where $v_0 = {1\over 2}\left(\sqrt{p\over p'}-\sqrt{p'\over p}\right) = -{1\over 2\sqrt{pp'}}$. The conformal weight of $e^{iv\cdot X}$ in the linear dilaton CFT is then
\ie\label{nsl}
h_{v-Q} = {1\over 2} (v - Q)^2 - {1\over 2}Q^2 = {1\over 2}v^2 - Q\cdot v .
\fe
Using
\ie
& u = v - Q = \sqrt{p'\over p}(\Lambda_+ + \rho) - \sqrt{p\over p'}(\Lambda_-+\rho),
\\
& Q^2 = {1\over pp'}\rho^2 = {1\over 12 pp'} N(N^2-1),
\fe
we see that indeed
\ie
h_{v-Q} = h_{(\Lambda_+,\Lambda_-)}.
\fe

We will denote by ${\cal O}_v$ a primary of the $W_N$ algebra and by $V_v$ the corresponding free {\it chiral} boson vertex operator $e^{iv\cdot X_L}$.
On a genus $g$ Riemann surface, correlators of the linear dilaton CFT are nontrivial only if the total charge is $(2-2g) Q$. For instance, the non-vanishing sphere two-point functions must involve a pair of operators $V_v$ and $V_{2Q-v}$, of equal conformal weights and total charge $2Q$. On the other hand, the fusion rule in the $W_N$ minimal model is such that the correlation function $\langle {\cal O}_{v_1}\cdots {\cal O}_{v_n}\rangle$ is nonvanishing only if $\sum_{i=1}^n v_i \in \Gamma_{p'\over p}+ \Gamma_{p\over p'} = \Gamma_{1\over pp'}$.

For each simple root $\A_i$, we have
\ie
\sqrt{p\over p'}\A_i \in \Gamma_{p\over p'},~~~~ \sqrt{p'\over p}\A_i \in \Gamma_{p'\over p}.
\fe
The vertex operators 
\ie
V_i^+ = V_{\sqrt{p\over p'}\A_i},~~~~V_i^- = V_{-\sqrt{p'\over p}\A_i}.
\fe
have conformal weight 1, and can be used as screening operators. By inserting screening charges, the contour integrals of these screening operators, we can obtain all correlators of $W_N$ primaries that obey the fusion rule. We can also absorb the background charge with screening charges. This relies on the fact
\ie
\rho = {1\over 2}\sum_{\A\in \Delta_+}\A,
\fe
where $\Delta_+$ is the set of all positive roots. So we can write
\ie
2Q = 4v_0 \rho = \sum_{\A\in \Delta_+} \left( \sqrt{p\over p'}\A - \sqrt{p'\over p}\A \right).
\fe
which may be further written as a sum of non-negative integer multiples of $\sqrt{p\over p'}\A_i$ and $-\sqrt{p'\over p}\A_i$, which are the screening operators.

As an example, consider ${\cal O}_v$ and its charge conjugate operator ${\cal O}_{\overline v}$. If $V_v$ is the Coulomb gas representation of ${\cal O}_v$, then $V_{2Q-v}$ has the correct dimension and charge (modulo root lattice) to represent ${\cal O}_{\overline v}$. Alternatively, one may take $V_{\overline v}$, which differs from $V_{2Q-v}$ by some screening charges. There is a Weyl reflection $w_0$ (the longest Weyl group element) such that
\ie
w_0 (\overline{v}) = -v, ~~~~ w_0(\rho) = -\rho.
\fe
The shifted Weyl transformation by $w_0$ acts as
\ie
w_0 \cdot \overline{v} & = \sqrt{p'\over p}(w_0(\overline\Lambda_+ + \rho) -\rho) -  \sqrt{p\over p'}(w_0(\overline\Lambda_- + \rho)-\rho)
\\
&= 2Q - v.
\fe
So indeed $\overline v$ and $2Q-v$ are identified by Weyl reflection and represent the same $W_N$ primary.

\section{Sphere three-point function}

On the sphere, $W_N$ conformal blocks can also be computed directly from affine Toda theory, by taking the residue of affine Toda conformal blocks as the vertex operators approach those of the $W_N$ minimal model \cite{Fateev:2007ab}. This spares us the messy screening integrals in the Coulomb gas approach, and allows for easy extraction of explicit three-point functions. Our computation closely follows that of \cite{Fateev:2007ab}. 

\subsection{Two point function and normalization}
The two and three point functions in $W_N$ minimal model can be obtained from those of the affine Toda theory, as follows. The affine Toda theory is given by the $N-1$ bosons with linear dilaton described in the previous section, with an additional potential 
\ie
\mu  \sum_{i=1}^{N-1} e^{ b \A_i \cdot X}
\fe
added to the Lagrangian. Following the convention of \cite{Fateev:2007ab}, the background charge ${\cal Q}$ is related to $b$ by ${\cal Q}=(b+b^{-1})\rho$, where $\rho$ is the Weyl vector. Note that ${\cal Q}$ will be related to $Q$ in the previous section by ${\cal Q}=iQ$. Normally, one considers the affine Toda theory with real $b$ and ${\cal Q}$. To obtain correlators of $W_N$ minimal model, analytic continuation on $b$ as well as a residue procedure will be applied, as we will describe later.

The primary operators in the affine Toda theory are given by
\ie\label{todaop}
V_\bv=e^{\bv\cdot X}.
\fe
$V_\bv$ and $V_{w\cdot\bv}$ represent the same operator (recall that $w\cdot {\bf v}$ is the shift Weyl transformation of ${\bf v}$ by $w\in W$), but generally come with different normalizations. They are related by
\ie
V_\bv=R_{w}(\bv)V_{w\cdot\bv},
\fe
where $R_{w}(\bv)$ is the reflection amplitude computed in \cite{Fateev:2001mj}:
\ie
R_{w}(\bv)={{\bf A}(w\cdot \bv)\over {\bf A}(\bv)}={{\bf A}(w(\bv-\cQ)+\cQ)\over {\bf A}(\bv)},
\fe
and
\ie
{\bf A}(\bv)&=\left[\pi\m\gamma(b^2)\right]^{(\bv-\cQ,\rho)\over b}\prod_{i>j}\Gamma(1-b(\bv-\cQ,\bh_j-\bh_i))\Gamma(-b^{-1}(\bv-\cQ,\bh_j-\bh_i))
\\
&=\left[\pi\m{-1\over\gamma(-b^2)}b^{-4}\right]^{(\bv-\cQ,\rho)\over b}\prod_{i>j}\Gamma(1-b\bP_{ij})\Gamma(-b^{-1}\bP_{ij}),
\fe
where ${\bf P}_{ij}\equiv (\cQ-{\bf v})\cdot(\bh_i-\bh_j)$.
In particular, applying this for the longest Weyl group element $w_0$, we obtain the relation
\ie
V_{\bar\bv}={{\bf A}(2\cQ-\bv)\over {\bf A}(\bv)}V_{2\cQ-\bv},
\fe
where $\bar\bv$ is the conjugate of $\bv$. Notice that the function ${\bf A}(\bv)$ has the property ${\bf A}(\bv)={\bf A}(\bar\bv)$. The operators $V_{\bf v}$ are such that the two point function between $V_{\bf v}$ and $V_{2\cQ-\bv}$ is canonically normalized,
\ie
\vev{V_{\bv}(x)V_{2\cQ-\bv}(0)}={1\over |x|^{2\Delta_{\bv}}}.
\fe
It follows that that two point function of $V_\bv$ and its charge conjugate is
\ie
\vev{V_{\bv}(x)V_{\bar\bv}(0)}={{\bf A}(2\cQ-\bv)\over {\bf A}(\bv)}{1\over |x|^{2\Delta_{\bv}}}.
\fe

In the $W_N$ minimal model, by (\ref{sub}), we have a similar relation (by a slight abuse of notation, we now denote by $V_v$ the primary operator in the $W_N$ minimal model that descends from the corresponding exponential operator in the free boson theory)
\ie
V_v=R_{w}(v)V_{w\cdot v},
\fe
where
\ie
R_{w}(v)={A(w\cdot v)\over A(v)}={A(w(v-Q)+Q)\over A(v)},
\fe
and
\ie
A(v)=\left[\pi\m{-1\over\gamma({p'\over p})}\left({p\over p'}\right)^2\right]^{-\sqrt{p\over p'}(v-Q,\rho)}\prod_{i>j}\Gamma(1+\sqrt{p'\over p}P_{ij})\Gamma(-\sqrt{p\over p'}P_{ij}),
\fe
where $P_{ij}=(v-Q)\cdot(\bh_i-\bh_j)$. The two point function between $V_v$ and its charge conjugate is then
\ie
\vev{V_{v}(x)V_{\bar v}(0)}^{\rm unnorm}={A(2Q-v)\over A(v)}{1\over |x|^{2\Delta_{v}}}.
\fe
In computing this in the Coulomb gas formalism, appropriated screening charges are inserted, to saturate the background charge. Consequently, the vacuum isn't canonically normalized. In fact, we have
\ie
\vev{1}^{\rm unnorm}={A(2Q)\over A(0)}.
\fe
The normalized correlators are related by
\ie
\vev{V_1\cdots V_n}={\vev{V_1\cdots V_n}^{\rm unnorm}\over \vev{1}^{\rm unnorm}}={ A(0)\over A(2Q)}\vev{V_1\cdots V_n}^{\rm unnorm}.
\fe
Here again the ``unnormalized" $n$-point function is understood to be computed with appropriated screening charges inserted. Next, we define the normalized operators $\tV_v$ by
\ie
\tV_{v}=\sqrt{A(v)A(2Q)\over A(2Q-v)A(0)}V_v\equiv B(v)V_v,
\fe
and then we have
\ie
\vev{\tV_{v}(x)\tV_{\bar v}(0)}
%={A(v)A(2Q)\over A(2Q-v)A(0)}\vev{V_{v}V_{v^*}}={A(v)\over A(2Q-v)}\vev{V_{v}V_{v^*}}^{\rm unnorm}
={1\over |x|^{2\Delta_v}}.
\fe

\subsection{Extracting correlation functions from affine Toda theory}

Let us proceed to the three point functions in the $W_N$ minimal model:
\ie
\vev{V_{v_1}V_{v_2}V_{v_3}}^{\rm unnorm}={C_{W_N}(v_1,v_2,v_3)\over |x_{12}|^{\Delta_1+\Delta_2-\Delta_3}|x_{23}|^{\Delta_2+\Delta_3-\Delta_1}|x_{13}|^{\Delta_1+\Delta_3-\Delta_2}}.
\fe
where $\Delta_i$ denotes the total scaling dimension of $V_{v_i}$. The normalized three point functions of the normalized operators $\tV_{v_i}$ are given by
\ie
&\vev{\tV_{v_1}\tV_{v_2}\tV_{v_3}}
&=B\left(v_1\right)B\left(v_2\right)B\left(v_3\right)^{-1}\vev{V_{v_1}V_{v_2}V_{2Q-\bar v_3}}^{\rm unnorm},
\fe
and the structure constants, with two-point functions normalized to unity, are
\ie\label{Cnor}
C_{nor}(v_1,v_2,v_3)=B\left(v_1\right)B\left(v_2\right)B\left(v_3\right)^{-1}C_{W_N}(v_1,v_2,2Q-\bar v_3).
\fe
Nontrivial data are contained in the structure constants $C_{W_N}(v_1,v_2,v_3)$, which we now compute. 

In the affine Toda theory, the three point functions of the operators (\ref{todaop}) are of the form
\ie
\vev{V_{\bv_1}V_{\bv_2}V_{\bv_3}}={C_{\rm Toda}(\bv_1,\bv_2,\bv_3)\over |x_{12}|^{\Delta_1+\Delta_2-\Delta_3}|x_{23}|^{\Delta_2+\Delta_3-\Delta_1}|x_{13}|^{\Delta_1+\Delta_3-\Delta_2}}.
\fe
The structure constants $C_{\rm Toda}(\bv_1,\bv_2,\bv_3)$ are computed in \cite{Fateev:2007ab}. They have poles when the relation 
\ie\label{cce}
\bv_1+\bv_2+\bv_3+b\sum^{N-1}_{k=1}s_k\A_k+{1\over b}\sum^{N-1}_{k=1}s'_k\A_k=2\cQ
\fe
is obeyed, where $s_k$ and $s'_k$ are nonnegative integers. The pole structure is as follows. For general ${\bf v}_i$'s, define a charge vector $\epsilon=\sum\epsilon_i\A_i$ through the following equation
\ie\label{ccep}
\bv_1+\bv_2+\bv_3+b\sum^{N-1}_{k=1}s_k\A_k+{1\over b}\sum^{N-1}_{k=1}s'_k\A_k+\epsilon=2\cQ.
\fe
The relation (\ref{cce}) is obeyed when $\epsilon_i=0$, $i=1,\cdots,N-1$. This is an order $N-1$ pole of the structure constant $C_{\rm Toda}(\bv_1,\bv_2,\bv_3)$, understood as a function of $\epsilon$. The $W_N$ minimal model structure constant, $C_{W_N}(v_1,v_2,v_3)$, is computed by taking $N-1$ successive residues,\footnote{The residue (\ref{residue}) can also be computed using a Coulomb gas integral. See (1.24) of \cite{Fateev:2007ab}. }
\ie\label{residue}
{\rm res}_{\epsilon_1\to 0}{\rm res}_{\epsilon_2\to\epsilon_1}\cdots {\rm res}_{\epsilon_{N-1}\to\epsilon_{N-2}} C_{\rm Toda}(\bv_1,\bv_2,\bv_3),
\fe
and then analytically continuing to the following imaginary values of $b$ and ${\bf v}_i$,
\ie\label{sub}
&b=-i\sqrt{p'\over p},~~~~\bv_j=iv_j.
\fe
The relation (\ref{cce}) is always satisfied by the $v_i$'s obeying the $W_N$ fusion rules in some Weyl chamber.
The overall normalization of the three point function can be then fixed by requiring
\ie
C_{W_N}(0,0,2Q)=1.
\fe

In \cite{Fateev:2007ab}, by bootstrapping the sphere four point function, the following class of three point function coefficients were computed in the affine Toda theory:
\ie\label{Ctoda}
&C_{\rm Toda}(\bv_1,\bv_2,\varkappa\omega^{N-1})\\
&=\left[\pi\mu\gamma(b^2)b^{2-2b^2}\right]^{(2\cQ-\sum\bv_i,\rho)\over b}
{\left(\Upsilon(b)\right)^{N}\Upsilon(\varkappa)\prod\limits_{\A\in\Delta_+}\Upsilon\Bigl((\cQ-\bv_1)\cdot\A\Bigr)\Upsilon\Bigl((\cQ-\bv_2)\cdot\A\Bigr)\over\prod\limits_{i,j=1}^{N-1}\Upsilon\Bigl({\varkappa\over N}+(\bv_1-\cQ)\cdot\bh_i+(\bv_2-\cQ)\cdot\bh_j\Bigr)},
\fe
where $\varkappa$ is a real number, $\omega^{N-1}$ is the fundamental weight vector associated to the anti-fundamental representation, and the $\bh_k$'s are charge vectors defined as
\ie\label{bhdef}
\bh_k=\omega^1-\sum^{k-1}_{i=1}\A_i,
\fe
where $\omega^1$ is the first fundamental weight, associated with the fundamental representation.
The function $\Upsilon$ is defined by
\ie
\log\Upsilon(x)=\int^\infty_0{dt\over t}\left[\left({\cQ\over 2}-x\right)^2e^{-t}-{\sinh^2\left({\cQ\over 2}-x\right){t\over 2}\over \sinh{bt\over 2}\sinh{t\over 2b}}\right].
\fe
It obeys the identities,
\ie\label{rupsilon}
\Upsilon(x+b)&=\gamma(bx)b^{1-2bx}\Upsilon(x),
\\
\Upsilon(x+1/b)&=\gamma(x/b)b^{2x/b-1}\Upsilon(x),
\\
\Upsilon(x)&=\Upsilon(b+1/b-x),
\fe
and has zeros at $x=-nb-m/b$ and at $x=(1+n)b+(1+m)/b$, for nonnegative integers $n,m$.

The procedure of computing $C_{W_N}(v_1,v_2,v_3)$ from the residue of (\ref{Ctoda}), when $v_3$ is proportional to $\omega^{N-1}$, is carried out in Appendix A. The result is
\ie
&C_{W_N}\left(v_1,v_2,\left(\sqrt{p'\over p}n-\sqrt{p\over p'}m\right)\omega_{N-1}\right)
\\
&=\left({p'\over p}\right)^{\sum_{j=1}^{N-2}(s_j s'_{j+1}-s_{j+1}s'_{j})}\left[{-\m\pi\over\gamma({p'\over p})}\right]^{\sum\limits^{N-1}_{k=1}s_k}\left[{-\m'\pi\over\gamma({p\over p'})}\right]^{\sum\limits^{N-1}_{k=1}s'_k}\left(\prod_{k=0}^{s'_{N-1}-1}\prod^{s_{N-1}-1}_{l=0} {-1\over \left(\sqrt{p'\over p}(n-l)-\sqrt{p\over p'}(m-k)\right)^2}\right)
\\
&~~~~\times\left[\prod^{s_{N-1}-1}_{l=0}\gamma(1+m-{p'\over p}(n-l))\right]\left[\prod^{s'_{N-1}-1}_{k=0}\gamma(1+n-{p\over p'}(m-k))\right]\prod^{N-1}_{j=1}R^{s_{j,j-1},s'_{j,j-1}}_{j,0},
\fe
where $R^{s_{j,j-1},s'_{j,j-1}}_{j,0}$ is the $\epsilon=0$ value of
\ie
R^{s_{j,j-1},s'_{j,j-1}}_{j,\epsilon}=&\left(\prod_{k=1}^{s'_{j,j-1}}\prod^{s_{j,j-1}}_{l=1}{-1\over (\epsilon\cdot\bh_j+\sqrt{p\over p'}k-\sqrt{p'\over p}l)^2} \prod_{i=j+1}^N {1\over (P^1_{ij}-\sqrt{p\over p'}k+\sqrt{p'\over p}l)^2}{1\over (P^2_{ij}-\sqrt{p\over p'}k+\sqrt{p'\over p}l)^2}\right)
\\
&\times\left[\prod^{s_{j,j-1}}_{l=1}\gamma(\epsilon\cdot\bh_j+{p'\over p}l)\prod_{i=j+1}^N\gamma(\sqrt{p'\over p}P^1_{ij}+{p'\over p}l)\gamma(\sqrt{p'\over p}P^2_{ij}+{p'\over p}l)\right]
\\
&\times\left[\prod^{s'_{j,j-1}}_{k=1}\gamma(\epsilon\cdot\bh_j+{p\over p'}k)\prod_{i=j+1}^N\gamma(-\sqrt{p\over p'}P^1_{ij}+{p\over p'}k)\gamma(-\sqrt{p\over p'}P^2_{ij}+{p\over p'}k)\right].
\fe
$P^1_{ij}$ and $P^2_{ij}$ are defined as $P^r_{ij}=(v_r-Q)\cdot(\bh_i-\bh_j)$, $r=1,2$, and the function $\gamma(x)$ is defined as $\gamma(x)=\Gamma(x)/\Gamma(1-x)$. $\m'$ is the dual cosmological constant, which is related to the cosmological constant $\m$ by
\ie
\m'={1\over \pi\gamma\left(-{p\over p'}\right)}\left[\pi\m\gamma\left(-{p'\over p}\right)\right]^{-{p\over p'}}.
\fe

In the special case of $s'_i=0$ for all $i=1,\cdots,N-1$, the expressions simplify:
\ie
&C_{W_N}\left(v_1,v_2,\left(\sqrt{p'\over p}n-\sqrt{p\over p'}m\right)\omega_{N-1}\right)
\\
&=\left[{-\m\pi\over\gamma({p'\over p})}\right]^{\sum\limits^{N-1}_{k=1}s_k}\left[\prod^{s_{N-1}-1}_{l=0}\gamma(1+m-{p'\over p}(n-l))\right]\prod^{N-1}_{j=1}R^{s_{j,j-1},0}_{j,0},
\fe
and
\ie
R^{s_{j,j-1},0}_{j,0}=&\left[\prod^{s_{j,j-1}}_{l=1}\gamma({p'\over p}l)\prod_{i=j+1}^N\gamma(\sqrt{p'\over p}P^1_{ij}+{p'\over p}l)\gamma(\sqrt{p'\over p}P^2_{ij}+{p'\over p}l)\right].
\fe

\subsection{Large $N$ factorization}

In this section, we compute three point functions of $W_N$ primaries $({\bf f},0)$, $({\bf f},{\bf f})$, and/or their charge conjugates, with the primary $(\Lambda_+,\Lambda_-)$ where $\Lambda_\pm$ are the symmetric or antisymmetric tensor products of ${\bf f}$ or ${\bf \bar f}$. While the former are thought to be dual to elementary scalar fields in the bulk $AdS_3$ theory, the latter are expected to be composite particles, or bound states, of the former. If this interpretation is correct, then the three point functions in the large $N$ limit must factorize into products of two-point functions, as the bound states become unbound at zero bulk coupling. We will see that this is indeed the case. Our method can be carried out more generally to identify all elementary particles and their bound states in the bulk at large $N$, including the light states.

\subsubsection{Massive scalars and their bound states}

To begin with, let us consider the three point function of $({\bf \bar f},0)$, $({\bf \bar f},0)$, and $(A,0)$, where $A$ is the antisymmetric tensor product of two $\bf f$'s. Note that in the large $N$ limit, $({\bf f},0)$ has scaling dimension $\Delta_{(\bff,0)}=1+\lambda$, while $(A,0)$ has twice the dimension, and is expected to be the lowest bound state of two $({\bf f},0)$'s. The charge vectors are
\ie
v_1=v_2=\sqrt{p'\over p}\omega_{N-1},~~~~v_3=\sqrt{p'\over p}\omega_{2}.
\fe
%We compute:
%\ie
%&{ A(2Q-\sqrt{p'\over p}w_{N-1})\over A(\sqrt{p'\over p}w_{N-1})}{A(0)\over A(2Q)}
%\\
%&=\left[\pi\m{-1\over\gamma({p'\over p})}\left({p\over p'}\right)^2\right]^{(N-1)}\prod^N_{i=2}{-1\over (i-1)^2\left({p\over p'}-1\right)^2}{\gamma\left({p'\over p}i-i+1\right)\over\gamma\left(\left({p'\over p}-1\right)(i-1)\right)}
%\fe
%and
%\ie
%&{ A(2Q-\sqrt{p'\over p}w_2)\over A(\sqrt{p'\over p}w_2)}{A(0)\over A(2Q)}
%\\
%&=\left[\pi\m{-1\over\gamma({p'\over p})}\left({p\over p'}\right)^2\right]^{2(N-2)}\left[\prod^N_{i=3}{-1\over (i-1)^2\left({p\over p'}-1\right)^2}{\gamma\left({p'\over p}i-i+1\right)\over\gamma\left(\left({p'\over p}-1\right)(i-1)\right)}\right]
%\\
%&~~~~\times\left[\prod^N_{i=3}{-1\over (i-2)^2\left({p\over p'}-1\right)^2}{\gamma\left({p'\over p}(i-1)-i+2\right)\over\gamma\left(\left({p'\over p}-1\right)(i-2)\right)}\right]
%\\
%&=\left[\pi\m{-1\over\gamma({p'\over p})}\left({p\over p'}\right)^2\right]^{2(N-2)}\left[\prod^N_{i=3}{-\left({p'\over p}\right)^2\over (i-2)^2\left({p\over p'}-1\right)^2}{\gamma\left({p'\over p}i-i+1\right)\over\gamma\left(\left({p'\over p}-1\right)(i-2)\right)}\right]
%\fe
The structure constant, extracted using affine Toda theory, is
\ie
&
C_{W_N}\left(\sqrt{p'\over p}\omega_{N-1},\sqrt{p'\over p}\omega_{N-1},2Q-\sqrt{p'\over p}\omega_{N-2}\right)
=\left[{-\m\pi\over\gamma({p'\over p})}\right]\gamma\left(1-{p'\over p}\right)\gamma\left(2{p'\over p}-1\right).
\fe
By (\ref{Cnor}), the normalized structure constant are computed to be
\ie
C_{nor}
%&=\sqrt{2}\sqrt{-{(N-1)p^3\Gamma\left(-{p'\over p}\right)\Gamma\left(-2\left(1-{p'\over p}\right)\right)\Gamma\left((1-N)\left(1-{p'\over p}\right)\right)\Gamma\left(N\left(1-{p'\over p}\right)\right)\over N p'^3\Gamma\left(-1+{p'\over p}\right)\Gamma\left(2\left(1-{p'\over p}\right)\right)\Gamma\left((-1+N)\left(1-{p'\over p}\right)\right)\Gamma\left(-N\left(1-{p'\over p}\right)\right)}}
%\\
&=\sqrt{2} \left[ - {(1-{1\over N}) \Gamma(-\lambda) \Gamma({2\lambda\over N})\Gamma(\lambda-{\lambda\over N}) \Gamma(-1-{\lambda\over N}) \over (1+{\lambda\over N})^3 \Gamma(\lambda) \Gamma(-\lambda + {\lambda\over N}) \Gamma(-{2\lambda\over N}) \Gamma({\lambda\over N})} \right]^{1\over 2}
\\
&=\sqrt{2}-{2+4\lambda+\pi\lambda\cot\pi\lambda+2\lambda(\gamma+\psi(\lambda))\over \sqrt{2}N}+\cO({1\over N^2}),
\fe
where $\gamma$ is the Euler-Mascheroni constant, and the $\psi(\lambda)$ is the digamma function. 

In the infinite $N$ limit, the bulk theory is expected to become free. If we denote $({\bf f},0)$ by $\phi$, the OPE of $\phi$ should behave like that of a free field of dimension $\Delta_{(\bff,0)}$. Given the two-point function
\ie
\vev{\phi(x)\bar\phi(0)}={1\over |x|^{2\Delta_{(\bff,0)}}},
\fe
the product of two $\phi$'s, normalized as ${1\over \sqrt{2}}\phi^2$, has the two point function $1/|x|^{4\Delta_{(\bff,0)}}$. With the identification
\ie
(A,0) \sim {1\over \sqrt{2}}\phi^2,
\fe
i.e. $(A,0)$ as a bound state of two $\phi$'s that becomes free in the large $N$ limit, the three-point function coefficient is indeed $\sqrt{2}$, agreeing with the free correlator $\langle \bar\phi(x_1) \bar\phi(x_2)\, {1\over \sqrt{2}}:\phi^2(x_3):\rangle$.

\bigskip

\centerline{\begin{fmffile}{factortwo}
        \begin{tabular}{c}
            \begin{fmfgraph*}(90,90)
            \fmfi{plain}{fullcircle scaled .95w shifted (.5w,.5h)}
                \fmfleft{L}
                \fmfright{R}
                \fmffixed{(-.05w,0)}{vO,L}
                \fmffixed{(.19w,0.35h)}{vJ1,R}
                \fmffixed{(.19w,-0.35h)}{vJ2,R}
                %\fmffixed{(-.5w,0)}{v,L}
                \fmf{dashes}{vO,vJ2}
                \fmf{dashes}{vO,vJ1}
                \fmflabel{$(A,0)$}{vO}
                \fmflabel{$(\bar\bff,0)$}{vJ1}
                \fmflabel{$(\bar\bff,0)$}{vJ2}
                \end{fmfgraph*}
        \end{tabular}
        \end{fmffile}
}
%\bigskip
\noindent

The next example we consider is the three point function of two $({\bf \bar f},0)$'s and $(S,0)$, where $S$ is the symmetric tensor product of two $\bf f$'s. In the large $N$ limit, $(S,0)$ has dimension $2\Delta_{(\bff,0)}+2$, and may be expected to be an excited resonance of two $({\bf f},0)$'s. The charge vectors of the three primaries are
\ie
v_1=v_2=\sqrt{p'\over p}\omega_{N-1},~~~~v_3=\sqrt{p'\over p}\,2\omega_{1}.
\fe
The structure constant computed from Coulomb integral is very simple:
\ie
C_{W_N}\left(\sqrt{p'\over p}\omega_{N-1},\sqrt{p'\over p}\omega_{N-1},2Q-\sqrt{p'\over p}2\omega_{N-1}\right)=1,
\fe
and the normalized structure constant is
\ie\label{nornnn}
C_{nor}
%&=\sqrt{2\Gamma\left({-2{p'\over p}}\right)\Gamma\left(-1+{p'\over p}\right)\Gamma\left(N\left(1-{p'\over p}\right)\right)\Gamma\left(1-N+(1+N){p'\over p}\right)\over N\Gamma\left(-{p'\over p}\right)\Gamma\left(2{p'\over p}\right)\Gamma\left(-N\left(1-{p'\over p}\right)\right)\Gamma\left(N-(N+1){p'\over p}\right)},
%\\
&= \left[ {2\Gamma(-\lambda) \Gamma({\lambda\over N}) \Gamma(-2-{2\lambda\over N}) \Gamma(2+\lambda+{\lambda\over N}) \over N \Gamma(\lambda) \Gamma(-1-{\lambda\over N}) \Gamma(2+{2\lambda\over N}) \Gamma(-1-\lambda-{\lambda\over N}) } \right]^{1\over 2}
\\
&={1+\lambda\over \sqrt{2}}+{\lambda(1+\lambda)(-4+2\gamma+\psi(-1-\lambda)+\psi(2+\lambda))\over 2\sqrt{2}N}+\cO({1\over N^2}).
\fe
Let us compare $(S,0)$ with the primary that appears in the OPE of two free fields $\phi$'s at level $(1,1)$, with normalized two-point function,
\ie\label{resnn}
{1\over \sqrt{2} \Delta_{(\bff,0)}} (\phi \partial\bar\partial\phi - \partial\phi \bar\partial\phi).
\fe
The structure constant of (\ref{resnn}) with two $\bar\phi$'s is $\Delta_{(\bff,0)}/ \sqrt{2}$, precisely agreeing with (\ref{nornnn}) in the large $N$ limit, as $\Delta_{(\bff,0)}=1+\lambda$. This leads us to identify
\ie
(S,0) \sim {1\over \sqrt{2} \Delta_{(\bff,0)}} (\phi \partial\bar\partial\phi - \partial\phi \bar\partial\phi).
\fe

Next, we consider the three point function of $({\bf f},0)$, $(\bar{\bf f},0)$, and $(adj,0)$, where $adj$ is the adjoint representation of $SU(N)$. A similar computation gives\footnote{Here and from now on, we write $C_{nor}(v_1,v_2,v_3)$ in terms of the three pairs of representations rather than charge vectors.}
%\ie
%v_1=\sqrt{p'\over p}w_1,~~~~v_2=\sqrt{p'\over p}w_{N-1},~~~~v_3=\sqrt{p'\over p}(w_1+w_{N-1}).
%\fe
%Applying the formula (\ref{Cnor}), we obtain the structure constant:
\ie
&
%C_{nor}(\sqrt{p'\over p}\omega_1,\sqrt{p'\over p}\omega_{N-1},\sqrt{p'\over p}(\omega_1+\omega_{N-1}))
C_{nor}((\bff,0),(\bar\bff,0),(adj,0))
%=\sqrt{(N-1)p^2\Gamma\left(1-N+(N-1){p'\over p}\right)\Gamma\left(N-N{p'\over p}\right)\over N p'^2\Gamma\left((N-1)\left(1-{p'\over p}\right)\right)\Gamma\left(-N\left(1-{p'\over p}\right)\right)},
=\left[ {(1-{1\over N}) \Gamma(-\lambda) \Gamma(\lambda-{\lambda\over N}) \over (1+{\lambda\over N})^2 \Gamma(\lambda) \Gamma(-\lambda + {\lambda\over N}) } \right]^{1\over 2}
\\
&=1-{1+\lambda+{1\over 2}\pi\lambda\cot\pi\lambda-\lambda\psi(\lambda)\over N}+\cO({1\over N^2}).
\fe
This allows us to identify
\ie
(adj,0)\sim \phi\bar \phi,
\fe
in large $N$ limit.

As a simple check of our identification, we can compute the three point function of $(A,0)$, $(\overline S,0)$, and $(adj,0)$, which is expected to factorize into three two-point functions (i.e. $\sim \langle \phi\bar\phi\rangle^3$) in the large $N$ limit. Indeed, with the three charge vectors
\ie
v_1=\sqrt{p'\over p}\omega_2,~~~~v_2=\sqrt{p'\over p}2\omega_{N-1},~~~~v_3=\sqrt{p'\over p}(\omega_1+\omega_{N-1}),
\fe
%We compute
%\ie
%&{ A(2Q-\sqrt{p'\over p}w_1+w_{N-1}))\over A(\sqrt{p'\over p}(w_1+w_{N-1}))}{A(0)\over A(2Q)}
%\\
%&=\left[\pi\m{-1\over\gamma({p'\over p})}\left({p\over p'}\right)^2\right]^{2(N-1)}\left[\prod^{N-1}_{i=2}{-1\over (i-1)^2\left({p\over p'}-1\right)^2}{\gamma\left({p'\over p}i-i+1\right)\over\gamma\left(\left({p'\over p}-1\right)(i-1)\right)}\right]
%\\
%&~~~~\times\left[\prod_{j=2}^{N} {-1\over (N-j)^2(1-{p\over p'})^2}{\gamma\left({p'\over p}(N+1-j)-N+j\right)\over\gamma\left(\left({p'\over p}-1\right)(N-j)\right)}\right],
%\fe
we have
\ie
&C_{W_N}\left(\sqrt{p'\over p}\omega_2,\sqrt{p'\over p}2\omega_{N-1},2Q-\sqrt{p'\over p}(\omega_1+\omega_{N-1})\right)
\\
&=\left[{-\m\pi\over\gamma({p'\over p})}\right]^{N-2}\gamma(1-2{p'\over p})\gamma({p'\over p})\left[\prod_{i=3}^N\gamma\left(\left({p'\over p}-1\right)(2-i)\right)\gamma\left({p'\over p}(\delta_{i,N}-1+i)+(2-i)\right)\right],
\fe
and for the normalized structure constant,
\ie
&
%C_{nor}\left(\sqrt{p'\over p}\omega_2,\sqrt{p'\over p}2\omega_{N-1},\sqrt{p'\over p}(\omega_1+\omega_{N-1})\right)
C_{nor}((A,0),(\overline S,0),(adj,0))
%\\
%&=\sqrt{p^2\left(1-N+{p'\over p}\right)^3\Gamma\left(N-(N+1){p'\over p}\right)\Gamma\left(1-N\left(1-{p'\over p}\right)\right)\over p'^2\left(1-2{p'\over p}\right)^2\Gamma\left(-1+N-N{p'\over p}\right)\Gamma\left(1-N+(1+N)\left(1+{p'\over p}\right)\right)}
=\left[{N^4(1+\lambda)^3\Gamma(1+\lambda)\Gamma\left(-1+\lambda+{\lambda\over N}\right)\over (N+\lambda)^2(N+2\lambda)^2\Gamma(-1-\lambda)\Gamma\left(2+\lambda+{\lambda\over N}\right)}\right]^{1\over 2}
\\
&=(1+\lambda)-{\lambda(1+\lambda)(6+\psi(-1-\lambda)+\psi(2+\lambda))\over 2N}+\cO({1\over N^2}),
\fe
which is indeed reproduced in the large $N$ limit by the three point function of free field products ${1\over \sqrt{2}}\phi\phi$, ${1\over \sqrt{2} \Delta_{(\bff,0)}}(\bar\phi \partial\bar\partial \bar\phi - \partial\bar \phi \bar\partial\bar \phi)$, and $\phi\bar \phi$.

%\bigskip

\centerline{\begin{fmffile}{factorthree}
        \begin{tabular}{c}
            \begin{fmfgraph*}(90,90)
            \fmfi{plain}{fullcircle scaled .95w shifted (.5w,.5h)}
                \fmfleft{L}
                \fmfright{R}
                \fmffixed{(-.05w,0)}{vO,L}
                \fmffixed{(.19w,0.35h)}{vJ1,R}
                \fmffixed{(.19w,-0.35h)}{vJ2,R}
                %\fmffixed{(-.5w,0)}{v,L}
                \fmf{dashes}{vO,vJ1,vJ2,vO}
                \fmflabel{$(adj,0)$}{vO}
                \fmflabel{$(A,0)$}{vJ1}
                \fmflabel{$(\bar S,0)$}{vJ2}
                \end{fmfgraph*}
        \end{tabular}
        \end{fmffile}
}
%\bigskip
\noindent

\subsubsection{Light states}

The bound states of basic primaries discussed so far can be easily guessed by comparison the scaling dimensions in the large $N$ limit. This is less obvious with the light states, which are labeled by a pair of identical representations, i.e. of the form $(R,R)$.

To begin with, consider the light state $(\bff,\bff)$, whose dimension in the large $N$ limit is $\Delta_{(\bff,\bff)}=\lambda^2/N$. The OPE of two $(\bff,\bff)$'s contains $(A,A)$ and $(S,S)$, whose dimensions in the large $N$ limit are both $2\Delta_{(\bff,\bff)}$, as well as $(A,S)$ and $(S,A)$, whose dimensions are $2\Delta_{(\bff,\bff)}+2$. A linear combination of $(A,A)$ and $(S,S)$ is thus expected to be the lowest bound state of two $(\bff,\bff)$'s. This linear combination can be determined by inspecting the three-point functions of two $(\bar\bff,\bar\bff)$'s with $(A,A)$ and $(S,S)$.

The normalized structure constant of two $(\bar\bff,\bar\bff)$'s with $(A,A)$ is computed to be
%. The charge vector of the three primaries are
%\ie
%v_1=v_2=\left(\sqrt{p'\over p}-\sqrt{p\over p'}\right)w_{N-1},~~~~v_3=\left(\sqrt{p'\over p}-\sqrt{p\over p'}\right)w_{2}.
%\fe
%We compute
%\ie
%&{ A(2Q-\left(\sqrt{p'\over p}-\sqrt{p\over p'}\right)w_2)\over A(\left(\sqrt{p'\over p}-\sqrt{p\over p'}\right)w_2)}{A(0)\over A(2Q)}
%\\
%&=\left[\pi\m{-1\over\gamma({p'\over p})}\left({p\over p'}\right)^2\right]^{2\left(1-{p\over p'}\right)(N-2)}{N^2\gamma\left(\left({p'\over p}-1\right)N\right)\gamma\left(\left({p\over p'}-1\right)N\right)\over\gamma\left(\left({p'\over p}-1\right)\right)\gamma\left(\left({p\over p'}-1\right)\right)}
%\\
%&~~~~\times{(N-1)^2\gamma\left(\left({p'\over p}-1\right)(N-1)\right)\gamma\left(\left({p\over p'}-1\right)(N-1)\right)\over 4\gamma\left(\left({p'\over p}-1\right)2\right)\gamma\left(\left({p\over p'}-1\right)2\right)},
%\fe
%and
%\ie
%&C_{W_N}\left(\left(\sqrt{p'\over p}-\sqrt{p\over p'}\right)w_{N-1},\left(\sqrt{p'\over p}-\sqrt{p\over p'}\right)w_{N-1},2Q-\left(\sqrt{p'\over p}-\sqrt{p\over p'}\right)w_{N-2}\right)
%\\
%&=\left[{-\m\pi\over\gamma({p'\over p})}\left(p\over p'\right)^2\right]^{1-{p\over p'}}\left( {-1\over \left(\sqrt{p'\over p}-\sqrt{p\over p'}\right)^2}\right)\left[\gamma(2-{p'\over p})\right]\left[\gamma(2-{p\over p'})\right]
%\\
%&~~~~\times\left({-1\over 36(\sqrt{p\over p'}-\sqrt{p'\over p})^6} \right)\left[\gamma({p'\over p})\gamma(3{p'\over p}-2)\gamma(2{p'\over p}-1)\right]\left[\gamma({p\over p'})\gamma(-2+3{p\over p'})\gamma(-1+2{p\over p'})\right].
%\fe
%Applying the formula (\ref{Cnor}), we obtain the structure constant:
\ie
&C_{nor}((\bar\bff,\bar\bff),(\bar\bff,\bar\bff),(A,A))%&=\sqrt{-p'\Gamma\left(-1+{p\over p'}\right)\Gamma\left(-2+2{p\over p'}\right)\Gamma\left(-3+3{p\over p'}\right)^2\Gamma\left({p'\over p}\right)\Gamma\left(-2+2{p'\over p}\right)\Gamma\left(-3+3{p'\over p}\right)^2\over p\Gamma\left(2-{p'\over p}\right)\Gamma\left(2-2{p\over p'}\right)\Gamma\left(3-3{p\over p'}\right)^2\Gamma\left(-{p\over p'}\right)\Gamma\left(2-2{p'\over p}\right)\Gamma\left(3-3{p'\over p}\right)^2}
%\\
%&~~~~\times\sqrt{\Gamma\left(N-N{p\over p'}\right)\Gamma\left(2-N+(N-1){p\over p'}\right)\Gamma\left(1+N-N{p'\over p}\right)\Gamma\left(2-N+(N-1){p'\over p}\right)\over \Gamma\left(-N+N{p\over p'}\right)\Gamma\left(N-(N-1){p\over p'}\right)\Gamma\left(N-(N-1){p'\over p}\right)\Gamma\left(1-N+N{p'\over p}\right)}
%\\
=\left[{(N+\lambda)\Gamma\left(1-\lambda\right)\Gamma\left({2\lambda\over N}\right)\Gamma\left(3\lambda\over N\right)^2\Gamma\left(-3\lambda\over N+\lambda\right)^2\Gamma\left(-2\lambda\over N+\lambda\right)\over N\Gamma\left(-3\lambda\over N\right)^2\Gamma\left(-2\lambda\over N\right)\Gamma\left(1+\lambda\right)\Gamma\left(-N\over N+\lambda\right)\Gamma\left(2\lambda\over N+\lambda\right)}\right.
\\
&~~~~~~~\times\left.{\Gamma\left(-\lambda\over N+\lambda\right)\Gamma\left(N\lambda\over N+\lambda\right)\Gamma\left(N+\lambda\over N\right)\Gamma\left(1+\lambda-{\lambda\over N}\right)\Gamma\left(N+2\lambda-N\lambda\over N+\lambda\right)\over \Gamma\left(3\lambda\over N+\lambda\right)^2\Gamma\left(-N\lambda\over N+\lambda\right)\Gamma\left(N(1+\lambda)\over N+\lambda\right)\Gamma\left(N+\lambda-N\lambda\over N\right)\Gamma\left(N-\lambda\over N\right)}\right]^{1\over 2}
\\
&=1+{\lambda^2(-\pi\cot\pi\lambda+\pi^2\lambda\cot^2\pi\lambda-18\gamma-2\psi(\lambda)-2\lambda\psi^{(1)}(\lambda))\over 2N^2}+\cO({1\over N^3}),
\fe
and with $(S,S)$,
%. The charge vectors are
%\ie
%v_1=v_2=\left(\sqrt{p'\over p}-\sqrt{p\over p'}\right)w_{N-1},~~~~v_3=\left(\sqrt{p'\over p}-\sqrt{p\over p'}\right)2w_{1}.
%\fe
%We compute
%\ie
%&{A(2Q-\left(\sqrt{p'\over p}-\sqrt{p\over p'}\right)2w_{N-1})A(0)\over A(\left(\sqrt{p'\over p}-\sqrt{p\over p'}\right)2w_{N-1})A(2Q)}
%\\
%&=\left[\pi\m{-1\over\gamma({p'\over p})}\left({p\over p'}\right)^2\right]^{2\left(1-{p\over p'}\right)(N-1)}{N^2\gamma\left(\left({p'\over p}-1\right)N\right)\gamma\left(\left({p\over p'}-1\right)N\right)\over\gamma\left(\left({p'\over p}-1\right)\right)\gamma\left(\left({p\over p'}-1\right)\right)}
%\\
%&~~~~\times{(N+1)^2\gamma\left(\left({p'\over p}-1\right)(N+1)\right)\gamma\left(\left({p\over p'}-1\right)(N+1)\right)\over4\gamma\left(2\left({p'\over p}-1\right)\right)\gamma\left(2\left({p\over p'}-1\right)\right)}.
%\fe
%Applying the formula (\ref{Cnor}), we obtain the structure constant:
\ie
&C_{nor}((\bar\bff,\bar\bff),(\bar\bff,\bar\bff),(S,S))%&=\sqrt{2^{8-4{p'\over p}-4{p\over p'}}(N+1)^2\Gamma\left({3\over 2}-{p\over p'}\right)\Gamma\left({3\over 2}-{p'\over p}\right)\Gamma\left(1+N-N{p\over p'}\right)\Gamma\left(1+N-N{p'\over p}\right)\over N^2\Gamma\left(-{1\over 2}+{p\over p'}\right)\Gamma\left(-{1\over 2}+{p\over p'}\right)\Gamma\left(N\left(-1+{p\over p'}\right)\right)\Gamma\left(N\left(-1+{p\over p'}\right)\right)}
%\\
%&~~~~\times\sqrt{\Gamma\left(-(1+N)\left(1-{p'\over p}\right)\right)\Gamma\left(-(1+N)\left(-1+{p\over p'}\right)\right)\over \Gamma\left(2+N-(N+1){p\over p'}\right) \Gamma\left(2+N-(N+1){p'\over p}\right)},
%\\
\\
&=\left[{2^{-{4\lambda^2\over N^2+N\lambda}}(N+1)^2\Gamma\left(1-\lambda\right)\Gamma\left(\lambda+N\lambda\over N\right)\Gamma\left(-\lambda-N\lambda\over N+\lambda\right)\Gamma\left(N+3\lambda\over 2N+2\lambda\right)\Gamma\left({1\over 2}-{\lambda\over N}\right)\Gamma\left(N+\lambda+N\lambda\over N+\lambda\right)\over N^2\Gamma\left(\lambda\right)\Gamma\left(N-\lambda\over 2(N+\lambda)\right)\Gamma\left(-N\lambda\over N+\lambda\right)\Gamma\left({1\over 2}+{\lambda\over N}\right)\Gamma\left(N+2\lambda+N\lambda\over N+\lambda\right)\Gamma\left(N-\lambda-N\lambda\over N\right)}\right]^{1\over 2}
\\
&=1+{\lambda^2(\pi\cot\pi\lambda-\pi^2\lambda\csc^2\pi\lambda+2(\gamma+\psi(\lambda)+\lambda\psi^{(1)}(\lambda)))\over 2N^2}+\cO({1\over N^3}),
\fe
where $\psi^{(1)}(\lambda)$ is the trigamma function. We will denote the operator $(\bff,\bff)$ by $\omega$, and the lowest nontrivial operator in the OPE of two such light operators by $\omega^2$. Anticipating large $N$ factorization, if $\omega$ were a free field, then the product operator with correctly normalized two-point function is ${1\over \sqrt{2}}\omega^2$. The structure constant fusing two $\omega$'s into their bound state ${1\over \sqrt{2}}\omega^2$ is therefore $\sqrt{2}$ in the free limit. This is indeed the case: the three point function coefficient of two $(\bar\bff,\bar\bff)$'s and the linear combination ${1\over\sqrt{2}}((S,S)+(A,A))$ is
\ie
C_{nor}&=\sqrt{2}-{4\sqrt{2}\gamma\lambda^2\over N^2}+\cO({1\over N^3}).
\fe
This leads to the identification
\ie
{(S,S)+(A,A)\over\sqrt{2}}\sim {1\over \sqrt{2}}\omega^2.
\fe
The other linear combination
\ie
{(S,S)-(A,A)\over\sqrt{2}}
\fe
is orthogonal to $\omega^2$ and has vanishing three point function with two $(\bar\bff,\bar\bff)$'s in the large $N$ limit. It is therefore a new elementary light particle.

To identify the first excited composite state of two $(\bff,\bff)$'s as a linear combination of $(A,S)$ with $(S,A)$, we compute the structure constants
\ie\label{scAS}
&C_{nor}((\bar\bff,\bar\bff),(\bar\bff,\bar\bff),(A,S))%&=\sqrt{(N-1)\pi^2 p'^6 \csc2\pi {p'\over p}\csc\pi N\left(1-{p\over p'}\right)\Gamma\left(3-2{p\over p'}\right)\Gamma\left({p\over p'}\right)\Gamma\left(-{p'\over p}\right)\over N^2p^6\Gamma\left(N\left(-1+{p\over p'}\right)\right)^2\Gamma\left(-2+2{p\over p'}\right)\Gamma\left(1+N-(1+N){p\over p'}\right)\Gamma\left(-{p\over p'}\right)\Gamma\left(-1+{p'\over p}\right)}
%\\
%&~~~~\times \sqrt{\Gamma\left(-N+(1+N){p\over p'}\right)\Gamma\left(2-N+(N-1){p'\over p}\right)\Gamma\left(1+N-N{p'\over p}\right)\over \Gamma\left(5-2{p'\over p}\right)^2\Gamma\left(N\left(-1+{p'\over p}\right)\right)\Gamma\left(N+(1-N){p'\over p}\right)}
%\\
=\left[{\pi^2(N-1) (N + \lambda)^6 \csc{2 \pi \lambda\over N} \csc{
  N \pi \lambda\over N + \lambda} \Gamma (1 - \lambda)\Gamma \left(N\over N + \lambda\right) \Gamma \left({-N - \lambda\over N}\right) \over N^6 \Gamma \left(\lambda\right)  \Gamma \left(\lambda\over  N\right)  \Gamma \left(-N\over N + \lambda\right)  \Gamma \left(-2 \lambda \over N + \lambda\right)  \Gamma \left(-N \lambda\over N + \lambda\right)^2 \Gamma \left((1 + N) \lambda \over  N + \lambda\right)  }\right.
  \\
&~~~~~~\times\left.{\Gamma \left(N + 3 \lambda \over N + \lambda\right)  \Gamma \left( N - \lambda +N \lambda\over N\right)  \Gamma \left(N - N \lambda \over  N + \lambda\right) \over\Gamma \left(1 -\lambda+ {\lambda\over N} \right)  \Gamma \left( 3N - 2 \lambda\over N\right) ^2}\right]^{1\over 2}
\\
&={\lambda^2\over 2N}-{\lambda^2(1-3\lambda+\pi\lambda\cot\pi\lambda+2\lambda\gamma+2\lambda\psi(\lambda))\over 2N^3}+\cO({1\over N^4}),
\fe
%Next, we compute the structure constant of two $({\bf f},{\bf f})$'s and $(S,A)$. The charge vectors are
and
%\ie
%v_1=v_2=\left(\sqrt{p'\over p}-\sqrt{p\over p'}\right)w_{N-1},~~~~v_3=\sqrt{p'\over p}2w_{1}-\sqrt{p\over p'}w_{2}.
%\fe
%The structure constant in this case would be the structure constant (\ref{scAS}) with $p$ and $p'$ exchanged.
\ie
&C_{nor}((\bar\bff,\bar\bff),(\bar\bff,\bar\bff),(S,A))%&=\sqrt{(N-1)\pi^2 p^6 \csc2\pi {p\over p'}\csc\pi N\left(1-{p'\over p}\right)\Gamma\left(3-2{p'\over p}\right)\Gamma\left({p'\over p}\right)\Gamma\left(-{p\over p'}\right)\over N^2p'^6\Gamma\left(N\left(-1+{p'\over p}\right)\right)^2\Gamma\left(-2+2{p'\over p}\right)\Gamma\left(1+N-(1+N){p'\over p}\right)\Gamma\left(-{p'\over p}\right)\Gamma\left(-1+{p\over p'}\right)}
%\\
%&~~~~\times \sqrt{\Gamma\left(-N+(1+N){p'\over p}\right)\Gamma\left(2-N+(N-1){p\over p'}\right)\Gamma\left(1+N-N{p\over p'}\right)\over \Gamma\left(5-2{p\over p'}\right)^2\Gamma\left(N\left(-1+{p\over p'}\right)\right)\Gamma\left(N+(1-N){p\over p'}\right)}
%\\
=\left[{\pi^2(N-1)N^6\csc\pi\lambda\csc{2N\pi\over N+\lambda}\Gamma\left(-N\over N+\lambda\right)\Gamma\left(N+\lambda\over N\right)\Gamma\left(1-{2\lambda\over N}\right)\over (N+\lambda)^6\Gamma\left(\lambda\right)^2\Gamma\left(2\lambda\over N\right)\Gamma\left(-\lambda-N\lambda\over N\right)\Gamma\left(-\lambda\over N+\lambda\right)\Gamma\left(-N\lambda\over N+\lambda\right)}\right.
\\
&~~~~~~\times\left.{\Gamma\left(1+\lambda+{\lambda\over N}\right)\Gamma\left(N+2\lambda-N\lambda\over N+\lambda\right)\Gamma\left(N+\lambda+N\lambda\over N+\lambda\right)\over\Gamma\left(N(1+\lambda)\over N+\lambda\right)\Gamma\left(-N-\lambda\over N\right)\Gamma\left(3N+5\lambda\over N+\lambda\right)^2}\right]^{1\over 2}
\\
&={\lambda^2\over 2N}+{\lambda^2(1-5\lambda+\pi\lambda\cot\pi\lambda+2\lambda\gamma+2\lambda\psi(\lambda))\over 2N^3}+\cO({1\over N^4}).
\fe
Comparing its large $N$ limit with the free field products leads to the identification of ${1\over\sqrt{2}}((A,S)+(S,A))$ as the two-particle state,
\ie\label{abvv}
{(A,S)+(S,A)\over\sqrt{2}}\sim{1\over \sqrt{2} \Delta_{(\bff,\bff)}} (\omega \partial\bar\partial\omega - \partial\omega \bar\partial\omega).
\fe 
Note that the RHS of (\ref{abvv}) has the correctly normalized two-point function provided that the dimension of $\omega$ is $\Delta_{(\bff,\bff)}=\lambda^2/N$. 
The orthogonal linear combination ${1\over\sqrt{2}}((A,S)-(S,A))$ has vanishing three point function with two $(\bar{\bf f},\bar{\bf f})$'s at infinite $N$.

There is an important subtlety, pointed out in \cite{Papadodimas:2011pf}: while ${1\over\Delta_{(\bff,\bff)}}\partial\bar\partial\omega$ is a descendant of $\omega$, it is not truly an elementary particle. In fact, direct inspection of three-point functions at large $N$ shows that it should be identified with the bound state of $\phi=(\bff,0)$ and $\widetilde\phi=(0,\bff)$, i.e.
\ie\label{omegark}
{1\over\Delta_{(\bff,\bff)}}\partial\bar\partial\omega \sim \phi\,\widetilde\phi.
\fe
This is not in conflict with the statement that $\omega$ itself is an elementary particle, since in the large $N$ limit $\partial\bar\partial\omega$ (without the normalization factor $1/\Delta_{(\bff,\bff)}$) becomes null. With the identification (\ref{omegark}), we can also express (\ref{abvv}) as
\ie
{(A,S)+(S,A)\over\sqrt{2}}\sim{1\over \sqrt{2} }\left( \omega \phi \widetilde\phi -  {1\over \Delta_{(\bff,\bff)}} \partial\omega \bar\partial\omega\right).
\fe
In the next subsection, we will see a nontrivial consistency check of this identification.

\subsubsection{Light states bound to massive scalars}

So far we have seen that the massive elementary particles and the light particles interact weakly among themselves at large $N$. One can also see that the bound state between a massive scalar and a light state becomes free in the large $N$ limit. We will consider the example of $(\bff,0)$ and $(\bff,\bff)$ fusing into $(A,\bff)$ or $(S,\bff)$. At infinite $N$, the operators $(A,\bff)$ and $(S,\bff)$ have the same dimension as that of the basic primary $(\bff,0)$, namely $\Delta_{(\bff,0)}=1+\lambda$, and the light state $(\bff,\bff)$ has dimension zero. A linear combination of $(A,\bff)$ and $(S,\bff)$ should be identified with the lowest bound state of $(\bff,0)$ and $(\bar\bff,\bar\bff)$.
This is seen from the three point function coefficients
%Again, we find out the right combination by looking at the three-point functions. First, let us compute the three point function of $(\bar\bff,0)$, $(\bar\bff,\bar\bff)$, and $(A,\bff)$. The charge vector of the primaries are
%\ie
%v_1=\sqrt{p'\over p}w_{N-1},~~~~v_2=\left(\sqrt{p'\over p}-\sqrt{p\over p'}\right)w_{N-1},~~~~v_3=\sqrt{p'\over p}w_2-\sqrt{p\over p'}w_{1}.
%\fe
%The normalized structure constant is
\ie
&C_{nor}((\bar\bff,0),(\bar\bff,\bar\bff),(A,\bff))=\left[-{\pi(N-1)^2\csc  {2\pi\lambda\over N+\lambda}\csc{N\pi\lambda\over N+\lambda}\sin{N\pi\over N+\lambda}\Gamma\left({\lambda-N\lambda\over N+\lambda}\right)\Gamma\left(1+{\lambda\over N+\lambda}\right)^2\over N^2\Gamma\left(-N\lambda\over N+\lambda\right)^2\Gamma\left(N(1+\lambda)\over N+\lambda\right)\Gamma\left(N+3\lambda\over N+\lambda\right)^2}\right]^{1\over 2}
\\
&={1\over \sqrt{2}}+{\lambda\over 2\sqrt{2}N}(\pi\cot\pi\lambda+2\gamma+2\psi(\lambda))+\cO({1\over N^2}),
\fe
and
%We also compute the three point function of $(\bar\bff,0)$, $(\bar\bff,\bar\bff)$, and $(S,\bar\bff)$. The charge vector of the primaries are
%\ie
%v_1=\sqrt{p'\over p}w_{N-1},~~~~v_2=\left(\sqrt{p'\over p}-\sqrt{p\over p'}\right)w_{N-1},~~~~v_3=\sqrt{p'\over p}2w_1-\sqrt{p\over p'}w_{1}.
%\fe
%The normalized structure constant is
\ie
&C_{nor}((\bar\bff,0),(\bar\bff,\bar\bff),(S,\bff))=\left[{2^{-1+{4\lambda\over N+\lambda}}(N+1)\Gamma\left(N-N\lambda\over N+\lambda\right)\Gamma\left(N+\lambda+N\lambda\over N+\lambda\right)\Gamma\left({N+2\lambda\over 2(N+\lambda)}\right)\over N\Gamma\left(N+2\lambda+N\lambda\over N+\lambda\right)\Gamma\left(-\lambda\over 2(N+\lambda)\right)\Gamma\left(N+\lambda-N\lambda\over N+\lambda\right)}\right]^{1\over 2}
\\
&={1\over \sqrt{2}}-{\lambda\over 2\sqrt{2}N}(\pi\cot\pi\lambda+2\gamma+2\psi(\lambda))+\cO({1\over N^2}).
\fe
By comparing with the free field product of the elementary massive scalar $\phi$ with the light field $\omega$, we can identify
\ie
{(A,\bff)+(S,\bff)\over \sqrt{2}}\sim \phi\omega.
\fe
%i.e. bound state of $(\bff,0)$ and $(\bff,\bff)$ is $((A,\bff)+(S,\bff))/\sqrt{2}$, which with $(\bff,0)$ and $(\bff,\bff)$ has the same structure constant in the large $N$ limit as the structure constant of the operators $\phi$, $\omega$, and $\phi\omega$. 
The orthogonal linear combination ${1\over \sqrt{2}}((A,\bff)-(S,\bff))$ has vanishing three point function with $(\bff,0)$ and $(\bff,\bff)$ in the infinite $N$ limit. This is a new elementary particle, with the same mass as that of $\phi$ in the infinite $N$ limit.\footnote{We thank S. Raju for emphasizing this point. Note that on dimensional grounds, if ${1\over \sqrt{2}}((A,\bff)-(S,\bff))$ were a bound state, it could only be that of $(\bff,0)$ with a light state of the form $(R,R)$, but by fusion rule $R$ must be $\bff$, and we already know that ${1\over \sqrt{2}}((A,\bff)-(S,\bff))$ is orthogonal to the bound state of $(\bff,0)$ with $(\bff,\bff)$ in the large $N$ limit. }

One can further study the fusion of $(0,\bff)$ with $(A,\bff)$ into $(A,S)$, and the fusion of $(0,\bff)$ with $(S,\bff)$ into $(S,A)$. The normalized structure constants for both three-point functions are $1/\sqrt{2}$ in the infinite $N$ limit. In particular,
\ie\label{cnno}
C_{nor}\left( (0,\bar\bff) , {(\bar A,\bar\bff)+(\bar S,\bar\bff)\over \sqrt{2}} , {(A,S)+(S,A)\over\sqrt{2}}\right) = {1\over \sqrt{2}} + {\cal O}({1\over N}).
\fe
This is precisely consistent with the identifications
\ie
& (0,\bff)\sim {\widetilde\phi},~~~~ {(A,\bff)+(S,\bff)\over\sqrt{2}} \sim \phi\omega,
~~~~ {(A,S)+(S,A)\over\sqrt{2}}\sim{1\over \sqrt{2} }\left( \omega \phi \widetilde\phi -  {1\over \Delta_{(\bff,\bff)}} \partial\omega \bar\partial\omega\right).
\fe
The leading ${\cal O}(N^0)$ contribution to (\ref{cnno}) comes from the free field contraction of 
\ie
\left\langle \overline{\tilde\phi} ~ :\overline{\phi}\overline{\omega}: \; {:\omega\phi\widetilde\phi:\over \sqrt{2}} \right\rangle.
\fe
This is shown in the following (bulk) picture
\bigskip
\bigskip

\centerline{\begin{fmffile}{factorfour}
        \begin{tabular}{c}
            \begin{fmfgraph*}(110,110)
            \fmfi{plain}{fullcircle scaled .95w shifted (.5w,.5h)}
                \fmfleft{L}
                \fmfright{R}
                \fmffixed{(-.05w,0)}{vO,L}
                \fmffixed{(.19w,0.35h)}{vJ1,R}
                \fmffixed{(.19w,-0.35h)}{vJ2,R}
                \fmffixed{(.4w,-.18h)}{vO,p1}
                \fmffixed{(.48w,0.01h)}{vO,p2}
                \fmffixed{(.37w,0.26h)}{vO,p3}
                %\fmffixed{(-.5w,0)}{v,L}
                \fmf{dashes}{vO,vJ1}
                \fmf{dashes,right=.3}{vO,vJ2}
                \fmf{dbl_dots,left=.2}{vO,vJ2}
                \fmflabel{$\widetilde\phi$}{p1}
                \fmflabel{$\phi$}{p2}
                \fmflabel{$\omega$}{p3}
                \fmflabel{$(A,S)+(S,A)$}{vO}
                \fmflabel{$(0,\bar\bff)$}{vJ1}
                \fmflabel{$(\bar A,\bar\bff)+(\bar S,\bar\bff)$}{vJ2}
                \end{fmfgraph*}
        \end{tabular}
        \end{fmffile}
}
\bigskip
\noindent

As the last example of this section, let us also observe the following three-point function:
\ie\label{cnaag}
C_{nor}\left( (0,\bar\bff) , {(\bar A,\bar\bff)-(\bar S,\bar\bff)\over \sqrt{2}} , {(A,S)-(S,A)\over\sqrt{2}}\right) = {1\over \sqrt{2}} + {\cal O}({1\over N}).
\fe
As argued earlier, the operator ${1\over \sqrt{2}}((A,\bff)-(S,\bff))$ is an elementary particle state; denote it by $\Psi$. We have $\Delta_\Psi = \Delta_{(\bff,0)}$ in the large $N$ limit. Analogously, ${1\over \sqrt{2}}((\bff,A)-(\bff,S))=\widetilde\Psi$, with $\Delta_{\widetilde\Psi}=\Delta_{(0,\bff)}$ at large $N$. There is a similar three-point function, fusing $\phi=(\bff,0)$ and $\widetilde\Psi$ into ${1\over \sqrt{2}}((S,A)-(A,S))$. Combining this with (\ref{cnaag}), we conclude that ${1\over \sqrt{2}}((A,S)-(S,A))$ is a bound state of two elementary massive particles, namely
\ie
{(A,S)-(S,A)\over\sqrt{2}} \,\sim\, {\Psi \widetilde\phi - \widetilde\Psi \phi\over \sqrt{2}}.
\fe

\section{Sphere four-point function}

In the section, we investigate the sphere four-point function in the $W_N$ minimal model, of the primary operators $(\bff, 0)$, $(\bar \bff,0)$, with a general primary $(\Lambda_+, \Lambda_-)$ and its charge conjugate. The main purpose of this exercise is to set things up for the torus two-point function in section 6.
We consider two different approaches in computing the sphere four-point function: the Coulomb gas formalism, and null state differential equations. In subsections 5.1 through 5.3, we illustrate the screening charge contour integral and its relation with conformal blocks in various channels, primarily in the $N=3$ example, i.e. the $W_3$ minimal model. In this case, the conformal blocks are computed by a two-fold contour integral on a sphere with four punches. More generally, the conformal blocks in the $W_N$ minimal model are given by $(N-1)$-fold contour integrals. 
The identification of the correct contour for each conformal block, however, is not obvious for general $N$.
%However, because of lacking the general prescription on determining the contours for $W_N$ conformal blocks, the method describes in these two section cannot be directly generalized to the general $W_N$ minimal model. 
In subsection 5.4, we recall the null state differential equations of \cite{Fateev:2007ab}, which applies to all $W_N$ minimal models. The conformal blocks are given by the $N$ linearly independently solutions of the null state differential equation. One observes that the $N$ distinct $t$-channel conformal blocks (to be defined below) are permuted under the action of the Weyl group. This motivates an identification of the Coulomb gas screening integral contours for the $t$-channel conformal blocks for all values of $N$, which we describe in subsection 5.5. 
%In particular, we will compute one of the $t$-channel conformal blocks in the general $W_N$ minimal model using the Coulomb gas formalism, and applying the actions of the Weyl group, we generate all the other $N-1$ conformal blocks. 
The monodromy invariance of our four-point functions is shown in Appendix D.

\subsection{Screening charges}

Let us illustrate the screening charge integral in the $W_3$ minimal model.
Consider the sphere four-point function of the primary operators $(\bff, 0)$, $(\bar\bff,0)$, with a general primary $(\Lambda_+, \Lambda_-)$ and its charge conjugate. The highest weight vectors of $\bff$ and $\bar\bff$ are the two fundamental weights $\omega^1$ and $\omega^2$ of $SU(3)$. In the Coulomb gas approach, we first replace the four $W_3$ primaries by the corresponding chiral boson vertex operators $e^{iv_i\cdot X_L}$, $i=1,2,3,4$, where the charge vectors $v_i$ are taken to be
\ie
v_1 = \sqrt{p'\over p} \omega^1, ~~~~ v_2 = \sqrt{p'\over p}\omega^2,~~~~ v_3 = \sqrt{p'\over p}\Lambda_+ - \sqrt{p\over p'}\Lambda_-, ~~~~ v_4 = 2Q - v_3.
\fe
There is some freedom in choosing the charge vectors, since different charge vectors related by the shifted Weyl transformations are identified with the same $W$-algebra primary. For instance, here we have chosen $v_4$ to be $2Q-v_3$ rather than $\bar v_3=\sqrt{p'\over p}\bar\Lambda_+ - \sqrt{p\over p'}\bar\Lambda_-$. Indeed these two ways to represent the primary $(\bar\Lambda_+, \bar\Lambda_-)$ are related by the longest Weyl reflection, as explained at the end of section 3.
In terms of Dynkin labels, we write
\ie
\omega^1 = (1,0),~~~\omega^2 = (0,1),~~~ Q = -{1\over \sqrt{pp'}}(1,1),~~~\Lambda_+ = (n_+,m_+),
~~~\Lambda_- = (n_-,m_-),
\fe
where $n_\pm, m_\pm$ are nonnegative integers that obey $n_++m_+\leq k=p-3$, $n_-+m_-\leq k+1=p-2$. The two simple roots are $\A_1 = (2,-1)$, $\A_2 = (-1,2)$. The corresponding simple Weyl reflections $s_1, s_2$ act on the weight vector $(n,m)$ by
\ie
s_1(n,m)=(-n,n+m),~~~s_2(n,m) = (n+m,-m).
\fe
%We have also
%\ie
%\lambda = \sqrt{p'\over p} (n_++1,m_++1),~~~~\lambda'=\sqrt{p\over p'} (n_-+1,m_-+1).
%\fe

To compute the sphere four point function of the $W_N$ primaries, we must insert screening charges so that the total charge is $2Q$. In our example, a total screening charge $-v_1-v_2 = - \sqrt{p'\over p} (\A_1+\A_2)$ is inserted. This is done by inserting two screening operators, $V_1^-$ and $V_2^-$, both of which have conformal weight 1. So we expect
\ie
\langle {\cal O}_{v_1}(x_1){\cal O}_{v_2}(x_2){\cal O}_{v_3}(x_3){\cal O}_{v_4}(x_4)\rangle
= \int_{C} ds_1 ds_2 \langle V_{v_1}(x_1)V_{v_2}(x_2)V_{v_3}(x_3)V_{v_4}(x_4) V_1^-(s_1) V_2^-(s_2)\rangle,
\fe
for some appropriate choice of the contour $C$ for the $(s_1,s_2)$-integral. 
In fact, by choosing the appropriate contour $C$, we can pick out the three independent conformal blocks in this case. One may allow the contours to start and end on one of the $x_i$'s where the vertex operator is inserted, but we will demand the contours are closed on the four-punctured sphere.\footnote{Strictly speaking, due to the branch cuts connecting the vertex operators $V_{v_i}$, the contour $C$ lies on a covering Riemann surface of the punctured sphere.} This will allow for a straightforward generalization to the torus two-point function later.

Without loss of generality, we will choose $x_3=0$, $x_4=\infty$, while keeping $x_1, x_2$ two general points on the complex plane. Write $V'_{v_4}(\infty) = \lim_{x_4\to \infty} x_4^{2 h_{v_4}} V_{v_4}(x_4)$. The correlation function with screening operators is computed in the free boson theory (with linear dilaton) as
\ie\label{integrand}
&\langle V_{v_1}(x_1)V_{v_2}(x_2)V_{v_3}(0)V_{v_4}'(\infty) V_1^-(s_1) V_2^-(s_2)\rangle
\\
&= x_{12}^{v_1\cdot v_2} s_{12}^{{p'\over p}\A_1\cdot\A_2} \prod_{i=1}^2 x_i^{v_i\cdot v_3} s_i^{-\sqrt{p'\over p}v_3\cdot \A_i} \prod_{i,j=1}^2 (x_i-s_j)^{-\sqrt{p'\over p}v_i\cdot \A_j}
\\
&= x_1^{{p'\over p}({2\over 3}n_+ + {1\over 3}m_+) - ({2\over 3}n_-+{1\over 3}m_-)} x_2^{{p'\over p}({1\over 3}n_+ + {2\over 3}m_+) - ({1\over 3}n_-+{2\over 3}m_-)}
s_1^{-{p'\over p}n_+ + n_-} s_2^{-{p'\over p}m_+ + m_-}
\\
&~~~~\times x_{12}^{p'\over 3p}  s_{12}^{-{p'\over p}} (x_1-s_1)^{-{p'\over p}}(x_2-s_2)^{-{p'\over p}}.
\fe
Note that as a function in $s_1$, (\ref{integrand}) has branch points at $s_1=0,\infty, x_1, s_2$. As a function in $s_2$, it has branch points at $s_2=0,\infty, x_2,s_1$. The property that there are 4, rather than 5, branch points in each $s_i$, will be important in the construction of the contour $C$.

\subsection{Integration contours}

We will consider the following type of the two-dimensional integration contour $C$. First integrate $s_2$ along a contour $C_2(s_1)$ which depends on $s_1$, and then integrate $s_1$ along a contour $C_1$. $C_2(s_1)$ is chosen to avoid the four branch points $s_2=0,\infty, x_2, s_1$, and $C_1$ is chosen to avoid the branch points $0,\infty, x_1, x_2$ ($x_2$ will be a branch point in $s_1$ after the integration over $s_2$). To ensure that one comes back to the same sheet by going once around the contour, we demand that $C_1,C_2$ have no net winding number around any branch point.\footnote{This is sufficient because the monodromies involved are abelian.}

For $C_2(s_1)$ to be well defined the entire time as $s_1$ moves along $C_1$, we also demand the following property of $C_1$: upon removal of the $s_1$-branch point $x_1$, $C_1$ becomes contractible. Since $x_1$ is not a branch point of the $s_2$-integrand, this makes it possible to choose $C_2(s_1)$ to avoid all branch points of $s_2$ and comes back to itself as $s_1$ goes around $C_1$, ensuring that the full contour integral is well defined.

Let us denote by $L(z_1,z_2)$ the following contour that goes around two points $z_1, z_2$ on the complex plane:

\centerline{\begin{fmffile}{Lzz}
        \begin{tabular}{c}
            \begin{fmfgraph*}(120,90)
                \fmfleft{i1,i2,i3}
                \fmfright{o1,o2,o3}
                \fmf{phantom}{i2,t1,a2,v1,c,v2,b2,t2,o2}
                \fmf{phantom}{i3,a3,x,c,b1,u,o1}
                \fmf{phantom}{i1,v,a1,c,y,b3,o3}
                \fmf{plain,tension=0,left=.1}{x,c}
                \fmf{plain,tension=0,right=.1}{c,b1}
                \fmf{plain,tension=0,right=.1}{y,c}
                \fmf{plain,tension=0,left=.1}{c,a1}
                \fmf{plain_arrow,tension=0,right=.8}{b1,b2}
                \fmf{plain_arrow,tension=0,right=.8}{a2,a1}
                \fmf{plain,tension=0,left=.45}{a2,b2}
                \fmf{plain_arrow,tension=0,left=.12}{u,v}
                \fmf{plain,tension=0,right=1.2}{x,v}
                \fmf{plain,tension=0,left=1.2}{y,u}
                \fmfdot{v1,v2}
                \fmfv{label=$z_1$}{v1}
                \fmfv{label=$z_2$}{v2}
             \end{fmfgraph*}
        \end{tabular}
        \end{fmffile}
}

This contour is well defined when there are branch cuts coming out of $z_1$ and $z_2$, and the monodromies around $z_1$ and $z_2$ commute. It is also nontrivial only when $z_1$ and $z_2$ are both branch points. If we integrate (\ref{integrand}) along a contour $L(z_1,z_2)$ where $z_1,z_2$ are two of the branch points of the integrand, the contour may be collapsed to a line interval connecting $z_1$ and $z_2$, namely

\centerline{\begin{fmffile}{Szz}
        \begin{tabular}{c}
            \begin{fmfgraph*}(120,50)
                \fmfleft{i1}
                \fmfright{o1}
                \fmf{phantom}{i1,v1,a2,c,b2,v2,o1}
                \fmf{plain_arrow,tension=0}{v1,v2}
                \fmfdot{v1,v2}
                \fmfv{label=$z_1$}{v1}
                \fmfv{label=$z_2$}{v2}
             \end{fmfgraph*}
        \end{tabular}
        \end{fmffile}
}
\noindent in the following sense. 
Let $g_{z_1}$ and $g_{z_2}$ be the action by the monodromy around $z_1$ and $z_2$ respectively. Then we can write
\ie
\int_{L(z_1,z_2)} \cdots = (1-g_{z_2}+g_{z_1}g_{z_2}-g_{z_2}^{-1}g_{z_1}g_{z_2}) \int_{z_1}^{z_2}\cdots
\fe
where an appropriate branch is chosen for the integral from $z_1$ to $z_2$ on the RHS.

The two-dimensional contour $C$ will be constructed as follows: we first integrate $s_2$ along a contour $C_2(s_1)$ of the form $L(z_1,z_2)$, where $z_1,z_2$ are two out of the four branch points $0,\infty, x_2, s_1$, and then integrate $s_1$ along a contour $C_1$ that is of the form $L(x_1,z)$ (so that it becomes contractible upon removal of $x_1$). We must then investigate the transformation of the contour integral under the monodromies associated with $s$- and $t$-channel Dehn twists:
\ie
&T_s:~ x_1~ {\rm going~around}~ x_2, ~{\rm and}
\\
& T_t:~ x_1~ {\rm going~around}~ 0.
\fe
These are analyzed in detail in Appendix B. We only describe the results below.

Among the following four $L$-contours for the $s_2$-integral: $L(x_2,\infty)$, $L(0,s_1)$, $L(0,\infty)$, and $L(x_2,s_1)$, only two are linearly independent. In fact, the basis $(L(x_2,\infty), L(0,s_1))$ is convenient for analyzing $t$-channel monodromies, whereas the basis $(L(0,\infty), L(x_2,s_1))$ is convenient for analyzing $s$-channel monodromies. The linear transformation between the two basis is given by
\ie\label{lcont}
\begin{pmatrix} L(0,\infty) \\ L(x_2,\infty) \end{pmatrix} = \begin{pmatrix} {1-g_{s_1}g_{x_2}\over 1-g_{s_1}} & {1-g_{0}\over 1-g_{s_1}} \\  -g_{s_1} {1-g_{x_2}\over 1-g_{s_1}} & - {1-g_0g_{s_1}\over 1-g_{s_1}} \end{pmatrix}
\begin{pmatrix} L(0,s_1) \\ L(s_1,x_2) \end{pmatrix}.
\fe
Using the basis for the $s_2$-integral adapted to the $t$-channel, namely $(L(x_2,\infty), L(0,s_1))$, we may consider the following four candidates for the two-dimensional contour $C$,
\ie\label{cicontours}
&\int_{C^{(1)}} = \int_{L(x_1,x_2)} ds_1\int_{L(x_2,\infty)}ds_2, ~~~~\int_{C^{(2)}} = \int_{L(x_1,x_2)} ds_1\int_{L(0,s_1)}ds_2,\\
&\int_{C^{(3)}} = \int_{L(0,x_1)} ds_1\int_{L(x_2,\infty)}ds_2, ~~~~\int_{C^{(4)}} = \int_{L(0,x_1)} ds_1\int_{L(0,s_1)}ds_2.
\fe
These contours are shown in the figures below:

\bigskip
\bigskip
\bigskip

\centerline{
\begin{fmffile}{Cone}
        \begin{tabular}{c}
            \begin{fmfgraph*}(80,60)
                \fmfleft{i1,i2,i3,ex1,i4}
                \fmfright{o1,o2,o3,ex2,o4}
%                \fmf{phantom}{i1,zero,i4}
%                \fmf{phantom}{o1,x2,o4}
                \fmf{phantom}{i2,zero,v1,x1,v2,x2,o2}
                \fmf{phantom}{i4,a,b,c,inf,o4}
                \fmf{phantom}{i3,mm,o4}
                \fmfv{label=$C^{(1)}$}{mm}
                \fmf{plain,tension=0}{x2,inf}
                \fmf{dashes,tension=0}{x1,x2}
%                \fmf{plain,tension=0}{i1,i4,o4,o1,i1}
                \fmfdot{zero,x1,x2,inf}
                \fmfv{label=$0$}{zero}
                \fmfv{label=$x_1$}{x1}
                \fmfv{label=$x_2$}{x2}
                \fmfv{label=$\infty$}{inf}
             \end{fmfgraph*}
        \end{tabular}
        \end{fmffile}
\begin{fmffile}{Ctwo}
        \begin{tabular}{c}
            \begin{fmfgraph*}(80,60)
                \fmfleft{i1,i2,i3,ex1,i4}
                \fmfright{o1,o2,o3,ex2,o4}
%                \fmf{phantom}{i1,zero,i4}
%                \fmf{phantom}{o1,x2,o4}
                \fmf{phantom}{i2,zero,v1,x1,v2,x2,o2}
                \fmf{phantom}{i4,a,b,c,inf,o4}
                \fmf{phantom}{i3,mm,o4}
                \fmfv{label=$C^{(2)}$}{mm}
                \fmf{plain,left=.5,tension=0}{zero,v2}
                \fmf{dashes,tension=0}{x1,x2}
                \fmfdot{zero,x1,x2,inf}
                \fmfv{label=$0$}{zero}
                \fmfv{label=$x_1$}{x1}
                \fmfv{label=$x_2$}{x2}
                \fmfv{label=$\infty$}{inf}
             \end{fmfgraph*}
        \end{tabular}
        \end{fmffile}        
\begin{fmffile}{Cthree}
        \begin{tabular}{c}
            \begin{fmfgraph*}(80,60)
                \fmfleft{i1,i2,i3,ex1,i4}
                \fmfright{o1,o2,o3,ex2,o4}
%                \fmf{phantom}{i1,zero,i4}
%                \fmf{phantom}{o1,x2,o4}
                \fmf{phantom}{i2,zero,v1,x1,v2,x2,o2}
                \fmf{phantom}{i4,a,b,c,inf,o4}
                \fmf{phantom}{i3,mm,o4}
                \fmfv{label=$C^{(3)}$}{mm}
                \fmf{plain,tension=0}{x2,inf}
                \fmf{dashes,tension=0}{zero,x1}
                \fmfdot{zero,x1,x2,inf}
                \fmfv{label=$0$}{zero}
                \fmfv{label=$x_1$}{x1}
                \fmfv{label=$x_2$}{x2}
                \fmfv{label=$\infty$}{inf}
             \end{fmfgraph*}
        \end{tabular}
        \end{fmffile}
\begin{fmffile}{Cfour}
        \begin{tabular}{c}
            \begin{fmfgraph*}(80,60)
                \fmfleft{i1,i2,i3,ex1,i4}
                \fmfright{o1,o2,o3,ex2,o4}
%                \fmf{phantom}{i1,zero,i4}
%                \fmf{phantom}{o1,x2,o4}
                \fmf{phantom}{i2,zero,v1,x1,v2,x2,o2}
                \fmf{phantom}{i4,a,b,c,inf,o4}
                \fmf{phantom}{i3,mm,o4}
                \fmfv{label=$C^{(4)}$}{mm}
                \fmf{plain,left=.5,tension=0}{zero,v1}
                \fmf{dashes,tension=0}{zero,x1}
                \fmfdot{zero,x1,x2,inf}
                \fmfv{label=$0$}{zero}
                \fmfv{label=$x_1$}{x1}
                \fmfv{label=$x_2$}{x2}
                \fmfv{label=$\infty$}{inf}
             \end{fmfgraph*}
        \end{tabular}
        \end{fmffile}        
}
\bigskip
\noindent The solid lines represents the interval onto which the $s_2$-contour collapses (as opposed to the contour itself), whereas the dashed lines represent the corresponding collapsing interval of the $s_1$-contour.

We will denote the integral of (\ref{integrand}) along $C^{(i)}$ by ${\cal J}_i$, $i=1,2,3,4$.
The $t$-channel monodromy $T_t$ then acts on the basis vector $(\cJ_1,\cJ_2,\cJ_3,\cJ_4)$ by the matrix
\ie\label{mtaa}
M_t = g_0(x_1)\begin{pmatrix} ~{\bf 1}~~ & {\bf 1}-g_{x_2}(s_1) \\~ 0~~ & g_0(s_1)g_{x_1}(s_1) \end{pmatrix},
\fe
while the $s$-channel monodromy $T_s$ acts by the matrix
\ie\label{msaa}
M_s = g_{x_2}(x_1)\begin{pmatrix} g_{x_1}(s_1) g_{x_2}(s_1) & ~~0~ \\ g_{x_1}(s_1)-g_{x_1}(s_1) g_0(s_1)& ~~{\bf 1} ~\end{pmatrix}.
\fe
In both (\ref{mtaa}) and (\ref{msaa}), $g_x(z)$ denotes the $2\times 2$ monodromy matrix that acts on the $s_1$-integrand (after having done the $s_2$-integral) by taking the point $z$ around $x$. The explicit form of $g_0(x_1)$, $g_{x_2}(x_1)$, $g_0(s_1)$, $g_{x_1}(s_1)$, $g_{x_2}(s_1)$ are given in Appendix B.

\subsection{The conformal blocks for $N=3$}

While we have constructed four candidates for the two-dimensional contour $C$ (out of many possibilities), there are only three linearly independent conformal blocks for the four-point function considered in section 5.1. Indeed, only three out of the four ${\cal J}_i$'s are linearly independent, as shown in Appendix C. They are
\ie
& \begin{pmatrix} \widetilde \cJ_2 \\ \cJ_3 \\ \cJ_4 \end{pmatrix} =  \begin{pmatrix} \int_{L(x_1,x_2)}ds_1 \int_{L(s_1,x_2)}ds_2\cdots \\ \int_{L(0,x_1)}ds_1\int_{L(x_2,\infty)}ds_2\cdots \\  \int_{L(0,x_1)}ds_1\int_{L(0,s_1)}ds_2\cdots \end{pmatrix},
\fe
where the integrand $\cdots$ is given by (\ref{integrand}).

There are three $s$-channel conformal blocks, corresponding to fusing the $(\bff,0)$ and $(\bar\bff,0)$ into $(0,0)$, $(adj,0)$, and $(adj',0)$, where $adj$ stands for the adjoint representation of $SU(3)$, and $adj'$ refers to a second adjoint $W_3$-conformal block whose lowest weight channel is the $(W^3)_{-1}$ descendant of $(adj,0)$. We denote these conformal blocks by
\ie
{\cal F}^s = \left({\cal F}^s(0), {\cal F}^s(adj),{\cal F}^s(adj')\right).
\fe
The lowest conformal weights in these channels are (computed using (\ref{confdim}))
\ie
&h_{(\bff,0)} = h_{(\bar\bff,0)} = {N-1\over 2N}(1+{N+1\over N+k}) = {4p'\over 3p}-1,
\\
&h_{(adj,0)} =1+{N\over N+k} = {3p'\over p}-2 ,~~~h_{(adj',0)}={3p'\over p}-1.
\fe
By comparing the $s$-channel monodromies, one finds that ${\cal F}^s$ is expressed in terms of the contour integrals via the linear transformation
\ie
{\cal F}^s ={\cal A}_s \begin{pmatrix} \widetilde {\cal J}_2 \\ {\cal J}_3 \\{\cal J}_4
\end{pmatrix},~~~~{\cal A}_s =  \begin{pmatrix} 1 & 0 & 0 \\ -{\zeta^{2 - m_+ - n_+} (1 - \zeta^{2 + m_+)} (1 - \zeta^{n_+})\over (1 - \zeta)^2 (1 + \zeta) (1 + \zeta + \zeta^2)} & 1 & 0 \\ -{\zeta^{3 - 2 m_+ - n_+} (1 - \zeta^{m_+}) (1 - \zeta^{1 + m_+ + n_+})\over (1 - \zeta)^2 (1 + \zeta) (1 + \zeta + \zeta^2)}& 0& 1 \end{pmatrix},
\fe
where $\zeta\equiv e^{2\pi i p'/p}$.

Similarly, in the $t$-channel, there are three conformal blocks, associated with three distinct primaries $(\Lambda_+ + \omega^1,\Lambda_-)$, $(\Lambda_+ - \omega^1+\omega^2,\Lambda_-)$, and $(\Lambda_+ - \omega^2,\Lambda_-)$. The conformal blocks are denoted
\ie
{\cal F}^t = \left({\cal F}^t(\omega^1), {\cal F}^t(-\omega^1+\omega^2),{\cal F}^t(-\omega^2)\right).
\fe
The lowest conformal weights in the respective channels are
\ie
& h_{(\Lambda_++\omega^1,\Lambda_-)} = h_{(\Lambda_+,\Lambda_-)} + {p'\over p}({2\over 3}n_+ + {1\over 3}m_+ +{4\over 3}) - {2\over 3} n_- - {1\over 3}m_-  - 1 ,
\\
& h_{(\Lambda_+-\omega^1+\omega^2,\Lambda_-)} =  h_{(\Lambda_+,\Lambda_-)} + {p'\over p}(-{1\over 3}n_+ + {1\over 3}m_+ +{1\over 3}) + {1\over 3} n_- - {1\over 3}m_-  ,
\\
& h_{(\Lambda_+-\omega^2,\Lambda_-)} =  h_{(\Lambda_+,\Lambda_-)} + {p'\over p}(-{1\over 3}n_+ - {2\over 3}m_+ -{2\over 3}) + {1\over 3} n_- + {2\over 3}m_- + 1.
\fe
By comparing with the $t$-channel monodromy, we find that ${\cal F}^t$ is expressed in terms of the contour integrals as
\ie
{\cal F}^t = {\cal A}_t \begin{pmatrix} \widetilde \cJ_2 \\ \cJ_3 \\ \cJ_4
\end{pmatrix} ,~~~~{\cal A}_t=  \begin{pmatrix} 1 & -{(1 - \zeta)^2  (1 + \zeta)\zeta^{-1 + m_+ + n_+}\over (1 - \zeta^{1 + m_+}) (1 - \zeta^{1 + n_+})}& -{(1 - \zeta)^2 (1 + \zeta)\zeta^{-1 + 2 m_+ + n_+} \over(1 - \zeta^{1 + m_+}) (1 - \zeta^{2 + m_+ + n_+})}  \\ 0 & 1 & 0 \\ 0 & 0 & 1 \end{pmatrix}.
\fe
Finally, the four point function is obtained by summing over either the $s$-channel or the $t$-channel conformal blocks,
\ie\label{fmfrel}
&\langle {\cal O}_{v_1}(x_1,\bar x_1){\cal O}_{v_2}(x_2,\bar x_2){\cal O}_{v_3}(0){\cal O}_{v_4}'(\infty)\rangle
= ({\cal F}^s)^\dagger {\cal M}^s {\cal F}^s
= ({\cal F}^t)^\dagger {\cal M}^t {\cal F}^t.
\fe
Here ${\cal M}^s$ and ${\cal M}^t$ are ``mass" matrices, and obey
\ie\label{tschange}
({\cal A}^s)^\dagger {\cal M}^s {\cal A}^s
= ({\cal A}^t)^\dagger {\cal M}^t {\cal A}^t.
\fe
${\cal M}^t$ is diagonal, while ${\cal M}^s$ is only block diagonal a priori, since there are two adjoint conformal blocks in the $s$-channel. The mass matrices are computed explicitly in Appendix C, up to the overall normalization which can be fixed by the identity $s$-channel. In this way, the four point function is entirely determined.

\subsection{Null state differential equations}

In this section, we describe a different method of computing the sphere four-point function of the $W_N$ primaries $(\bff,0)$, $(\overline\bff,0)$ with $(\Lambda_+,\Lambda_-)$ and its charge conjugate, following \cite{Fateev:2007ab}. Analogously to section 5.1, now for general $N$, the four operator on the sphere are $\cO_{v_i}$ with the charge vectors $v_i$ given by
\ie\label{vis}
v_1 = \sqrt{p'\over p} \omega^1, ~~~~ v_2 = \sqrt{p'\over p}\omega^{N-1},~~~~ v_3 = v \equiv \sqrt{p'\over p}\Lambda_+ - \sqrt{p\over p'}\Lambda_-, ~~~~ v_4 = 2Q -v.
\fe
To compare with the formulae in section 3, we also write
\ie
u = \lambda+\lambda' = v-Q,
\fe
where $\lambda$ and $\lambda'$ lie in the lattices $\Gamma_{p/p'}^*$ and $\Gamma_{p'/p}^*$, and are defined modulo simultaneous shifts by lattice vectors of $\Gamma_{pp'}$ with the opposite signs.
As shown in \cite{Fateev:2007ab}, the primary states $(\bff,0)$ and $(\overline\bff,0)$ are complete degenerate. They obey a set of null state equations. For instance, in the $W_3$ minimal model, the vertex operators $\cO_{v_1}$ gives rise to the null states
\ie\label{nse}
&\left(W_{-1}-{3w\over 2\Delta}L_{-1}\right)\cO_{v_1}=0,
\\
&\left(W_{-2}-{12w\over \Delta(5\Delta+1)}L_{-1}^2+{6w(\Delta+1)\over \Delta(5\Delta+1)}L_{-2}\right)\cO_{v_1}=0,
\\
&\left(W_{-3}-{16w\over \Delta(\Delta-1)(5\Delta+1)}L_{-1}^3+{12w\over \Delta(5\Delta+1)}L_{-1}L_{-2}+{3w(\Delta-3)\over 2\Delta(5\Delta+1)}L_{-3}\right)\cO_{v_1}=0.
\fe
Here $\Delta$ and $w$ are the conformal weight and spin-3 charge of ${\cal O}_{v_1}$. Explicitly, they are given by
\ie
\Delta={4p'\over 3p}-1,~~~~w^2=-{2\Delta^2\over 27}{5p'-3p\over 3p-5p}.
\fe
Similar relations hold for $\cO_{v_2}$. Using the null state equations, one finds that in the $W_3$ minimal model the conformal blocks obey hypergeometric differential equation of $(3,2)$-type. 

The null state method applies straightforwardly to the $W_N$ minimal model with general $N$, and the conformal blocks therein obey the following hypergeometric differential equation of $(N,N-1)$-type:
\ie\label{hgeom}
&\bigg[x\prod^N_{k=1}\left(x{d\over dx}+{p'\over p}+\sqrt{p'\over p}P_{1,k}\right)-\prod^N_{k=1}\left(x{d\over dx}+\sqrt{p'\over p}P_{1,k}\right)\bigg]{\cal G}(x)=0,
\fe
where $x$ is the conformally invariant cross ratio of the four $x_i$'s, and $P_{i,j}$ are defined in terms of the charge vectors as
\ie
P_k=u\cdot \bh_k,~~~~P_{ij}=P_i-P_j.
\fe
The vectors ${\bf h}_k$ were defined in (\ref{bhdef}).
The solutions to (\ref{hgeom}) are
\ie\label{GF}
{\cal G}_k(x)=x^{\sqrt{p'\over p}P_{k,1}}{}_N F_{N-1}(\vec\m_k;\widehat{\vec\n}_k|x)\equiv x^{-\sqrt{p'\over p}P_{1}}G_k(x).
\fe
where $\vec\m_k$ and $\vec\n_k$ are the following $N$-dimensional vectors:
\ie\label{munuv}
&\vec\m_k=\sqrt{p'\over p}(P_{k,1},\cdots,P_{k,N})+{p'\over p}(1,\cdots,1),
\\
&\vec\n_k=\sqrt{p'\over p}(P_{k,1},\cdots,P_{k,N})+(1,\cdots,1),
\fe
and $\widehat{\vec\n}_k$ is the $(N-1)$-dimensional vector defined by dropping the $k$-th entry of $\vec\nu_k$. ${}_N F_{N-1}(a_1,\cdots, a_N;b_1,\cdots,b_{N-1}|x)$ is the generalized hypergeometric function.

One observes that, the action of shifted Weyl transformations on $v$ (or equivalently, ordinary Weyl transformation on $u$) permutes the $N$ $t$-channel conformal blocks. One may define a Weyl group action on $P_k$ as
\ie
w(P_k)=w(u)\cdot \bh_k=u\cdot w^{-1}(\bh_k).
\fe
The Weyl group acts as permutations on $\bh_k$, and hence permutes $P_k$ and $G_k(x)$ as well. Diagrammatically, the $t$-channel conformal blocks can be represented as 

\bigskip
\bigskip
\bigskip

\centerline{\begin{fmffile}{tchannel}
        \begin{tabular}{c}
            \begin{fmfgraph*}(180,75)
                \fmfleft{i1,i2}
                \fmfright{o1,o2}
                \fmf{plain,tension=1}{i1,v1}
                \fmf{plain,tension=1}{i2,v1}
                \fmf{plain,tension=1}{o1,v2}
                \fmf{plain,tension=1}{o2,v2}
                \fmf{plain,tension=0.3,label=$u+\sqrt{p'\over p}\bh_k$}{v1,v2}
                \fmfdot{v1,v2}
                \fmfv{label=$\sqrt{p'\over p}w^{1}-Q$}{i2}
                \fmfv{label=$u$}{i1}
                \fmfv{label=$\sqrt{p'\over p}w^{N-1}-Q$}{o2}
                \fmfv{label=$-u$}{o1}
             \end{fmfgraph*}
        \end{tabular}
        \end{fmffile}
}

\bigskip
\bigskip

\noindent The shifted Weyl transformation on $v$ permutes the diagrams with different internal lines.

In terms of the conformal blocks ${\cal G}_k(x)$ or $G_k(x)$, the four-point function is given by
\ie\label{fivethirty}
&\vev{\cO_{v_1}(x_1)\cO_{v_2}(x_2)\cO_{v_3}(0)\cO'_{v_4}(\infty)}
\\
&=
%|\infty|^{-2u^2+2Q^2}
|x_1-x_2|^{{2p'\over Np}} |x_1|^{2\sqrt{p'\over p}Q\cdot \bh_1} |x_2|^{-2\sqrt{p'\over p}Q\cdot \bh_{N}-2{p'\over p}}G\left({x_1\over x_2},{\bar x_1\over \bar x_2}\right).
\fe
where $G(x,\bar x)$ sums up the product of holomorphic and anti-holomorphic conformal blocks,
\ie\label{G}
G(x,\bar x)=\sum^N_{j=1}(\cM_{u})_{jj} G_j(x)G_j(\bar x).
\fe
$\cM_u$ is a diagonal ``mass matrix". We indicated here the explicit $u$-dependence of $\cM_u$, though $G_j(x)$ depend on $u$ as well. $\cM_{u}$ can be expressed in terms of the structure constants (three point function coefficients) via
\ie
(\cM_u)_{jj}%&=C^{u+Q+\sqrt{p'\over p} \bh_j}_{\sqrt{p'\over p}w^1,u+Q}C^{Q-u-\sqrt{p'\over p} \bh_j}_{\sqrt{p'\over p}w^{N-1},Q-u}
%\\
&=B\left(\sqrt{p'\over p}w^1\right)^2C_{W_N}\left(\sqrt{p'\over p}w^1,u+Q,Q-u-\sqrt{p'\over p} \bh_j\right)
\\
&~~~~\times C_{W_N}\left(Q+u+\sqrt{p'\over p} \bh_j,\sqrt{p'\over p}w^{N-1},Q-u\right)
\\
&=\gamma\left({p'\over p}\right)\gamma\left(N\left(1-{p'\over p}\right)\right)\prod^{N}_{i=1,i\neq j}{\gamma\left(\sqrt{p'\over p}P_{ij}\right)\gamma\left({p'\over p}-\sqrt{p'\over p}P_{ij}\right)}.
\fe
In deriving the last line, we used the results of $B$ and $C_{W_N}$ computed in section 4.
%\ie
%C^{Q-u-\sqrt{p'\over p} \bh_j}_{\sqrt{p'\over p}w^{N-1},Q-u}&=C_{W_N}\left(Q-u,Q+u+\sqrt{p'\over p} \bh_j,\sqrt{p'\over p}nw^{N-1}\right)
%\\
%&=\left(-{\pi\m\over \gamma({p'\over p})}\right)^{N-j}\prod^{N}_{i=j+1}{\gamma(\sqrt{p'\over p}P_{ij})\over \gamma(1-{p'\over p}+\sqrt{p'\over p}P_{ij})},
%\\
%C^{u+Q+\sqrt{p'\over p} \bh_j}_{\sqrt{p'\over p}w^1,u+Q}&=C_{W_N}\left(u+Q,Q-u-\sqrt{p'\over p} \bh_j,\sqrt{p'\over p}nw^{1}\right)
%\\
%&=\left(-{\pi\m\over \gamma({p'\over p})}\right)^{j-1}\prod^{j-1}_{i=1}{\gamma(\sqrt{p'\over p}P_{ij})\over \gamma(1-{p'\over p}+\sqrt{p'\over p}P_{ij})}.
%\fe
%\ie
%&C^{u+Q+\sqrt{p'\over p} \bh_j}_{\sqrt{p'\over p}w^1,u+Q}C^{Q-u-\sqrt{p'\over p} \bh_j}_{\sqrt{p'\over p}w^{N-1},Q-u}=\left(-{\pi\m\over \gamma({p'\over p})}\right)^{N-1}\prod^{N}_{i=1,i\neq j}{\gamma(\sqrt{p'\over p}P_{ij})\over \gamma(1-{p'\over p}+\sqrt{p'\over p}P_{ij})}
%\\
%&=\left(-{\pi\m\Gamma(1-{p'\over p})\over \Gamma({p'\over p})}\right)^{N-1}\prod^{N}_{i=1,i\neq j}{\Gamma(\sqrt{p'\over p}P_{ij})\Gamma({p'\over p}-\sqrt{p'\over p}P_{ij})\over \Gamma(1-{p'\over p}+\sqrt{p'\over p}P_{ij})\Gamma(1-\sqrt{p'\over p}P_{ij})},
%\fe
Note that, expectedly, the Weyl transformations on $u$ also permutes the $N$ diagonal entries of $\cM_{u}$. For later use, we also define
\ie\label{Cu}
C^2_u \equiv (\cM_u)_{N,N}=\gamma\left({p'\over p}\right)\gamma\left(N\left(1-{p'\over p}\right)\right)\prod^{N-1}_{i=1}\gamma\left(\sqrt{p'\over p}P_{i,N}\right) \gamma\left({p'\over p}-\sqrt{p'\over p}P_{i,N}\right).
\fe

\subsection{The contour for general $N$}

Let us return to the Coulomb gas formalism, and we are now ready to present a contour prescription for the four-point conformal blocks in $W_N$ minimal models with general $N$.
It may appear rather difficult to directly identify the $N$ contours that give precisely
the $N$ linearly independent conformal blocks. But once we find the contour that gives one of the $N$ $t$-channel conformal blocks, we can apply Weyl transformations on the charge vector $u$ and generate the remaining $N-1$ $t$-channel conformal blocks. 

The screening charge integral that computes the four point function, or rather, a conformal block, takes the form
\ie\label{contint}
\bG_u\left({x_1\over x_2}\right)= &x_2^{{p'\over p}-\sqrt{p'\over p}P_{N}}x_1^{\sqrt{p'\over p}P_{1}}\oint ds_1~s_1^{-\sqrt{p'\over p}(u+Q)\cdot \A_1}\left(x_1-s_1\right)^{-{p'\over p}}
\\
&\times\left(\prod_{i=1}^{N-2}\oint ds_{i+1}~s_{i+1}^{-\sqrt{p'\over p}(u+Q)\cdot \A_{i+1}}\left(s_{i}-s_{i+1}\right)^{-{p'\over p}}\right)\left(x_2-s_{N-1}\right)^{-{p'\over p}}
\fe
where $s_1,s_2,\cdots,s_{N-1}$ are integrated along the following choice of contour:
\ie
\prod_{i=1}^{N-1} \oint ds_i=\int_{L(0,x_1)} ds_1\int_{L(0,s_1)} ds_2\cdots \int_{L(0,s_{N-2})} ds_{N-1}.
\fe
Pictorially, this is represented as

\bigskip
\bigskip

\centerline{
\begin{fmffile}{Cgen}
        \begin{tabular}{c}
            \begin{fmfgraph*}(240,80)
                \fmfleft{i1,i2,i3,ex1,i4}
                \fmfright{o1,o2,o3,ex2,o4}
%                \fmf{phantom}{i1,zero,i4}
%                \fmf{phantom}{o1,x2,o4}
                \fmf{phantom}{i2,zero,v3,v2,v1,x1,x2,o2}
                \fmf{phantom}{i4,a,b,c,inf,o4}
                \fmf{dashes,left=.4,tension=0}{zero,v1}
                \fmf{dbl_dots,left=.6,tension=0}{zero,w2}
                \fmf{plain,left=.8,tension=0}{zero,w3}
                \fmf{dots,tension=0}{zero,x1}
                \fmffixed{(0,-.23h)}{w2,v2}
                \fmffixed{(0,-.38h)}{w3,v3}
                \fmfdot{zero,x1,x2,inf}
                \fmfv{label=$0$}{zero}
                \fmfv{label=$x_1$}{x1}
                \fmfv{label=$x_2$}{x2}
                \fmfv{label=$\infty$}{inf}
                \fmfv{label=$s_1$}{v1}
                \fmfv{label=$s_2$}{w2}
                \fmfv{label=$s_3$}{w3}
             \end{fmfgraph*}
        \end{tabular}
        \end{fmffile}        
}
%\bigskip

\noindent where the various lines represent the collapsing intervals of the $L$-contours of $s_1,s_2,s_3,\cdots$. In the $N=3$ case, this is the last contour of (\ref{cicontours}), denoted by $C^{(4)}$ in section 5.2.

The integral (\ref{contint}) can be computed by collapsing the prescribed contour to successive integrations over straight lines,
\ie
&\int_{L(0,x_1)} ds_1\int_{L(0,s_1)} ds_2\cdots \int_{L(0,s_{N-2})} ds_{N-1}=\cN_u\int^{x_1}_0 ds_1\int^{s_1}_0 ds_2\cdots\int^{s_{N-2}}_0 ds_{N-1},
\fe
where the factor $\cN_u$ is obtained by taking the differences of line integrals related by monodromies, similarly to the derivation in Appendix B. The result is
\ie
&\cN_u= \prod^{N-1}_{i=1}(1-g_{s_i})(1-g_{0,i}),
\\
&g_{s_i}=e^{-2\pi i {p'\over p}},
\\
&g_{0,N-i}=e^{-2\pi i \sqrt{p'\over p}u\cdot \sum^{i-1}_{j=1}\A_{N-j}-2\pi i \sqrt{p'\over p}(u+Q)\cdot \A_{N-i}}=e^{2\pi i(-\sqrt{p'\over p}P_{i,N}+{p'\over p})}.
\fe
The integral expression $\bG_u$ is related to the conformal block $G_N(x)$ derived in the previous subsection as
\ie
\bG_u\left({x_1\over x_2}\right)=&\cN_u x_2^{{p'\over p}-\sqrt{p'\over p}P_{N}}x_1^{\sqrt{p'\over p}P_{1}}\int^{x_1}_0 ds_1~s_1^{-\sqrt{p'\over p}(u+Q)\cdot \A_1}\left(x_1-s_1\right)^{-{p'\over p}}
\\
&\times\left(\prod_{i=1}^{N-2}\int^{s_{i}}_0 ds_{i+1}~s_{i+1}^{-\sqrt{p'\over p}(u+Q)\cdot \A_{i+1}}\left(s_{i}-s_{i+1}\right)^{-{p'\over p}}\right)\left(x_2-s_{N-1}\right)^{-{p'\over p}}
%\\
%=&x_2^{{p'\over p}-\sqrt{p'\over p}P_{N}}x_1^{\sqrt{p'\over p}P_{1}}\int^{x_1}_0 ds_1~s_1^{-\sqrt{p'\over p}(u+Q)\cdot \A_1-\sqrt{p'\over p}u\cdot\sum^{N-1}_{j=2}\A_j}\left(x_1-s_1\right)^{-{p'\over p}}
%\\
%&\times\left(\prod_{i=1}^{N-2}\int^{1}_0 d\xi_{i+1}~\xi_{i+1}^{-\sqrt{p'\over p}(u+Q)\cdot \A_{i+1}-\sqrt{p'\over p}u\cdot\sum^{N-1}_{j=i+2}\A_j}\left(1-\xi_{i+1}\right)^{-{p'\over p}}\right)\left(x_2-s_{N-1}\right)^{-{p'\over p}}
%\\
%=&x_2^{{p'\over p}-\sqrt{p'\over p}P_{N}}x_1^{\sqrt{p'\over p}P_{N}}\left(\prod_{i=1}^{N-1}\int^{1}_0 d\xi_{i}~\xi_{i}^{-\sqrt{p'\over p}(u+Q)\cdot \A_{i}-\sqrt{p'\over p}u\cdot\sum^{N-1}_{j=i+1}\A_j}\left(1-\xi_{i}\right)^{-{p'\over p}}\right)\left(x_2-x_1\prod^{N-1}_{i=1} \xi_{i}\right)^{-{p'\over p}}
%\\
%=&\cN_u x_2^{-\sqrt{p'\over p}P_{N}}x_1^{\sqrt{p'\over p}P_{N}}\left(\prod_{i=1}^{N-1}\int^{1}_0 d\xi_i~\xi_i^{-1+{p'\over p}-\sqrt{p'\over p}P_{i,N}}\left(1-\xi_i\right)^{-{p'\over p}}\right)\left(1-{x_1\over x_2}\prod^{N-1}_{i=1} \xi_{i}\right)^{-{p'\over p}}
%\\
%=&\cN_u x_2^{-\sqrt{p'\over p}P_{N}}x_1^{\sqrt{p'\over p}P_{N}} { \prod^{N}_{k=1}\Gamma(\sqrt{p'\over p}P_{N,k}+{p'\over p})\over \prod^{N}_{k=1}\Gamma(\sqrt{p'\over p}P_{N,k}+1)} {\Gamma(1-{p'\over p})^{N-1}\over \Gamma({p'\over p})} {}_{N}F_{N-1}(\vec\m_{N};\vec\n_N|{x_1\over x_2})
\\
=&\cN_u { \prod^{N-1}_{k=1}\Gamma(\sqrt{p'\over p}P_{N,k}+{p'\over p})\over \prod^{N-1}_{k=1}\Gamma(\sqrt{p'\over p}P_{N,k}+1)} {\Gamma(1-{p'\over p})^{N-1}} G_N\left({x_1\over x_2}\right)
\\
\equiv& \cN_u L_u G_N\left({x_1\over x_2}\right),
\fe
i.e. they differ only by the normalization constant ${\cal N}_u L_u$. Here we made use of the integral representation of the generalized hypergeometric function:
\ie
&{}_{N}F_{N-1}(a_1,\cdots,a_N;b_1,\cdots,b_{N-1}|x)
\\
&=\left(\prod^{N-1}_{k=1}{\Gamma(b_k)\over \Gamma(a_k)\Gamma(b_k-a_k)}\right)\int^1_0\cdots\int^1_0\prod^{N-1}_{k=1}\xi_k^{a_k-1}(1-\xi_k)^{b_k-a_k-1}\left(1-x\prod^{N-1}_{k=1}\xi_k\right)^{-a_N}d\xi_1\cdots d\xi_{N-1}.
\fe
Now we have obtained the $N$-th $t$-channel conformal block of section 5.4. To produce the other $t$-channel conformal blocks, we act the Weyl transformation on $u$, and obtain
\ie
G_i\left({x_1\over x_2}\right)=G_N\left({x_1\over x_2}\right)\bigg|_{u\rightarrow w(u)}=\cN_{w(u)}^{-1}L^{-1}_{w(u)}\bG_{w(u)}\left({x_1\over x_2}\right).
\fe
In terms of the contour integral $\bG_u(x)$, the four-point function (\ref{G}) can be written as
\ie\label{fourptcont}
G(x,\bar x)={1\over (N-1)!}\sum_{w\in W}|\cC_{w(u)}\bG_{w(u)}(x)|^2,
\fe
where we defined the normalization constant $\cC_u$ as
\ie\label{cudeff}
\cC_{u}=C_{u}L^{-1}_{u}\cN_{u}^{-1}.
\fe
A useful formula, derived using (\ref{Cu}), is
\ie\label{stcts}
C^2_{u}L^{-2}_{u}%=& \Gamma(1-{p'\over p})^{2-2N}\gamma\left({p'\over p}\right)\gamma\left(N\left(1-{p'\over p}\right)\right)
%\\
%&\times\prod^{N-1}_{k=1} { \Gamma(\sqrt{p'\over p}P_{N,k}+1)^2\over \Gamma(\sqrt{p'\over p}P_{N,k}+{p'\over p})^2}\gamma\left(\sqrt{p'\over p}P_{i,N}\right) \gamma\left({p'\over p}-\sqrt{p'\over p}P_{i,N}\right)
%\\
=&- \Gamma(1-{p'\over p})^{2-2N}\gamma\left({p'\over p}\right)\gamma\left(N\left(1-{p'\over p}\right)\right)
\prod^{N-1}_{k=1}  \csc\pi\sqrt{p'\over p}P_{k,N}\sin\pi\left(\sqrt{p'\over p}P_{k,N}-{p'\over p}\right).
\fe
The representation of the four-point function (\ref{fourptcont}) is the main result of this section. It may seen rather unnecessary given that we already know the relatively simple expression for the conformal blocks as generalized hypergeometric functions. But as discussed in the next section, our $t$-channel contour prescription allows for a straightforward generalization to torus two-point functions.

\section{Torus two-point function}

\subsection{Screening integral representation}

We now consider the torus two-point function of a fundamental primary and an anti-fundamental primary operator in the $W_N$ minimal model, ${\cal O}_{v_1}$ and ${\cal O}_{v_2}$. The relevant genus one conformal blocks will be constructed using free bosons on the Narain lattice $\Gamma^{N-1,N-1}$, with insertions of vertex operators $V_{v_1}$ and $V_{v_2}$, along with screening operators $V_1^-,V_2^-,\dots,V_{N-1}^-$. Note that the set of screening operators is the same as in the earlier computation of sphere four point function, now the total charge being 0 on the torus (as opposed to $2Q$ on the sphere).

Our starting point is the torus correlation function in the free boson theory with screening operators insertions,
\ie
& Z^{bos}_{\Gamma^{N-1,N-1}} \langle V_{v_1}(z_1) V_{v_2}(z_2) V_1^-(t_1)\cdots V_{N-1}^-(t_{N-1}) \rangle_\tau
\\
&= {1\over |\eta(\tau)|^{2N-2}} \left| {\theta_1(z_{12}|\tau)\over \partial_z\theta_1(0|\tau)} \right|^{2v_1\cdot v_2}
\left| {\theta_1(z_1-t_1|\tau)\over \partial_z\theta_1(0|\tau)} \right|^{-2{p'\over p}}
\left| {\theta_1(z_2-t_{N-1}|\tau)\over \partial_z\theta_1(0|\tau)} \right|^{-2{p'\over p}}
\prod^{N-2}_{i=1} \left| {\theta_1(t_{i,i+1}|\tau)\over \partial_z\theta_1(0|\tau)} \right|^{2{p'\over p}\A_i\cdot \A_{i+1}}
\\
&~~~~\times \sum_{(v,\bar v)\in \Gamma^{N-1,N-1}} q^{{1\over 2}v^2}\bar q^{{1\over 2}\bar v^2}
\exp\left[2\pi i \left( v\cdot (v_1z_1+v_2z_2-\sqrt{p'\over p}\sum_{i=1}^{N-1}\A_i t_i) \right.\right.
\\
&\left.\left.~~~~~~~~~~~~~~~~~~~~~~~~~~~~~~~~~~~~~~~~~~~~~~~~- \bar v\cdot (v_1 \bar z_1+v_2\bar z_2-\sqrt{p'\over p}\sum_{i=1}^{N-1}\A_i \bar t_i)\right)\right]
\\
&= \sum_{u\in \Gamma_{pp'}^*/\Gamma_{pp'}} \left|G^{bos}_u(z_1,z_2,t_1,\cdots,t_{N-1}|\tau)\right|^2.
\fe
Our convention is that the coordinate $z$ on the torus of modulus $\tau$ is identified under $z\sim z+1\sim z+\tau$. The lattice $\Gamma^{N-1,N-1}$ is defined as in (\ref{gammann}).
$G^{bos}_u$ is a genus one character of the free boson with $N+1$ vertex operator insertions,
\ie\label{bostpf}
& G^{bos}_u(z_1,z_2,t_1,\cdots,t_{N-1}|\tau) 
\\
&= 
{1\over \eta(\tau)^{N-1}} \left( {\theta_1(z_{12}|\tau)\over \partial_z\theta_1(0|\tau)} \right)^{{p'\over pN}}
\left( {\theta_1(z_1-t_1|\tau)\over \partial_z\theta_1(0|\tau)} \right)^{-{p'\over p}}
\left( {\theta_1(z_2-t_{N-1}|\tau)\over \partial_z\theta_1(0|\tau)} \right)^{-{p'\over p}}
 \prod_{i=1}^{N-2}\left( {\theta_1(t_{i,i+1}|\tau)\over \partial_z\theta_1(0|\tau)} \right)^{-{p'\over p}}
\\
&~~~\times \sum_{n\in\Gamma_{pp'}} q^{{1\over 2}(u+n)^2} \exp\left[2\pi i\sqrt{p'\over p} \left((u+n )\cdot (\omega_1 z_1+\omega_{N-1}z_2-\A_i t_i) \right)\right].
\fe
%As seen previously, $u$ can be represented as $u=\lambda+\lambda'$, with double Weyl group $W\times W$ acting on $(\lambda,\lambda')$. The $W_N$ primaries are labeled by the non-degenerate orbits of $W\times W$. 
%Let
%\ie
%G^{W_3}_u(x_1,x_2,t_1,t_2|\tau) = \sum_{w\in W} \epsilon(w) G^{bos}_{w(\lambda)+\lambda'}(x_1,x_2,t_1,t_2|\tau).
%\fe
Recall that in the formula for the $W_N$ minimal character (\ref{WNcharacter}), an alternating sum over Weyl orbits is perfomed in order to cancel the contribution from null states in the conformal family of $u=\lambda+\lambda'$ at the level $h_{w(\lambda)+\lambda'}-h_{\lambda+\lambda'}$ and higher. A similar procedure is applied here to produce the correct minimal $W_N$ torus correlation function. A $t$-channel conformal block for the torus two-point function can be represented by the following diagram:

\bigskip
\bigskip
\bigskip

\centerline{\begin{fmffile}{torustchannel}
        \begin{tabular}{c}
            \begin{fmfgraph*}(150,85)
                \fmfleft{i1}
                \fmfright{o1}
                \fmf{plain,tension=1}{i1,v1}
                \fmf{plain,tension=1}{o1,v2}
                \fmf{plain,tension=0.25,left=1,label=$\lambda+\lambda'+\sqrt{p'\over p}\bh_k$}{v1,v2}
                \fmf{plain,tension=0.25,right=1,label=$\lambda+\lambda'$}{v1,v2}
                \fmfdot{v1,v2}
                \fmfv{label=$\sqrt{p'\over p}w^{1}-Q$}{i1}
                \fmfv{label=$\sqrt{p'\over p}w^{N-1}-Q$}{o1}
             \end{fmfgraph*}
        \end{tabular}
        \end{fmffile}
}
\bigskip
\bigskip

\noindent On the lower arc, there are null states at the level $h_{\lambda+w(\lambda')}-h_{\lambda+\lambda'}$ that are included by the free boson character. On the upper arc, there are null states at the level\footnote{Similar to (\ref{nsl}), one can show that $h_{\lambda+w(\lambda')+\sqrt{p'\over p}\bh_k}-h_{\lambda+\lambda'+\sqrt{p'\over p}\bh_k}$ is always a nonnegative integer, when $\lambda+\sqrt{p'\over p}\bh_k$ and $\lambda'$ sit in the identity affine Weyl chamber of $\Gamma_{p\over p'}^*$ and $\Gamma_{p'\over p}^*$.} $h_{\lambda+w(\lambda')+\sqrt{p'\over p}\bh_k}-h_{\lambda+\lambda'+\sqrt{p'\over p}\bh_k}$.
%\ie
%h_{\lambda+w(\lambda')+\sqrt{p'\over p}h_k}-h_{\lambda+\lambda'+\sqrt{p'\over p}h_k}=(\A_w\cdot\lambda)(-\A_w\cdot\lambda')+\sqrt{p'\over p}(-\A_w\cdot\lambda')(\A_w\cdot h_k),
%\fe
To cancel the contribution from these null states, we consider the alternating sum:\footnote{The reason that we are summing over the Weyl orbits of $\lambda'$ (rather than, say $\lambda$) has to do with the inserted vertex operator being $(\bff,0)$ rather than $(0,\bff)$.
%One may think that we also have to alternatingly sum over the orbit of the Weyl action on $\lambda$. However, the number $h_{w(\lambda)+\lambda'+\sqrt{p'\over p}h_k}-h_{\lambda+\lambda'+\sqrt{p'\over p}h_k}$ is in general not an integer. 
%\ie
%h_{w(\lambda)+\lambda'+\sqrt{p'\over p}h_k}=h_{\lambda+\lambda'+\sqrt{p'\over p}h_k}+(\A_w\cdot\lambda)(-\A_w\cdot\lambda')-\sqrt{p'\over p}(\A_w\cdot\lambda)(\A_w\cdot h_k),
%\fe
Also note that normalization factors involving the structure constants, e.g. (\ref{stcts}) are needed to obtain the full correlator. 
%One may also think that we need to perform this sum after doing the contour integral and correctly includes also the structure constants, but one can show that the contour integral commutes with the alternating sum (\ref{alttsum}), and the term 
In fact, (\ref{stcts}) is invariant under the Weyl transformation acting on $\lambda'$, i.e. $C_{\lambda+w(\lambda')}L^{-1}_{\lambda+w(\lambda')}=C_{\lambda+\lambda'}L^{-1}_{\lambda+\lambda'}$. This is consistent with the $W_N$ primaries being labelled by $u=\lambda+\lambda'$ up to the double Weyl action.}
\ie\label{alttsum}
\cG^{bos}_{\lambda+\lambda'}(z_1,z_2,t_1,\cdots,t_{N-1}|\tau)=\sum_{w\in W}\epsilon(w)G^{bos}_{\lambda+w(\lambda')}(z_1,z_2,t_1,\cdots,t_{N-1}|\tau).
\fe
Next, we integrate the positions $t_{i}$ of the screening operators on an $(N-1)$-dimensional contour. Different appropriate contour choices may give different conformal blocks, say in the $t$-channel or $s$-channel. 

\bigskip

\centerline{\begin{fmffile}{Ts}
        \begin{tabular}{c}
            \begin{fmfgraph*}(120,85)
                \fmfleft{i1}
                \fmfright{o1}
                \fmf{plain,tension=1}{i1,v1}
                \fmf{plain,tension=1}{o1,v2}
                \fmf{plain,tension=0.25,left=1}{v1,v2}
                \fmf{plain,tension=0.25,right=1}{v1,v2}
                \fmfdot{v1,v2}
%                \fmfv{label=$\sqrt{p'\over p}w^{1}-Q$}{i1}
%                \fmfv{label=$\sqrt{p'\over p}w^{N-1}-Q$}{o1}
             \end{fmfgraph*}
        \end{tabular}
        \end{fmffile}
        ~~~~~~~~
        \begin{fmffile}{Tt}
        \begin{tabular}{c}
            \begin{fmfgraph*}(100,100)
                \fmfleft{i1,i2,i3,i4}
                \fmfright{o1,o2,o3,o4}
                \fmf{phantom}{i2,v1,z,v2,o2}
                \fmf{phantom}{i4,w1,z1,w2,o4}
                \fmf{plain,tension=0.1}{w1,x,w2}
                \fmf{plain,tension=0.3}{x,y}
                \fmffixed{(0,-.22h)}{y,z}
                \fmf{plain,tension=0,left=1}{v1,v2}
                \fmf{plain,tension=0,right=1}{v1,v2}
                \fmfdot{x,y}
%                \fmfv{label=$\sqrt{p'\over p}w^{1}-Q$}{i1}
%                \fmfv{label=$\sqrt{p'\over p}w^{N-1}-Q$}{o1}
             \end{fmfgraph*}
        \end{tabular}
        \end{fmffile}
}
\centerline{~~$t$-channel~~~~~~~~~~~~~~~~~~~~~~~~~~~~~~~~$s$-channel}
\bigskip

\noindent As in the case of sphere four-point function, we will construct the integration contour by composing one-dimensional contours with no net winding numbers, which ensures that the integral is well defined despite the branch cuts in the integrand. To go from the four-punctured sphere to the two-punctured torus, we can simply cut out holes around the points $0$ and $\infty$ on the complex plane, and glue the two boundaries of resulting annulus to form the torus. The annulus coordinate $x$ to the torus coordinate $z$ are related by the exponential map $x=e^{2\pi i z}$. The $L$-contours introduced in section 5.2 are closed contours that avoids the branch cuts including $0$ and $\infty$, and thus are readily extended to the case of the torus under the exponential map. In particular, the part of the contour that winds around $0$ or $\infty$ now winds around cycles of the torus.

\centerline{\begin{fmffile}{zerox}
        \begin{tabular}{c}
            \begin{fmfgraph*}(120,90)
                \fmfleft{i1,i2,i3}
                \fmfright{o1,o2,o3}
                \fmf{phantom}{i2,t1,a2,v1,c,v2,b2,t2,o2}
                \fmf{phantom}{i3,a3,x,c,b1,u,o1}
                \fmf{phantom}{i1,v,a1,c,y,b3,o3}
                \fmf{plain,tension=0,left=.1}{x,c}
                \fmf{plain,tension=0,right=.1}{c,b1}
                \fmf{plain,tension=0,right=.1}{y,c}
                \fmf{plain,tension=0,left=.1}{c,a1}
                \fmf{plain_arrow,tension=0,right=.8}{b1,b2}
                \fmf{plain_arrow,tension=0,right=.8}{a2,a1}
                \fmf{plain,tension=0,left=.45}{a2,b2}
                \fmf{plain_arrow,tension=0,left=.12}{u,v}
                \fmf{plain,tension=0,right=1.2}{x,v}
                \fmf{plain,tension=0,left=1.2}{y,u}
                \fmfdot{v1,v2}
                \fmfv{label=$0$}{v1}
                \fmfv{label=$x$}{v2}
             \end{fmfgraph*}
        \end{tabular}
        \end{fmffile}
        ~~~~~~$\Longrightarrow$~~~~~~~
        \begin{fmffile}{torusx}
        \begin{tabular}{c}
            \begin{fmfgraph*}(80,150)
                \fmfstraight
                \fmfleft{i1,i2,i3,i4,i5}
                \fmfright{o1,o2,o3,o4,o5}
                \fmf{plain,tension=0}{i1,i2,i3,i4}
                \fmf{plain,tension=0}{o1,o2,o3,o4}
                \fmf{plain,left=.3}{i4,o4}
                \fmf{plain,right=.3}{i4,o4}
                \fmf{phantom}{i2,a,v1,x,v2,b,o2}
                \fmf{phantom}{i2,s1,i3}
                \fmf{phantom}{o2,t1,o3}
                \fmf{phantom,tension=2}{s1,t1}
                \fmf{plain_arrow,tension=0,right=.8}{v1,o2}
                \fmf{plain_arrow,tension=0,left=.8}{v2,i2}
                \fmf{plain,tension=0,right=.3}{i3,s1}
                \fmf{plain,tension=0,left=.3}{s1,v2}
                \fmf{plain,tension=0,right=.3}{t1,v1}
                \fmf{plain,tension=0,left=.3}{o3,t1}
                \fmf{dashes,tension=0,right=.45}{o2,i3}
                \fmf{dashes,tension=0,left=.45}{i2,o3}
                \fmfdot{x}
                \fmfv{label=$z$}{x}
             \end{fmfgraph*}
        \end{tabular}
        \end{fmffile}
}

\noindent We will still use $L(0,x)$ or $L(\infty,x)$ to denote the contour on the torus related by the exponential map, with the understanding that when the $L$-contour winds around $0$ or $\infty$ on the plane, it now winds around the spatial cycle either above or below $z={1\over 2\pi i}\log x$ on the torus.

Let us consider the following contour integral:
\ie\label{tccbtt}
&{\cal G}_u^{t}(z_1,z_2|\tau) =\int_{L(0,z_1)} dt_1\int_{L(0,t_1)} dt_2\cdots \int_{L(0,t_{N-2})} dt_{N-1}
 \cG^{bos}_u(z_1,z_2,t_1,\cdots,t_{N-1}|\tau),
\fe
which, as in the case of sphere four-point function, is a conformal block in $t$-channel. The contours $L(0,z_1)$, $L(0,t_1)$, $\cdots$, $L(0,t_{N-2})$, for $t_1,\cdots,t_{N-1}$ integrals, are now contours on the torus of the type shown in the right figure above. The positions of the two primaries, $z_1$, $z_2$ and the positions of the screening charges $t_i$, are in cylinder coordinates. They are related to $x_1,x_2$ and $s_i$ described in section 5.5, now annulus coordinates, by the conformal map
\ie
x_i=e^{2\pi i z_i},~~~~s_i=e^{2\pi i t_i}.
\fe
%Naively, this gives a $t$-channel conformal blocks which can be represented as the following diagram. However, this is not a conformal block in the $W_N$ minimal model. As we have seen in subsection 3.2, the dimension $h_{\lambda+w(\lambda')}$ differs from the dimension $h_{\lambda+\lambda'}$ by a nonnegative integer. The dimension $h_{\lambda+w(\lambda')+\sqrt{p'\over p}\bh_k}$ also differs from the dimension $h_{\lambda+\lambda'+\sqrt{p'\over p}\bh_k}$ by a nonnegative integer:
%\ie
%h_{\lambda+w(\lambda')+\sqrt{p'\over p}\bh_k}=h_{\lambda+\lambda'+\sqrt{p'\over p}\bh_k}+(\A_w\cdot\lambda)(-\A_w\cdot\lambda')+\sqrt{p'\over p}(-\A_w\cdot\lambda')(\A_w\cdot \bh_k),
%\fe
%where the sum of the last two terms is always a nonnegative integer when $\lambda$ and $\lambda'$ sitting in the identity Weyl chamber of $\Gamma_{p\over p'}^*$ and $\Gamma_{p'\over p}^*$. Following the same rationale in subsection 3.2, to make the theory smaller, we perform the following alternating sum:
%\ie\label{altsum}
%{\cal G}^t_{\lambda+\lambda'}(z_1,z_2|\tau) =\sum_{w\in W}\epsilon(w)C_{\lambda+w(\lambda')}L^{-1}_{\lambda+w(\lambda')}\cN^{-1}_{\lambda+w(\lambda')}{\cal G}_{\lambda+w(\lambda')}^{sc,t}(z_1,z_2|\tau),
%\fe
%which gives us the $W_N$ conformal blocks, and notice that we have normalized the conformal block and included the factor $C_{\lambda+w(\lambda')}$ defined in (\ref{Cu}). . 
Generally, it appears rather difficult to explicitly identify a set of contours that gives all the conformal blocks in one channel. Instead, we use the trick described in section 5.5, starting from (\ref{tccbtt}) and obtain the other $N-1$ $t$-channel contours by Weyl transformation on $u=\lambda+\lambda'$. Note that in arriving at (\ref{tccbtt}) we have already performed an alternating sum on $\lambda'$, so the Weyl transformations that permute the different $t$-channel conformal blocks really only act on $\lambda$.

%We know that the primary operator $\cO_{\lambda+\lambda'-Q}$ has null states at the level $(\A_w\cdot\lambda)(-\A_w\cdot\lambda')$, which can be represented by $w(\lambda)+\lambda'$ or $\lambda+w(\lambda')$. We also look at the other internal line $\lambda+\lambda'+\sqrt{p'\over p}\bh_k$. Only the $\lambda+w(\lambda')+\sqrt{p'\over p}\bh_k$ can represent a null state, but not $w(\lambda)+\lambda'+\sqrt{p'\over p}\bh_k$. This can be seen by computing the dimensions.

%It is convenient to work in the basis of $s$-channel conformal blocks, related to ${\cal G}_u$ via
%\ie
%{\cal F}^t_{u}(x_1,x_2|\tau) = \begin{pmatrix} 1 & 0 & 0 \\ * & * & * \\ * & * & * \end{pmatrix}  {\cal G}_{u}(x_1,x_2|\tau).
%\fe
The torus two-point function of the primaries $(\bff,0)$ and $(\bar \bff,0)$ is then given by
%\ie
%\langle {\cal O}_{v_1}(z_1,\bar z_1){\cal O}_{v_2}(z_2,\bar z_2)\rangle_\tau =\sum_{u\in (\Gamma_{pp'}^*/\Gamma_{pp'}-{\rm fixed})/W\times W} \left({\cal F}^t_{u}(z_1,z_2|\tau)\right)^\dagger {\cal M}^t_u {\cal F}^t_{u}(z_1,z_2|\tau)
%\fe
%where ${\cal M}^t_u$ is a diagonal $3\times 3$ matrix, which can be determined using the sphere 3-point functions $\langle{\cal O}_{(adj,0)} {\cal O}_u \overline{{\cal O}_u}\rangle$ and $\langle{\cal O}_{(adj',0)} {\cal O}_u \overline{{\cal O}_u}\rangle$.
\ie\label{tpf}
\langle {\cal O}_{v_1}(z_1,\bar z_1){\cal O}_{v_2}(z_2,\bar z_2)\rangle_\tau &={1\over N!}\sum_{\lambda\in\Delta_1,~\lambda'\in\Delta_2,~w\in W} \big|\cC_{w(u)}{\cal G}^t_{w(\lambda+\lambda')}(z_1,z_2|\tau)\big|^2,
%\\
%&=\sum_{\lambda\in(\Gamma_{p'\over p}^*/\Gamma_{pp'}-\text{fixed}),~\lambda'\in(\Gamma_{p'\over p}^*/\Gamma_{pp'}-\text{fixed})/W} {\cal M}^t_{\lambda+\lambda'} \big|{\cal G}^t_{\lambda+\lambda'}(z_1,z_2|\tau)\big|^2
%\\
%& =\sum_{u\in \Gamma_{pp'}^*/\Gamma_{pp'}} {\cal M}^t_{u} \big|{\cal G}^t_{u}(z_1,z_2|\tau)\big|^2.
\fe
where $\Delta_1$ and $\Delta_2$ are the identity chambers of the shifted affine Weyl transformation in the lattices $\Gamma_{p\over p'}^*$ and $\Gamma_{p'\over p}^*$ respectively. In summing $\lambda$ and $\lambda'$ independently, we have overcounted, as $(\lambda,\lambda')$ are identified under (\ref{lltid}). This is compensated by including an extra factor of $1/N$, turning the factor ${1\over (N-1)!}$ in (\ref{fourptcont}) into ${1\over N!}$ in (\ref{tpf}). The normalization factor ${\cal C}_u$ was given in (\ref{cudeff}) and (\ref{stcts}).

\subsection{Monodromy and modular invariance}

On the torus with two operators inserted at $x_1$ and $x_2$, besides the $s$-monodromy ($x_1$ circling around $x_2$), $t$-monodromy ($x_1\rightarrow x_1+1$ below $x_2$), and $u$-monodromy ($x_1\rightarrow x_1+1$ above $x_2$), there are also what we may call the ``$v$-monodromy" which is $x_1\rightarrow x_1+\tau$ on the left of $x_2$, and ``$w$-monodromy" which is $x_1\rightarrow x_1+\tau$ on the right of $x_2$. Three of these five monodromies are independent. The two-point function should be invariant under these three monodromy transformations, as well as the modular transformations ($T:\tau\rightarrow \tau+1$ and $S:\tau\rightarrow -1/\tau$). 

The $t$-channel conformal blocks in (\ref{tpf}) are trivially invariant under the $t$-monodromy and $T$-modular transformation. The $s$- and $u$-monodromy, on the other hand, mix the different $t$-channel conformal blocks. The invariance of the full two-point function can be seen by expanding (\ref{tpf}) in powers of $q=e^{2\pi i\tau}$ with $z_1-z_2$ fixed, where each term in the expansion is a sphere four-point function of $\cO_{v_1}, \cO_{v_2}$ with a pair of conjugate $W_N$ primaries, or their decedents. The $s$- and $u$-monodromy invariance then follow from those of the sphere four-point functions.

The $S$-modular invariance is less obvious in terms of the $t$-channel conformal blocks. On the other hand, it acts in a simple way on the $s$-channel conformal blocks, and in particular leaves the identity channel invariant. The identity $s$-channel conformal block for the torus two-point function can be constructed by an easy generalization of the ${\widetilde J}_2$ contour in the $N=3$ case for the sphere four-point function.

\subsection{Analytic continuation to Lorentzian signature}

As a potential application of the exact torus two-point function, we wish to consider its analytic continuation (with $\tau=i\beta$) to Lorentzian signature. The result is the Lorentzian thermal two-point function $\langle {\cal O}_{v_1}(t){\cal O}_{v_2}(0)\rangle_\beta$ of the $W_N$ minimal model on the circle (in our convention of $z$-coordinate, of circumference 1), at temperature $T=1/\beta$. This Lorentzian two-point function measures the response of the system some time after the initial perturbation (by one of the two operators), and its decay in time would indicate thermalization of the perturbed system. Of course, since all operator scaling dimensions in the $W_N$ minimal model are multiples of ${1\over Npp'} = {1\over N(N+k)(N+k+1)}$, Poincar\'e recurrence must occur at time $t=Npp'\sim N^3$. In fact, we will see that it occurs at time $t=Np$ in the two-point function $\langle {\cal O}_{v_1}(t){\cal O}_{v_2}(0)\rangle_\beta$. Nonetheless, the behavior of the two-point function at time of order $N^0$ in the large $N$ limit should be a useful probe of the dual semi-classical bulk geometry.

For simplicity of notation, we will denote both ${\cal O}_{v_1}$ and ${\cal O}_{v_2}$ by ${\cal O}$ in most of the discussion below, thinking of ${\cal O}$ as a real operator. Starting with the Euclidean torus two-point function $\langle {\cal O}(z,\bar z) {\cal O}(0,0)\rangle_\tau$, we can write
\ie
z=x+iy,~~~~\bar z=x-iy,
\fe
and then at least locally make the Wick rotation $y$ to $-it$. In other words, we would like to make the replacement 
\ie
z\to x+t,~~~~\bar z\to x-t.
\fe
The resulting two-point function has a singularity at $x=\pm t$, when the two operators are light-like separated (as null rays go around the cylinder periodically, the two operators are light-like separated also when $x\pm t$ is an integer\footnote{If there is thermalization behavior at late time, the two-point function should decay in the distribution sense.}). 
%Locally, when $|z_{12}|\ll1$ and $|\tau|$, the torus two-point function reduces to the sphere two-point function taking the form as
%\ie\label{stpf}
%\vev{\cO(z)\cO(0)}={1\over |z|^{2\Delta}},
%\fe
%Naively, when we analytically continues to the Lorentz signature, we replace $iy$ by $t$. However, the two point function divergent at $x=\pm t$, and also for noninteger dimension, the holomorphic (or antiholomorphic) part $(x-t)^{-\Delta}$ of the two point function (\ref{stpf}) has multi sheets. 
One must then specify how one wishes to analytically continue from $t<|x|$ to $t>|x|$. If we are interested in the time-ordered two-point function at $t>0$,
\ie
\vev{{\cal T}\cO(x,t)\cO(0)}_\beta &=\sum_n e^{-\beta E_n}\bra{n}{\cal T}\cO(x,t)\cO(0)\ket{n}
\\
&=\sum_{n,m} e^{-(\beta-it) E_n-iE_m t}\bra{n}\cO(x,0)\ket{m}\bra{m}\cO(0)\ket{n},
\fe
then the correct prescription is to replace $iy$ by $t-i\epsilon$, where $\epsilon$ is a small positive number.
%For the sum over $m$ convergent, we need to add an infinitesimal imaginary term into the time $t$.

Now consider the analytic continuation of the conformal block (\ref{tccbtt}). We can set $z_2=0$ and $z_1=x+iy$, and applying our prescription, replacing $z_1$ by $x+t-i\epsilon$. Similarly, we will analytically continue the complex conjugate, anti-holomorphic conformal block by sending $\bar z_1\to x-t+i\epsilon$.

We are interested in the behavior of the two point function at time $t$ of order ${\cal O}(N^0)$ but parametrically large. For this purpose, we may consider simply integer values of $t$ and generic $x$. To obtain the values of the two-point function at integer time $t=n$, we can start at $(x,t=0)$, and apply the $t$-monodromy which moves $t\rightarrow t+1$ (with negative imaginary part so that ${\cal O}(x,t)$ goes below the insertion of ${\cal O}(0)$) $n$ times. The $t$-monodromy on the holomorphic conformal block is given by
%\footnote{This formula can be obtained by looking at (\ref{bG0}).}
\ie
{\cal G}^t_{\lambda+\lambda'}(x+t+1+i\epsilon,0|\tau) =e^{2\pi i\left(\sqrt{p'\over p}P_N+{p'(N-1)\over 2pN}\right)}{\cal G}^t_{\lambda+\lambda'}(x+t+i\epsilon,0|\tau).
\fe
The anti-holomorphic conformal block transforms with the same phase, due to the complex conjugation and the inverse $t$-monodromy.
The phase factor is simply due to the difference of the conformal weight of the primary operators labeled by $u=\lambda+\lambda'$ and $u+\sqrt{p'\over p}{\bf h}_k$ in the $t$-channel. The two-point function at $t=n$ is then given by
\ie\label{integertime}
\langle {\cal O}(x,t=n){\cal O}(0)\rangle_\beta &={1\over N!}\sum_{\lambda\in\Delta_1,~\lambda'\in\Delta_2,~w\in W} e^{2\pi i\left(2\sqrt{p'\over p}w(P_N)+{p'(N-1)\over pN}\right)} \big|\cC_{w(u)}{\cal G}^t_{w(\lambda+\lambda')}(x|i\beta)\big|^2.
\fe
Recall that $w(P_N)=w(u)\cdot {\bf h}_N=u\cdot w^{-1}({\bf h}_N)$, and $\sqrt{p'\over p}w(P_N)$ is always an integer multiple of $1/(Np)$. So in fact the two-point function  $\langle {\cal O}(x,t){\cal O}(0)\rangle_\beta$ has time periodicity at most $Np$ (this is simply a consequence of the fusion rule).

Unfortunately, we do not yet know a way to extract the large $N$ behavior of the analytically continued two-point function, or even simply the two-point function at integer times, (\ref{integertime}), for that matter. In the $N=2$ case, i.e. Virasoro minimal models,\footnote{The contour integral expression for the torus two-point function in the Virasoro minimal model has been derived in \cite{Jayaraman:1988ex}} the contour integral is one-dimensional, and we have computed (\ref{integertime}) numerically in Appendix F.

\section{Conclusion}

We have given in section 4 the explicit formulae for the coefficients of all three-point functions of primaries in the $W_N$ minimal model, subject to the condition that one of the primaries is of the form $(\otimes_{\rm sym}^n \bar\bff, \otimes_{\rm sym}^m\bar\bff)$, where $\otimes_{\rm sym}^n \bar\bff$ is the $n$-th symmetric product tensor of the anti-fundamental representation $\bar\bff$. This allows us to study the large $N$ factorization and identify the bound state structure of a large class of operators. Apart form the elementary massive scalars $(\bff,0)=\phi$, $(0,\bff)=\widetilde\phi$, and the obvious elementary light state $(\bff,\bff)=\omega$, there are additional elementary light states e.g. ${1\over \sqrt{2}} ((S,S)-(A,A))$, as well as additional elementary massive states e.g. ${1\over \sqrt{2}}((A,\bff)-(S,\bff))=\Psi$. On the other hand, we have identified the following operators as composite particles:
\ie
& (A,0) \sim {1\over \sqrt{2}}\phi^2,
\\
& (S,0) \sim {1\over \sqrt{2}\Delta_{(\bff,0)}} (\phi\partial\bar\partial\phi - \partial\phi \bar\partial\phi),
\\
& (adj,0) \sim \phi\bar\phi,
\\
& {(S,S)+(A,A)\over\sqrt{2}} \sim {1\over \sqrt{2}}\omega^2,
\\
& {(A,\bff)+(S,\bff)\over \sqrt{2}}\sim \phi\omega,
\\
& {(A,S)+(S,A)\over\sqrt{2}} \sim {1\over \sqrt{2}\Delta_{(\bff,\bff)}} (\omega\partial\bar\partial \omega - \partial\omega \bar\partial\omega)
\\
&~~~~~~~~~~~~~~~~~~~~
\sim {1\over \sqrt{2}}\left( \omega\phi\widetilde\phi - {1\over\Delta_{(\bff,\bff)}} \partial\omega\bar\partial\omega \right),
\\
&{(A,S)-(S,A)\over\sqrt{2}} \,\sim\, {\Psi \widetilde\phi - \widetilde\Psi \phi\over \sqrt{2}}.
\fe
We have also seen that the identification ${1\over\Delta_{(\bff,\bff)}}\partial\bar\partial\omega \sim \phi\widetilde\phi$ of \cite{Papadodimas:2011pf} is consistent with the large $N$ factorization of composite operators. It would be nice to have a systematic classification of all elementary states/particles among the $W_N$ primaries and their bound state structure. This should not be difficult using our approach.

The other main result of this paper is the exact torus two-point function of the basic primaries $(\bff,0)$ and $(\bar\bff,0)$, expressed explicitly as an $(N-1)$-fold contour integral. Direct evaluation of the contour integral appears difficult, but nonetheless feasible numerically at small $N$ (as demonstrated in the $N=2$ case in Appendix F). As our formulae are written for individual holomorphic conformal blocks, the analytic continuation to Lorentzian thermal two-point function is entirely straightforward. It would be very interesting to understand its large $N$ behavior, say at time of order $N^0$. We expect some sort of thermalization behavior (as already shown in the $N=2$ example at large $k$, in fact) reflected in the decay of the two-point function in time, and the precise nature of the decay contains information about the dual bulk geometry. If the BTZ black hole dominates the thermodynamics at some temperature (above the Hawking-Page transition temperature), then we expect to see exponential decay of the thermal two-point function. To the best of our knowledge, such an exponential decay of the two-point function has not been demonstrated directly in a CFT with a semi-classical gravity dual (the closest being the long string CFT\footnote{The long string picture a priori holds in the orbifold point, which is far from the semi-classical regime in the bulk. One may expect that a similar qualitative picture holds for the deformed orbifold CFT in the semi-classical gravity regime, but showing this appears to be a nontrivial problem.} of \cite{Maldacena:1996ds, Maldacena:2001kr} and in toy matrix quantum mechanics models
\cite{Iizuka:2008hg, Iizuka:2008eb}). The $W_N$ minimal model, being exactly solvable and has a weakly coupled gravity dual at large $N$ (though seemingly very different from ordinary semi-classical gravity), seems to be a good place to address this issue. To extract the answer to this question from our result on the torus two-point function, however, is left to future work.

\subsection*{Acknowledgments}

We are grateful to Matthias Gaberdiel, Rajesh Gopakumar, Thomas Hartman, Hong Liu, Shiraz Minwalla, Greg Moore for useful discussions, and especially to Kyriakos Papadodimas and Suvrat Raju for comments on a preliminary draft. C.C. would like to
thank the organizers of Simons Summer Workshop in Mathematics and Physics 2011 for their
hospitality during the course of this work. X.Y. would like to thank Aspen Center for Physics, Tata Institute of Fundamental Research, Harish-Chandra Institute, and Simons Center for Geometry and Physics, for their hospitality during the course of this work. 
This work is supported in part by the Fundamental Laws Initiative Fund at Harvard University, and by NSF Award PHY-0847457.

\appendix

\section{The residues of Toda structure constants}

Let us carry out the procedure of obtaining the structure constant $C_{W_N}(v_1,v_2,v_3)$ in the $W_N$ minimal model by taking the residues of correlators in the affine Toda theory. Firstly, using (\ref{rupsilon}), we derive the identities
\ie
&{\Upsilon(x)\over \Upsilon(x+nb+m/b)}
\\
&=(-1)^{mn}\left(\prod^{m-1}_{i=0}\prod^{n-1}_{j=0}{1\over(i/b+x+jb)^2}\right)\left[\prod^{n-1}_{j=0}{b^{-1+2bx+2jb^2}\over\gamma(bx+jb^2)}\right]\left[\prod^{m-1}_{j=0}{b^{-2x/b-2j/b^2+1}\over\gamma(x/b+j/b^2)}\right],
\fe
and
\ie
&{\Upsilon(x)\over \Upsilon(x-nb-m/b)}
\\
&=(-1)^{mn}\left(\prod_{i=1}^m\prod^n_{j=1} {1\over (x-{i\over b}-jb)^2}\right)\left[\prod^n_{j=1}\gamma(bx-jb^2)b^{1-2bx+2jb^2}\right]\left[\prod^m_{j=1}\gamma(x/b-j/b^2)b^{-1+2x/b-2j/b^2}\right].
\fe
Next, we factorize the denominator of (\ref{Ctoda}) into four groups, and substitute in (\ref{ccep}), and set $\epsilon=0$ in the factors that remains nonzero when $\epsilon=0$. The factors in the denominator of (\ref{Ctoda}) with $j>i$ become
\ie
&\Upsilon\Bigl({\varkappa\over N}+(\bv_1-\cQ)\cdot\bh_i+(\bv_2-\cQ)\cdot\bh_j\Bigr)
\\
&=\Upsilon\Bigl(b(s_{i-1}-s_i)+{1\over b}(s'_{i-1}-s'_i)+(\cQ-\bv_2)\cdot(\bh_i-\bh_j)\Bigr),
\fe
and for $j<i$ we have
\ie
&\Upsilon\Big({\varkappa\over N}+(\bv_1-\cQ)\cdot\bh_i+(\bv_2-\cQ)\cdot\bh_j\Big)
\\
&=\Upsilon\Bigl(b(s_{j-1}-s_j)+{1\over b} (s'_{j-1}-s'_j)+(\cQ-\bv_1)\cdot(\bh_j-\bh_i)\Bigr).
\fe
The denominator factors with $i=j=N$ become
\ie
&\Upsilon\Big({\varkappa\over N}+(\bv_1-\cQ)\cdot\bh_i+(\bv_2-\cQ)\cdot\bh_j\Big)
\\
&=\Upsilon\Bigl(\varkappa+bs_{N-1}+{1\over b}s'_{N-1}\Bigr),
\fe
and for $i=j\neq N$, we have
\ie\label{vd}
&\Upsilon\Big({\varkappa\over N}+(\bv_1-\cQ)\cdot\bh_j+(\bv_2-\cQ)\cdot\bh_j+\epsilon\cdot\bh_j\Big)
\\
&=\Upsilon\Bigl(b(s_{j-1}-s_j)+{1\over b}(s'_{j-1}-s'_j)+\epsilon_j-\epsilon_{j-1}\Bigr),
\fe
where $s_0=s'_0=\epsilon_0=0$.

Now, it is clear that (\ref{vd}) are the only factors in the denominator that vanish at $\epsilon=0$, and also they vanish only when $s_j\ge s_{j-1}$ and $s'_j\ge s'_{j-1}$, or $s_j< s_{j-1}$ and $s'_j< s'_{j-1}$. Let us first assume $s_j\ge s_{j-1}$ and $s'_j\ge s'_{j-1}$. We have
\ie
&{\Upsilon(b)\over \Upsilon\Bigl(b(s_{j-1}-s_j)+{1\over b}(s'_{j-1}-s'_j)+\epsilon\cdot\bh_j\Bigr)}
={1\over \epsilon\cdot\bh_j}(-1)^{s'_{j,j-1}s_{j,j-1}}\left(\prod_{k=1}^{s'_{j,j-1}}\prod^{s_{j,j-1}}_{l=1} {1\over (\epsilon\cdot\bh_j+{k\over b}+lb)^2}\right)
\\
&\times\left[\prod^{s_{j,j-1}}_{l=1}\gamma(\epsilon\cdot\bh_j-lb^2)\right]\left[\prod^{s'_{j,j-1}}_{k=1}\gamma(\epsilon\cdot\bh_j-k/b^2)\right]
\cdot b^{s_{j,j-1}-b^2(s_{j-1,j}-1)s_{j,j-1}-s'_{j,j-1}+b^2(s'_{j-1,j}-1)s'_{j,j-1}},
\fe
The prefactor $1\over\epsilon\cdot\bh_j$ is the only divergent piece in the $\epsilon \to 0$, and at this point we could take $\epsilon\rightarrow0$ on the remaining factor, but we will keep the formula with nonzero $\epsilon$ for later use. There are also
\ie
&{\Upsilon\Bigl((\cQ-\bv_2)\cdot(\bh_j-\bh_i)\Bigr)\over\Upsilon\Bigl(b(s_{j-1}-s_j)+{1\over b}(s'_{j-1}-s'_j)+(\cQ-\bv_2)\cdot(\bh_j-\bh_i)\Bigr)}
\\
&=(-1)^{s_{j,j-1}s'_{j,j-1}}\left(\prod_{k=1}^{s'_{j,j-1}}\prod^{s_{j,j-1}}_{l=1} {1\over (\bP^2_{ji}-{k\over b}-lb)^2}\right)
\times\left[\prod^{s_{j,j-1}}_{l=1}\gamma(b\bP^2_{ji}-lb^2)\right]\left[\prod^{s'_{j,j-1}}_{k=1}\gamma(\bP^2_{ji}/b-k/b^2)\right]
\\
&~~~~\times b^{s_{j,j-1}-b^2(s_{j-1,j}-1)s_{j,j-1}-s'_{j,j-1}+b^2(s'_{j-1,j}-1)s'_{j,j-1}-2b\bP^2_{ji}s_{j,j-1}+2\bP^2_{ji}s'_{j,j-1}/b},
\fe
and
\ie
&{\Upsilon\Bigl((\cQ-\bv_1)\cdot(\bh_j-\bh_i)\Bigr)\over\Upsilon\Bigl(b(s_{j-1}-s_j)+{1\over b}(s'_{j-1}-s'_j)+(\cQ-\bv_1)\cdot(\bh_j-\bh_i)\Bigr)}
\\
&=(-1)^{s_{j,j-1}s'_{j,j-1}}\left(\prod_{k=1}^{s'_{j,j-1}}\prod^{s_{j,j-1}}_{l=1} {1\over (\bP^1_{ji}-{k\over b}-lb)^2}\right)
\times\left[\prod^{s_{j,j-1}}_{l=1}\gamma(b\bP^1_{ji}-lb^2)\right]\left[\prod^{s'_{j,j-1}}_{k=1}\gamma(\bP^1_{ji}/b-k/b^2)\right]
\\
&~~~~\times b^{s_{j,j-1}-b^2(s_{j-1,j}-1)s_{j,j-1}-s'_{j,j-1}+b^2(s'_{j-1,j}-1)s'_{j,j-1}-2b\bP^1_{ji}s_{j,j-1}+2\bP^1_{ji}s'_{j,j-1}/b},
\fe
where we introduced the notation $s_{i,j}\equiv s_i-s_j$ and $\bP^a_{ij}=(\cQ-\bv_a)\cdot(\bh_i-\bh_j)$, $a=1,2$. Combing the above three terms, we have
\ie
&{\Upsilon(b)\over \Upsilon\Bigl(bs_{j-1,j}+{1\over b}s'_{j-1,j}+\epsilon\cdot\bh_j\Bigr)}\prod_{i=j+1}^N{\Upsilon\Bigl(\bP^1_{ji}\Bigr)\over\Upsilon\Bigl(bs_{j-1,j}+{1\over b}s'_{j-1,j}+\bP^1_{ji}\Bigr)}{\Upsilon\Bigl(\bP^2_{ji}\Bigr)\over\Upsilon\Bigl(bs_{j-1,j}+{1\over b}s'_{j-1,j}+\bP^2_{ji}\Bigr)}
\\
&={1\over \epsilon\cdot\bh_j}(-1)^{s'_{j,j-1}s_{j,j-1}}R^{s_{j,j-1},s'_{j,j-1}}_{j,\epsilon} b^{C_j},
\fe
where $R^{s_{j,j-1},s'_{j,j-1}}_{j,\epsilon}$ is defined to be
\ie
R^{s_{j,j-1},s'_{j,j-1}}_{j,\epsilon}=&\left(\prod_{k=1}^{s'_{j,j-1}}\prod^{s_{j,j-1}}_{l=1}{1\over (\epsilon\cdot\bh_j+{k\over b}+lb)^2} \prod_{i=j+1}^N {1\over (\bP^1_{ji}-{k\over b}-lb)^2}{1\over (\bP^2_{ji}-{k\over b}-lb)^2}\right)
\\
&\times\left[\prod^{s_{j,j-1}}_{l=1}\gamma(\epsilon\cdot\bh_j-lb^2)\prod_{i=j+1}^N\gamma(b\bP^1_{ji}-lb^2)\gamma(b\bP^2_{ji}-lb^2)\right]
\\
&\times\left[\prod^{s'_{j,j-1}}_{k=1}\gamma(\epsilon\cdot\bh_j-k/b^2)\prod_{i=j+1}^N\gamma(\bP^1_{ji}/b-k/b^2)\gamma(\bP^2_{ji}/b-k/b^2)\right].
\fe
The exponent $C_j$ of $b$ is given by
\ie
C_j
&=(2N-2j+1)(s_{j,j-1}-s'_{j,j-1})+2(s_{j-1}s'_{j}-s_{j}s'_{j-1})+b^2\left[(2N-2j+1)s_{j,j-1}+s_{j-1}^2-s_j^2\right]
\\
&~~~~-{1\over b^2}\left[(2N-2j+1)s'_{j,j-1}+s'^2_{j-1}-s'^2_j\right]-2bs_{j,j-1}\varkappa+2{1\over b}s'_{j,j-1}\varkappa,
\fe
where we have used
\ie
&\sum_{i=j+1}^N (\bP^1_{ji}+\bP^2_{ji})
&=\varkappa+b(N-j)s_{j,j-1}+bs_j+{1\over b}(N-j)s'_{j,j-1}+{1\over b}s'_j.
\fe
We also have
\ie
&{\Upsilon(\varkappa)\over \Upsilon(\varkappa+bs_{N-1}+s'_{N-1}/b)}
=(-1)^{s_{N-1}s'_{N-1}}\left(\prod_{k=0}^{s'_{N-1}-1}\prod^{s_{N-1}-1}_{l=0} {1\over (\varkappa+{k\over b}+lb)^2}\right)
\\
&~~~~~~~~\times\left[\prod^{s_{N-1}-1}_{l=0}\gamma(1-b\varkappa-lb^2)\right]\left[\prod^{s'_{N-1}-1}_{k=0}\gamma(1-\varkappa/b-k/b^2)\right]
\\
&~~~~~~~~\times b^{-s_{N-1}+2b\varkappa s_{N-1}+b^2s_{N-1}(s_{N-1}-1)+s'_{N-1}-2{1\over b}\varkappa s'_{N-1}-{1\over b^2}s'_{N-1}(s'_{N-1}-1)}.
\fe
Putting the above terms together, the total exponent of $b$ is
\ie
&\sum_{j=1}^{N-1}C_j-s_{N-1}+2b\varkappa s_{N-1}+b^2s_{N-1}(s_{N-1}-1)+s'_{N-1}-2{1\over b}\varkappa s'_{N-1}-{1\over b^2}s'_{N-1}(s'_{N-1}-1)
\\
&=2(1+b^2)\sum^{N-1}_{j=1}s_j-2(1+{1\over b^2})\sum^{N-1}_{j=1}s'_j+2\sum_{j=1}^{N-2}(s_j s'_{j+1}-s_{j+1}s'_{j}).
\fe
Finally, we rewrite the prefactor of (\ref{Ctoda}) in the form
\ie
&\left[\m\pi\gamma(b^2)b^{2-2b^2}\right]^{(2\cQ-\sum\bv_i,\rho)\over b}
&=\left[{-\m\pi\over\gamma(-b^2)}b^{-2-2b^2}\right]^{\sum\limits^{N-1}_{k=1}s_k}\left[{-\m'\pi\over\gamma(-{1\over b^2})}b^{{2\over b^2}+2}\right]^{\sum\limits^{N-1}_{k=1}s'_k}.
\fe
The residue of the three point function is then
\ie
&{\rm res}_{\epsilon_1\to 0}{\rm res}_{\epsilon_2\to\epsilon_1}\cdots {\rm res}_{\epsilon_{N-1}\to\epsilon_{N-2}} C_{toda}(\bv_1,\bv_2,\varkappa\omega_{n-1})
\\
&=(ib)^{2\sum_{j=1}^{N-2}(s_j s'_{j+1}-s_{j+1}s'_{j})}\left[{-\m\pi\over\gamma(-b^2)}\right]^{\sum\limits^{N-1}_{k=1}s_k}\left[{-\m'\pi\over\gamma(-{1\over b^2})}\right]^{\sum\limits^{N-1}_{k=1}s'_k}\left(\prod_{k=0}^{s'_{N-1}-1}\prod^{s_{N-1}-1}_{l=0} {1\over (\varkappa+{k\over b}+lb)^2}\right)
\\
&~~~~\times\left[\prod^{s_{N-1}-1}_{l=0}\gamma(1-b\varkappa-lb^2)\right]\left[\prod^{s'_{N-1}-1}_{k=0}\gamma(1-\varkappa/b-k/b^2)\right]\prod^{N-1}_{j=1}R^{s_{j,j-1},s'_{j,j-1}}_{j,\epsilon}.
\fe
The $\epsilon$ is the subscript of $R_{j,\epsilon}^{s,s'}$ is understand to be taken to zero in computing the residue, but we will leave it in the formula as we will make use of it below.

In the case $s_j< s_{j-1}$ and $s'_j< s'_{j-1}$, we can apply the following identity:
\ie
{\Upsilon(x)\over \Upsilon(x-nb-m/b)}={\Upsilon(b+1/b-x)\over \Upsilon(b+1/b-x+nb+m/b)},
\fe
and then the residue will be computed by the above formula with the replacement
\ie
\epsilon\rightarrow b+1/b-\epsilon,~~~~\bP^1_{ji}\rightarrow b+1/b-\bP^1_{ji},~~~~\bP^2_{ji}\rightarrow b+1/b-\bP^2_{ji},
\fe
and then set $\epsilon$ to zero. Finally, we obtain the structure constants in the $W_N$ minimal model by the analytic continuation (\ref{sub}).

\section{Monodromy of integration contours}

In this appendix, we analyze the $s$ and $t$ channel monodromy action on the contour integrals described in section 5.2.

Let us begin by considering the $s_2$-integral. The $s_2$-integrand has branch points at $0, s_1, x_2, \infty$. There are relations among the $L$ contours encircling a pair of the branch points. For instance,
\ie
&\int_{L(0,\infty)} = - \int_{L(0,\{s_1,x_2\})}
\\
&= \int_0^{s_1} + \int_{s_1}^{x_2} + g_{x_2}\int_{x_2}^{s_1} + g_{x_2}g_{s_1}\int_{s_1}^0
+g_{x_2}g_{s_1}g_0 \int_0^{s_1} + g_{x_2}g_{s_1}g_0g_{s_1}^{-1} \int_{s_1}^{x_2} 
+g_{x_2}g_{s_1}g_0g_{s_1}^{-1}g_{x_2}^{-1} \left(\int_{x_2}^{s_1} + \int_{s_1}^0\right)
\\
&= (1-g_{x_2}g_{s_1}+ g_{x_2}g_{s_1}g_0 - g_{x_2}g_{s_1}g_0g_{s_1}^{-1}g_{x_2}^{-1})\int_0^{s_1}
+(1-g_{x_2}+g_{x_2}g_{s_1}g_0g_{s_1}^{-1}-g_{x_2}g_{s_1}g_0g_{s_1}^{-1}g_{x_2}^{-1})\int_{s_1}^{x_2}.
\fe
Now since all the $g$'s are commuting phase factors, we can write
\ie
&\int_{L(0,\infty)} = (1-g_0)(1-g_{s_1}g_{x_2}) \int_0^{s_1}
+(1-g_0)(1-g_{x_2}) \int_{s_1}^{x_2}.
\fe
Naively, one may think that the integral over $L(x_2,\infty)$ is given by the same expression with $0$ and $x_2$ exchanged. This is not correct, however, due to the choice of branch in the line integrals. We have
\ie
&\int_{L(x_2,\infty)} = - \int_{L(x_2,\{s_1,x_0\})}
\\
&= \int_{x_2}^{s_1} + g_{s_1}\int_{s_1}^{0} + g_{s_1}g_0\left( \int_0^{s_1} + \int_{s_1}^{x_2}\right)
+ g_{s_1}g_0 g_{x_2} \left(\int_{x_2}^{s_1} + \int_{s_1}^0\right) 
+g_{s_1}g_0 g_{x_2}g_0^{-1} \int_{0}^{s_1} +g_{s_1}g_0 g_{x_2}g_0^{-1}g_{s_1}^{-1} \int_{s_1}^{x_2}
\\
& = - (1-g_{s_1}g_0)(1- g_{x_2})\int_{s_1}^{x_2}
- g_{s_1}(1-g_0)(1 - g_{x_2}) \int_0^{s_1}.
\fe
Together with using the following relation between the $L$-contour and the ``collapsed" line integral,
\ie
&\int_{L(0,s_1)} = (1-g_{s_1})(1- g_0)\int_0^{s_1},
\\
&\int_{L(s_1,x_2)} = (1-g_{s_1})(1-g_{x_2}) \int_{s_1}^{x_2},
\fe
we derive the formula (\ref{lcont}).
%\ie\label{lcont}
%\begin{pmatrix} L(0,\infty) \\ L(x_2,\infty) \end{pmatrix} = \begin{pmatrix} {1-g_{s_1}g_{x_2}\over 1-g_{s_1}} & {1-g_{0}\over 1-g_{s_1}} \\  -g_{s_1} {1-g_{x_2}\over 1-g_{s_1}} & - {1-g_0g_{s_1}\over 1-g_{s_1}} \end{pmatrix}
%\begin{pmatrix} L(0,s_1) \\ L(s_1,x_2) \end{pmatrix}.
%\fe
%So there are in fact only two linearly independent $L$-contours for the $s_2$-integral. The basis $(L(x_2,\infty), L(0,s_1))$ will be convenient for analyzing $t$-channel monodromies, whereas the basis $(L(0,\infty), L(x_2,s_1))$ will be convenient for analyzing $s$-channel monodromies.

Now consider the two-dimensional contours (\ref{cicontours}).
Let us denote by ${\cal I}^{(i)}$ the contours obtained from $C^{(i)}$ by collapsing $L(z_1,z_2)$ into straight lines, and by $J_i$ the integral of (\ref{integrand}) along ${\cal I}^{(i)}$, and also by $\cJ_i$ the integral of (\ref{integrand}) along $C^{(i)}$, $i=1,2,3,4$. $J_i$ and $\cJ_i$ are related via
\ie
&\begin{pmatrix} \cJ_1\\ \cJ_2 \end{pmatrix}=({\bf 1}-g_{x_2}(s_1))({\bf 1}-g_{x_1}(s_1))\begin{pmatrix} (1-g_{x_2})(1-g_{\infty})J_1\\ (1-g_0)(1-g_{s_1})J_2 \end{pmatrix},
\\
&\begin{pmatrix} \cJ_3\\ \cJ_4 \end{pmatrix}=({\bf 1}-g_{0}(s_1))({\bf 1}-g_{x_1}(s_1))\begin{pmatrix} (1-g_{x_2})(1-g_{\infty})J_3\\ (1-g_0)(1-g_{s_1})J_4 \end{pmatrix}.
\fe
$T_t$ and $T_s$ acts on $(\cJ_1,\cJ_2,\cJ_3,\cJ_4)$ via the monodromy matrices (\ref{mtaa}) and (\ref{msaa}).
%\ie
%M_t = g_0(x_1)\begin{pmatrix} ~{\bf 1}~~ & {\bf 1}-g_{x_2}(s_1) \\~ 0~~ & g_0(s_1)g_{x_1}(s_1) \end{pmatrix}
%\fe
%where $g_x(z)$ is the $2\times 2$ monodromy matrix on the $s_1$-integrand (after having performed the $s_2$-integral) by taking the point $z$ around $x$. $T_s$ acts via the monodromy matrix
%\ie
%M_s = g_{x_2}(x_1)\begin{pmatrix} g_{x_1}(s_1) g_{x_2}(s_1) & ~~0~ \\ g_{x_1}(s_1)-g_{x_1}(s_1) g_0(s_1)& ~~{\bf 1} ~\end{pmatrix}
%\fe
Define $\zeta \equiv e^{2\pi i {p'\over p}} $. We find
\ie
g_0(x_1) = \zeta^{{2\over 3}n_+ + {1\over 3}m_+ }e^{-2\pi i ({2\over 3}n_-+{1\over 3}m_-)}, ~~~~ g_{x_2}(x_1) = \zeta^{1\over 3},
\fe
and
\ie
g_0(s_1) = \begin{pmatrix} \zeta^{-n_+} & 0 \\ 0 & \zeta^{-n_+-m_+-1} \end{pmatrix},~~~~ g_{x_1}(s_1) = \zeta^{-1}{\bf 1},~~~~g_{x_2}(s_1) = A^{-1} \begin{pmatrix} 1 & 0 \\ 0 & \zeta^{-2} \end{pmatrix} A.
\fe
The matrix $A$ is the linear transformation of $L$-contours,
\ie
\begin{pmatrix} L(0,\infty) \\ L(s_1,x_2) \end{pmatrix} =A
\begin{pmatrix} L(x_2,\infty) \\ L(0,s_1) \end{pmatrix},
\fe
and from (\ref{lcont}) we know
\ie
A &=  -{1\over 1-g_0 g_{s_1}} \begin{pmatrix} 1-g_0~ & -1+g_0g_{x_2}g_{s_1} \\ 1-g_{s_1}~ & g_{s_1}(1-g_{x_2}) \end{pmatrix}.
\fe
Using the monodromy phases of the $s_2$-integrand,
\ie
g_0 = \zeta^{-m_+},~~~ g_{s_1}=g_{x_2}=\zeta^{-1},
\fe
we find
\ie
A &= -{1\over 1-\zeta^{-m_+-1}}\begin{pmatrix}  {1-\zeta^{-m_+}} ~& \zeta^{-m_+-2}-1 \\  {1-\zeta^{-1}}~ &  \zeta^{-1}-\zeta^{-2} \end{pmatrix}.
\fe

\section{Identifying the conformal blocks with contour integrals}

It is useful to work in instead of $({\cal J}_1,{\cal J}_2,{\cal J}_3,{\cal J}_4)$, the basis
\ie
& \begin{pmatrix} \widetilde \cJ_1 \\ \tilde \cJ_2 \end{pmatrix} = A \begin{pmatrix} \cJ_1 \\ \cJ_2 \end{pmatrix} = \int_{L(x_1,x_2)}ds_1 \begin{pmatrix} \int_{L(0,\infty)}ds_2\cdots \\  \int_{L(s_1,x_2)}ds_2\cdots \end{pmatrix} ,
\\
& \begin{pmatrix} \cJ_3 \\ \cJ_4 \end{pmatrix} = \int_{L(0,x_1)}ds_1 \begin{pmatrix} \int_{L(x_2,\infty)}ds_2\cdots \\  \int_{L(0,s_1)}ds_2\cdots \end{pmatrix}.
\fe
In fact, $\widetilde \cJ_1$ vanishes identically, as a consequence of the relation
\ie
A({\bf 1}-g_{x_2}(s_1))({\bf 1}-g_{x_1}(s_1))=-{\zeta^{4 - m_+} (1 - \zeta^{1 + m_+})\over (1 + \zeta)(1 - \zeta)^3}\begin{pmatrix}  0 ~& 0 \\  \zeta~ &  1 \end{pmatrix}.
\fe
Acting on $( \widetilde \cJ_2, \cJ_3, \cJ_4)$, the monodromy matrices are of the form
\ie
&\tilde M_s = \zeta^{1\over 3}  \begin{pmatrix} \zeta^{-3}~ &0 ~~&0 \\ {(1-\zeta^{2+m_+})(1-\zeta^{n_+})\over \zeta^{1+m_++n_+}(1-\zeta^2)}~ & 1~~ & 0 \\  {(1-\zeta^{m_+})(1-\zeta^{1+m_++n_+})\over \zeta^{2m_++n_+}(1-\zeta^2)}~ &0~~ & 1 \end{pmatrix} ,
\\
& \tilde M_t =   \zeta^{{2\over 3}n_+ + {1\over 3}m_+ }e^{-2\pi i ({2\over 3}n_-+{1\over 3}m_-)}
\begin{pmatrix} 1~~ &{(1-\zeta)^2(1+\zeta)\zeta^{-2+m_+ }\over 1-\zeta^{m_++1}} ~&{(1-\zeta)^2(1+\zeta)\zeta^{-3+ m_+}\over 1-\zeta^{m_++1}} \\0~~ & \zeta^{-1-n_+}~ & 0 \\ 0~~ &0~ & \zeta^{-2-n_+-m_+} \end{pmatrix}.
\fe

As described in section 5.3, the four point function is obtained by summing over either $s$ or $t$ channel conformal blocks (\ref{fmfrel}). 
%\ie
%&\langle {\cal O}_{v_1}(x_1,\bar x_1){\cal O}_{v_2}(x_2,\bar x_2){\cal O}_{v_3}(0){\cal O}_{v_4}'(\infty)\rangle
%= ({\cal F}^s)^\dagger {\cal M}^s {\cal F}^s
%= ({\cal F}^t)^\dagger {\cal M}^t {\cal F}^t,
%\fe
The mass matrices therein,
${\cal M}^t$ and ${\cal M}^s$, are of the form
\ie
{\cal M}^t= \begin{pmatrix} a &0&0  \\ 0 & b & 0 \\ 0 & 0 & c \end{pmatrix},~~~~{\cal M}^s= \begin{pmatrix} d &0&0  \\ 0 & * & * \\ 0 & * & * \end{pmatrix},
\fe
and obey (\ref{tschange}).
(\ref{tschange}) is solved with
\ie
&{a\over c}={ \zeta^{2 - 2 m_+ - n_+} (1 - \zeta^{m_+}) (1 - \zeta^{1 + m_+}) (1 - \zeta^{1 + n_+}) (1 - \zeta^{1 + m_+ + n_+}) (1 - \zeta^{2 + m_+ + n_+})^2\over(1 - \zeta)^4 (1 + \zeta)^2 (1 - \zeta^{2 + n_+}) (1 - \zeta^{3 + m_+ + n_+})},
\\
&{b\over c}={ \zeta^{-m_+} (1 - \zeta^{m_+}) (1 - \zeta^{1 + m_+ + n_+}) (1 - \zeta^{2 + m_+ + n_+})\over(1 - \zeta^{2 + m_+}) (1 - \zeta^{n_+}) (1 - \zeta^{1 + n_+})}.
\fe
The overall normalization can be fix by the identity $s$-channel, which then fixes the entire four point function. From this four point function one may also extract the coefficients of the sphere 3-point functions, $\langle {\cal O}_{(adj,0)} {\cal O}_u \overline{{\cal O}_u}\rangle$, $\langle {\cal O}_{(adj',0)} {\cal O}_u \overline{{\cal O}_u}\rangle$, etc., and reproduce some of the results in section 3.

\section{Monodromy invariance of the sphere four-point function}

In this section, we show that the formula (\ref{G}) for the four-point function is invariant under the $t$- and $u$-monodromy transformations, i.e. circling $x_1$ around $0$ and $\infty$. By (\ref{GF}), the $t$-monodromy acting as a phase on the $t$-channel conformal blocks; hence, the the four-point function (\ref{G}) is trivially invariant. To exhibit the $u$-monodromy, let us apply the following identity on the generalized hypergeometric function:
\ie\label{transgh}
&{}_{N}F_{N-1}(a_1,\cdots,a_N;b_1,\cdots,b_{N-1}|x)
\\
&={\prod^{N-1}_{k=1}\Gamma(b_k)\over \prod^N_{k=1}\Gamma(a_k)}\sum^N_{k=1}{\Gamma(a_k)\prod^N_{j=1,j\neq k}\Gamma(a_j-a_k)\over \prod^{N-1}_{j=1}\Gamma(b_j-a_k)}(-x)^{-a_k}
\\
&~~~~\times{}_{N}F_{N-1}(a_k,a_k-b_1+1,\cdots,a_k-b_{N-1}+1;1-a_1+a_k,\cdots,1-a_N+a_k|{1\over x}).
\fe
Via this identity, the conformal block $G_l(x)$ can be rewrited as
\ie\label{transcb}
G_l(x)&=x^{\sqrt{p'\over p}P_{l}}{}_{N} F_{N-1}(\vec\m_l;\widehat{\vec\n_l}|x)
\\
&={\prod^{N}_{i=1}\Gamma(\sqrt{p'\over p}P_{li}+1)\over \prod^N_{i=1}\Gamma(\sqrt{p'\over p}P_{li}+{p'\over p})}\sum^N_{k=1}\Gamma(\sqrt{p'\over p}P_{lk}+{p'\over p})\Gamma(1+\sqrt{p'\over p}P_{kl}-{p'\over p})
\\
&~~~~\times{\prod^N_{j=1,j\neq k}\Gamma(\sqrt{p'\over p}P_{kj})\over \prod^{N}_{j=1}\Gamma(\sqrt{p'\over p}P_{kj}+1-{p'\over p})}e^{i\pi\left(\sqrt{p'\over p}P_{kl}-{p'\over p}\right)}H_k(x),
\fe
where $H_k(x)$ are the $u$-channel conformal blocks, given by
\ie\label{ucb}
 H_k(x)=x^{\sqrt{p'\over p}P_{k}-{p'\over p}}{}_{N} F_{N-1}(\vec\m_k';\widehat{\vec\n}_k'|{1\over x}),
\fe
and $\vec\m_k'$, $\vec\n_k'$ are $N$-vectors defined as
\ie
&\vec\m_k'=\sqrt{p'\over p}(P_{1,k},\cdots,P_{N,k})+{p'\over p}(1,\cdots,1),
\\
&\vec\n_k'=\sqrt{p'\over p}(P_{1,k},\cdots,P_{N,k})+(1,\cdots,1).
\fe
Again $\widehat{\vec \nu}_k'$ is given by $\vec\nu_k'$ dropping the $k$-th entry.
In terms of the $u$-channel conformal blocks $H_l(x)$, the four-point function can be written as
\ie\label{transtpf}
&\sum_{l=1}^N(\cM_u)_{ll}|G_l(x)|^2
%\\
%&=\sum_{l=1}^N\left(-{\pi\m\Gamma(1-{p'\over p})\over \Gamma({p'\over p})}\right)^{N-1}\left(\prod^{N}_{i=1,i\neq l}{\Gamma(\sqrt{p'\over p}P_{il})\Gamma({p'\over p}-\sqrt{p'\over p}P_{il})\over \Gamma(1-{p'\over p}+\sqrt{p'\over p}P_{il})\Gamma(1-\sqrt{p'\over p}P_{il})}\right)\left({\prod^{N}_{i=1}\Gamma(\sqrt{p'\over p}P_{li}+1)\over \prod^N_{i=1}\Gamma(\sqrt{p'\over p}P_{li}+{p'\over p})}\right)^2
%\\
%&\times\sum^N_{k_1=1}\sum^N_{k_2=1}\Gamma(\sqrt{p'\over p}P_{lk_1}+{p'\over p})\Gamma(1-\sqrt{p'\over p}P_{l,k_1}-{p'\over p})\Gamma(\sqrt{p'\over p}P_{l,k_2}+{p'\over p})\Gamma(1-\sqrt{p'\over p}P_{l,k_2}-{p'\over p})(-1)^{\sqrt{p'\over p}P_{k_1k_2}}
%\\
%&\times{\prod^N_{j=1,j\neq k_1}\Gamma(\sqrt{p'\over p}P_{k_1 j})\over \prod^{N}_{j=1}\Gamma(\sqrt{p'\over p}P_{k_1 j}+1-{p'\over p})} {\prod^N_{j=1,j\neq k_2}\Gamma(\sqrt{p'\over p}P_{k_2 j})\over \prod^{N}_{j=1}\Gamma(\sqrt{p'\over p}P_{k_2 j}+1-{p'\over p})} 
%\\
%&\times x^{\sqrt{p'\over p}P_{k_1 1}-{p'\over p}}{}_{N}F_{N-1}(\vec\m_{k_1}';\vec\n_{k_1}'|{1\over x})\bar x^{\sqrt{p'\over p}P_{k_2 1}-{p'\over p}}{}_{N}F_{N-1}(\vec\m_{k_2}';\vec\n_{k_2}'|{1\over \bar x})
\\
&=\gamma\left({p'\over p}\right)\gamma\left(N\left(1-{p'\over p}\right)\right)\sum_{l=1}^N\left(\prod^{N}_{i=1,i\neq l}{\Gamma(\sqrt{p'\over p}P_{il})\Gamma(1-\sqrt{p'\over p}P_{il})\over \Gamma(1-{p'\over p}+\sqrt{p'\over p}P_{il})\Gamma({p'\over p}-\sqrt{p'\over p}P_{il})}\right){1\over \Gamma({p'\over p})^2}
\\
&~~~~\times\sum^N_{k_1=1}\sum^N_{k_2=1}\Gamma(\sqrt{p'\over p}P_{lk_1}+{p'\over p})\Gamma(1-\sqrt{p'\over p}P_{lk_1}-{p'\over p})\Gamma(\sqrt{p'\over p}P_{lk_2}+{p'\over p})\Gamma(1-\sqrt{p'\over p}P_{lk_2}-{p'\over p})e^{i\pi\sqrt{p'\over p}P_{k_1k_2}}
\\
&~~~~\times{\prod^N_{j=1,j\neq k_1}\Gamma(\sqrt{p'\over p}P_{k_1j})\over \prod^{N}_{j=1}\Gamma(\sqrt{p'\over p}P_{k_1j}+1-{p'\over p})} {\prod^N_{j=1,j\neq k_2}\Gamma(\sqrt{p'\over p}P_{k_2j})\over \prod^{N}_{j=1}\Gamma(\sqrt{p'\over p}P_{k_2j}+1-{p'\over p})} H_{k_1}(x)H_{k_2}(\bar x).
\fe
Using the following identity
\ie\label{key}
&\sum_{l=1}^N\left(\prod^{N}_{i=1,i\neq l}{\Gamma(\sqrt{p'\over p}P_{il})\Gamma(1-\sqrt{p'\over p}P_{il})\over \Gamma(1-{p'\over p}+\sqrt{p'\over p}P_{il})\Gamma({p'\over p}-\sqrt{p'\over p}P_{il})}\right)
\\
&\times\Gamma(\sqrt{p'\over p}P_{lk_1}+{p'\over p})\Gamma(1-\sqrt{p'\over p}P_{lk_1}-{p'\over p})\Gamma(\sqrt{p'\over p}P_{lk_2}+{p'\over p})\Gamma(1-\sqrt{p'\over p}P_{lk_2}-{p'\over p})e^{i\pi\sqrt{p'\over p}P_{k_1k_2}}
\\
&=\pi^2\sum_{l=1}^N\left(\prod^{N}_{i=1,i\neq l}{\sin\pi({p'\over p}-\sqrt{p'\over p}P_{il})\over \sin\pi(\sqrt{p'\over p}P_{il})}\right)\csc\pi(\sqrt{p'\over p}P_{lk_1}+{p'\over p})\csc\pi(\sqrt{p'\over p}P_{lk_2}+{p'\over p})e^{i\pi\sqrt{p'\over p}P_{k_1k_2}}
\\
&\propto \delta_{k_1,k_2},
\fe
(\ref{transtpf}) may be simplified to
\ie\label{H}
&\sum_{l=1}^N(\cM_u)_{ll} |G_l(x)|^2
\\
&=\gamma\left({p'\over p}\right)\gamma\left(N\left(1-{p'\over p}\right)\right)\sum^N_{k=1}\sum_{l=1}^N\left(\prod^{N}_{i=1,i\neq l}{\Gamma(\sqrt{p'\over p}P_{il})\Gamma(1-\sqrt{p'\over p}P_{il})\over \Gamma(1-{p'\over p}+\sqrt{p'\over p}P_{il})\Gamma({p'\over p}-\sqrt{p'\over p}P_{il})}\right){1\over \Gamma({p'\over p})^2}
\\
&~~~~\times\Gamma(\sqrt{p'\over p}P_{lk}+{p'\over p})^2\Gamma(1-\sqrt{p'\over p}P_{lk}-{p'\over p})^2 \left(\prod^N_{j=1,j\neq k}{\Gamma(\sqrt{p'\over p}P_{kj})\over \Gamma(\sqrt{p'\over p}P_{kj}+1-{p'\over p})} \right)^2 {1\over \Gamma(1-{p'\over p})^2}
\\
&~~~~\times|H_k(x)|^2
\\
&=\gamma\left({p'\over p}\right)\gamma\left(N\left(1-{p'\over p}\right)\right)\sum_{j=1}^N\prod^{N}_{i=1,i\neq j}{\Gamma(\sqrt{p'\over p}P_{ji})\Gamma({p'\over p}-\sqrt{p'\over p}P_{ji})\over \Gamma(1-{p'\over p}+\sqrt{p'\over p}P_{ji})\Gamma(1-\sqrt{p'\over p}P_{ji})}|H_j(x)|^2
\\
&=\sum_{j=1}^N (\widetilde\cM_u)_{jj}|H_j(x)|^2,
\fe
where the $u$-channel mass matrix $\widetilde\cM_u$ is given in terms of the structure constants as (here the subscript $u$ is the charge vector)
\ie
(\widetilde\cM_u)_{jj}%&=C^{Q-u+\sqrt{p'\over p} \bh_j}_{\sqrt{p'\over p}w^1,Q-u}C^{u+Q-\sqrt{p'\over p} \bh_j}_{\sqrt{p'\over p}w^{N-1},u+Q}
%\\
&=B\left(\sqrt{p'\over p}w^1\right)^2C_{W_N}\left(\sqrt{p'\over p}w^1,Q-u,Q+u-\sqrt{p'\over p} \bh_j\right)
\\
&~~~~\times C_{W_N}\left(Q-u+\sqrt{p'\over p} \bh_j,\sqrt{p'\over p}w^{N-1},u+Q\right)
\fe
%\ie
%&C^{Q-u+\sqrt{p'\over p} \bh_j}_{\sqrt{p'\over p}w^1,Q-u}=\left(-{\pi\m\over \gamma({p'\over p})}\right)^{j-1}\prod^{j-1}_{i=1}{\gamma(\sqrt{p'\over p}P_{ji})\over \gamma(1-{p'\over p}+\sqrt{p'\over p}P_{ji})},
%\\
%&C^{u+Q-\sqrt{p'\over p} \bh_j}_{\sqrt{p'\over p}w^{N-1},u+Q}=\left(-{\pi\m\over \gamma({p'\over p})}\right)^{N-j}\prod^{N}_{i=j+1}{\gamma(\sqrt{p'\over p}P_{ji})\over \gamma(1-{p'\over p}+\sqrt{p'\over p}P_{ji})}.
%\fe
%\ie
%&C^{Q-u+\sqrt{p'\over p} \bh_j}_{\sqrt{p'\over p}w^1,Q-u}C^{u+Q-\sqrt{p'\over p} \bh_j}_{\sqrt{p'\over p}w^{N-1},u+Q}=\left(-{\pi\m\over \gamma({p'\over p})}\right)^{N-1}\prod^{N}_{i=1,i\neq j}{\gamma(\sqrt{p'\over p}P_{ji})\over \gamma(1-{p'\over p}+\sqrt{p'\over p}P_{ji})}
%\\
%&=\left(-{\pi\m\Gamma(1-{p'\over p})\over \Gamma({p'\over p})}\right)^{N-1}\prod^{N}_{i=1,i\neq j}{\Gamma(\sqrt{p'\over p}P_{ji})\Gamma({p'\over p}-\sqrt{p'\over p}P_{ji})\over \Gamma(1-{p'\over p}+\sqrt{p'\over p}P_{ji})\Gamma(1-\sqrt{p'\over p}P_{ji})}
%\fe
The $u$-monodromy acts as a phase on the $u$-channel conformal blocks (\ref{ucb}). The four-point function (\ref{H}) is invariant.

\section{$q$-expansion of the torus two-point function}

In this section, we study the $q$-expansion of the torus conformal block (\ref{tccbtt}). Let us start by expanding (\ref{bostpf}) as
\ie
G_u^{bos}(z_1,z_2|\tau)= \sum_{n\in\Gamma_{pp'}} q^{-{N-1\over 24}+{1\over 2}(u+n)^2}\left[ G_{u+n}^{bos,(0)}(z_1,z_2)+ G_{u+n}^{bos,(1)}(z_1,z_2)q+\cO(q^2)\right],
\fe
where $G_u^{bos,(n)}(z_1,z_2)$ are obtained from the $q$-expansion of the $\theta_1$ and $\eta$ functions in (\ref{bostpf}). For simplicity, here we will assume that $N$ is sufficiently large, and examine only the first few terms in the $q$ expansion. For this purpose, we can ignore the sum over the lattice $\Gamma_{pp'}$ by setting $n=0$, while restricting $u\in \Gamma^*_{pp'}/\Gamma_{pp'}$ to take the value in the equivalence class that minimize $u^2$, since the effects of nonzero $n$ only come in of the order $q^{\sim N^2}$. Plugging this formula into (\ref{alttsum}) and (\ref{tccbtt}), we obtain
\ie\label{expaaa}
{\cal G}^t_{\lambda+\lambda'}(z_1,z_2|\tau)=\sum_{w\in W}q^{-{N-1\over 24}+{1\over 2}(\lambda+w(\lambda'))^2}\left[ \bG_{\lambda+w(\lambda')}^{(0)}(z_1,z_2)+ \bG_{\lambda+w(\lambda')}^{(1)}(z_1,z_2)q+\cO(q^2)\right].
\fe
Next, we expand the product of theta functions in (\ref{bostpf}),
\ie\label{thexpand}
&{1\over \eta(\tau)^{N-1}} \left( {\theta_1(z_{12}|\tau)\over \partial_z\theta_1(0|\tau)} \right)^{{p'\over pN}}
\left( {\theta_1(z_1-t_1|\tau)\over \partial_z\theta_1(0|\tau)} \right)^{-{p'\over p}}
\left( {\theta_1(z_2-t_{N-1}|\tau)\over \partial_z\theta_1(0|\tau)} \right)^{-{p'\over p}}
 \prod_{i=1}^{N-2}\left( {\theta_1(t_{i,i+1}|\tau)\over \partial_z\theta_1(0|\tau)} \right)^{-{p'\over p}}
 \\
 &= q^{-{N-1\over 24}}\left({i\over 4\pi}\right)^{{p'\over pN}-{p'\over p}N}\left( {x_{12}\over  \sqrt{x_1x_2}} \right)^{{p'\over pN}}
\left(  {x_{1}-s_1\over  \sqrt{x_1s_1}}\right)^{-{p'\over p}}
\left(  {x_{2}-s_{N-1}\over  \sqrt{s_{N-1}x_2}}\right)^{-{p'\over p}}
 \prod_{i=1}^{N-2}\left(  {s_{i,i+1}\over  \sqrt{s_is_{i+1}}}\right)^{-{p'\over p}}
 \\
 &~~~~\times\left[1+\left(N-1-{p'\over pN}{x_{12}^2\over x_1 x_2}\right)q+{p'\over p}\left(\sum_{k=1}^{N-1}{s_{k-1,k}^2\over s_{k-1}s_k}+{(x_2-s_{N-1})^2\over s_{N-1}x_2}\right)q+\cO(q^2)\right],
\fe
where $s_0\equiv x_1$, and we have made a conformal transformation $x_i=e^{2\pi i z_i}$ and $s_i=e^{2\pi i t_i}$. The zeroth order term in this expansion, after the contour integral, gives\footnote{Here a conformal factor of the form $x_1^{h_{\bff}}x_2^{h_{\bar\bff}}$, together with the factors in (\ref{fivethirty}), is included in rewriting ${\bf G}_u^{(0)}$ in terms of the sphere four-point conformal block ${\bf G}_u$.}
%\ie
%{\cal G}_u^{sc,t,(0)}(z_1,z_2|\tau)\propto \bG_u\left({x_1\over x_2}\right),
%\fe
%which is the sphere four-point conformal block. In this order, the torus two-point conformal block (\ref{altsum}) is the sum of different sphere four-point conformal blocks:
\ie\label{bG0}
\bG_{u}^{(0)}(z_1,z_2) =\left({i\over 4\pi}\right)^{{p'\over pN}-{p'\over p}N} (x_2-x_1)^{{p'\over Np}} x_1^{{p'(N-1)\over 2pN}} x_2^{{p'(N-1)\over 2pN}-{p'\over p}}\bG_{u}\left({x_1\over x_2}\right).
\fe
The first order terms in the expansion (\ref{expaaa}) can be split into three terms,
\ie\label{expppp}
\bG_{u}^{(1)}(z_1,z_2)=\bG_{u}^{(1),1}(z_1,z_2)+\bG_{u}^{(1),2}(z_1,z_2)+\bG_{u}^{(1),3}(z_1,z_2),
\fe
coming from the three terms of order $q$ in the second line of (\ref{thexpand}),
\ie
\left(N-1+{p'\over pN}{x_{12}^2\over x_1 x_2}\right),~~~~{p'\over p}\sum_{k=1}^{N-1}{s_{k-1,k}^2\over s_{k-1}s_k},~~~~{p'\over p}\sum_{k=1}^{N-1}{(x_2-s_{N-1})^2\over s_{N-1}x_2}.
\fe
The first term is independent of $s_i$ and its contribution is proportional to ${\bf G}_u^{(0)}$ after doing the contour integral. The second term of (\ref{expppp}) is computed as
\ie
&\bG^{(1),2}_u\left(z_1,z_2\right)=\left({i\over 4\pi}\right)^{{p'\over pN}-{p'\over p}N} {p'\over p}\sum_{k=1}^{N-1}{s_{k-1,k}^2\over s_{k-1}s_k} (x_2-x_1)^{{p'\over Np}} x_1^{\sqrt{p'\over p}P_{1}+{p'(N-1)\over 2pN}} x_2^{-\sqrt{p'\over p}P_{N}+{p'(N-1)\over 2pN}}
\\
&~~~~\times\int^{x_1}_0 ds_1~s_1^{-\sqrt{p'\over p}(u+Q)\cdot \A_1}\left(x_1-s_1\right)^{-{p'\over p}}
\\
&~~~~\times\left(\prod_{i=1}^{N-2}\int^{s_{i}}_0 ds_{i+1}~s_{i+1}^{-\sqrt{p'\over p}(u+Q)\cdot \A_{i+1}}\left(s_{i}-s_{i+1}\right)^{-{p'\over p}}\right)\left(x_2-s_{N-1}\right)^{-{p'\over p}}
%\\
%=&x_2^{{p'\over p}+P_1-P_N}\int^{x_1}_0 ds_1~s_1^{-\sqrt{p'\over p}(u+Q)\cdot \A_1-\sqrt{p'\over p}u\cdot\sum^{N-1}_{j=2}\A_j}\left(x_1-s_1\right)^{-{p'\over p}}
%\\
%&\times\left(\prod_{i=1}^{N-2}\int^{1}_0 dt_{i+1}~t_{i+1}^{-\sqrt{p'\over p}(u+Q)\cdot \A_{i+1}-\sqrt{p'\over p}u\cdot\sum^{N-1}_{j=i+2}\A_j}\left(1-t_{i+1}\right)^{-{p'\over p}}\right)\left(x_2-s_{N-1}\right)^{-{p'\over p}}
%\\
%=&x_2^{{p'\over p}+P_1-P_N}\left(\prod_{i=1}^{N-1}\int^{1}_0 dt_{i}~t_{i}^{-\sqrt{p'\over p}(u+Q)\cdot \A_{i}-\sqrt{p'\over p}u\cdot\sum^{N-1}_{j=i+1}\A_j}\left(1-t_{i}\right)^{-{p'\over p}}\right)\left(x_2-x_1\prod^{N-1}_{i=1} t_{i}\right)^{-{p'\over p}}
%\\
%&=\left({i\over 4\pi}\right)^{{p'\over pN}-{p'\over p}N}(x_2-x_1)^{{p'\over Np}} x_1^{\sqrt{p'\over p}P_{N}+{p'(N-1)\over 2pN}} x_2^{-\sqrt{p'\over p}P_{N}+{p'(N-1)\over 2pN}+{p'\over p}}
%\\
%&~~~~\times {p'\over p}\sum_{k=1}^{N-1}\left(\prod_{i=1}^{N-1}\int^{1}_0 dt_i~t_i^{-1+{p'\over p}+\sqrt{p'\over p}P_{N,i}-\delta_{i,k}}\left(1-t_i\right)^{-{p'\over p}+2\delta_{i,k}}\right)\left(1-{x_1\over x_2}\prod^{N-1}_{i=1} t_{i}\right)^{-{p'\over p}}
\\
&=\left({i\over 4\pi}\right)^{{p'\over pN}-{p'\over p}N}(x_2-x_1)^{{p'\over Np}} x_1^{\sqrt{p'\over p}P_{N}+{p'(N-1)\over 2pN}} x_2^{-\sqrt{p'\over p}P_{N}+{p'(N-1)\over 2pN}-{p'\over p}}
\\
&~~~~\times{p'\over p}\sum_{k=1}^{N-1} { \prod^{N}_{i=1}\Gamma(\sqrt{p'\over p}P_{N,i}+{p'\over p}-\delta_{i,k})\over \prod^{N}_{i=1}\Gamma(\sqrt{p'\over p}P_{N,i}+1+\delta_{i,k})} {\Gamma(1-{p'\over p})^{N-1}\over \Gamma({p'\over p})}(1-{p'\over p})(2-{p'\over p})
\\
&~~~~\times {}_{N} F_{N-1}(\vec\m_{N}-\delta_{k};\widehat{\vec\n}_N+\delta_{k}|{x_1\over x_2}),
\fe
where $\vec 1=(1,\cdots,1)$ and $(\vec \delta_k)_i=\delta_{k,i}$. The third term of (\ref{expppp}) is given by
%\ie
%&{(s_{N-1}-x_2)^2\over x_2 s_{N-1}}
%\\
%&={(1-{x_1\over x_2}\prod^{N-1}_{i=1}t_i)^2\over {x_1\over x_2}\prod^{N-1}_{i=1}t_i}
%\fe
\ie
&\bG^{(1),3}_u\left(z_1,z_2\right)=\left({i\over 4\pi}\right)^{{p'\over pN}-{p'\over p}N}{p'\over p}{(s_{N-1}-x_2)^2\over x_2 s_{N-1}} (x_2-x_1)^{{p'\over Np}} x_1^{\sqrt{p'\over p}P_{1}+{p'(N-1)\over 2pN}} x_2^{-\sqrt{p'\over p}P_{N}+{p'(N-1)\over 2pN}}
\\
&~~\times\int^{x_1}_0 ds_1~s_1^{-\sqrt{p'\over p}(u+Q)\cdot \A_1}\left(x_1-s_1\right)^{-{p'\over p}}\left(\prod_{i=1}^{N-2}\int^{s_{i}}_0 ds_{i+1}~s_{i+1}^{-\sqrt{p'\over p}(u+Q)\cdot \A_{i+1}}\left(s_{i}-s_{i+1}\right)^{-{p'\over p}}\right)\left(x_2-s_{N-1}\right)^{-{p'\over p}}
%\\
%=&x_2^{{p'\over p}+P_1-P_N}\int^{x_1}_0 ds_1~s_1^{-\sqrt{p'\over p}(u+Q)\cdot \A_1-\sqrt{p'\over p}u\cdot\sum^{N-1}_{j=2}\A_j}\left(x_1-s_1\right)^{-{p'\over p}}
%\\
%&\times\left(\prod_{i=1}^{N-2}\int^{1}_0 dt_{i+1}~t_{i+1}^{-\sqrt{p'\over p}(u+Q)\cdot \A_{i+1}-\sqrt{p'\over p}u\cdot\sum^{N-1}_{j=i+2}\A_j}\left(1-t_{i+1}\right)^{-{p'\over p}}\right)\left(x_2-s_{N-1}\right)^{-{p'\over p}}
%\\
%=&x_2^{{p'\over p}+P_1-P_N}\left(\prod_{i=1}^{N-1}\int^{1}_0 dt_{i}~t_{i}^{-\sqrt{p'\over p}(u+Q)\cdot \A_{i}-\sqrt{p'\over p}u\cdot\sum^{N-1}_{j=i+1}\A_j}\left(1-t_{i}\right)^{-{p'\over p}}\right)\left(x_2-x_1\prod^{N-1}_{i=1} t_{i}\right)^{-{p'\over p}}
%\\
%&=\left({i\over 4\pi}\right)^{{p'\over pN}-{p'\over p}N}  (x_2-x_1)^{{p'\over Np}} x_1^{\sqrt{p'\over p}P_{N}+{p'(N-1)\over 2pN}-1} x_2^{-\sqrt{p'\over p}P_{N}+{p'(N-1)\over 2pN}-{p'\over p}+1}
%\\
%&~~~~\times{p'\over p}\left(\prod_{i=1}^{N-1}\int^{1}_0 dt_i~t_i^{-2+{p'\over p}+\sqrt{p'\over p}P_{N,i}}\left(1-t_i\right)^{-{p'\over p}}\right)\left(1-{x_1\over x_2}\prod^{N-1}_{i=1} t_{i}\right)^{-{p'\over p}+2}
\\
&=\left({i\over 4\pi}\right)^{{p'\over pN}-{p'\over p}N} (x_2-x_1)^{{p'\over Np}} x_1^{\sqrt{p'\over p}P_{N}+{p'(N-1)\over 2pN}-1} x_2^{-\sqrt{p'\over p}P_{N}+{p'(N-1)\over 2pN}-{p'\over p}+1}
\\
&~~~~\times{p'\over p}{ \prod^{N}_{k=1}\Gamma(\sqrt{p'\over p}P_{N,k}+{p'\over p}-1)\over \prod^{N-1}_{k=1}\Gamma(\sqrt{p'\over p}P_{N,k})} {\Gamma(1-{p'\over p})^{N-1}\over \Gamma({p'\over p}-1)} {}_{N} F_{N-1}(\vec\m_{N}-\vec 1-\vec\delta_{N};\widehat{\vec\n}_N-\vec 1 |{x_1\over x_2}).
\fe
Using the identity (\ref{transgh}), $\bG^{(1),2}_u$ and $\bG^{(1),3}_u$ can be combined into
\ie
&\bG^{(1),2}_u+\bG^{(1),3}_u
\\
&=\left({i\over 4\pi}\right)^{{p'\over pN}-{p'\over p}N} (x_2-x_1)^{{p'\over Np}} x_1^{{p'(N-1)\over 2pN}} x_2^{{p'(N-1)\over 2pN}-{p'\over p}}{p'\over p}(1-{p'\over p})(2-{p'\over p})\sum_{k=1}^{N}{\prod^{N-1}_{i=1}\Gamma(\sqrt{p'\over p}P_{N,i}+1)\over \prod^N_{i=1}\Gamma(\sqrt{p'\over p}P_{N,i}+{p'\over p})}
\\
&\times\sum^N_{m=1}{\Gamma(\sqrt{p'\over p}P_{N,m}+{p'\over p})\Gamma(\sqrt{p'\over p}P_{m,N}+1-{p'\over p})\prod^N_{j=1,j\neq m}\Gamma(\sqrt{p'\over p}P_{m,j}+\delta_{m,k}-\delta_{j,k})\over \prod^{N}_{j=1}\Gamma(\sqrt{p'\over p}P_{m,j}+1-{p'\over p}+\delta_{j,k}+\delta_{m,k})}
\\
&\times e^{i\pi\left(\sqrt{p'\over p}P_{m,l}-{p'\over p}\right)}\left({x_1\over x_2}\right)^{\sqrt{p'\over p}P_m-{p'\over p}+\delta_{m,k}}{}_N F_{N-1}(\vec\m_m'-\delta_{k,m}\vec 1-\vec\delta_k;\widehat{\vec\n}_m'-\delta_{k,m}\vec 1+(1-\delta_{k,m})\vec\delta_k |{x_2\over x_1}).
\fe

\section{Thermal two-point function in Virasoro minimal models}

In this appendix, we study numerically the torus two-point function of $(\bff,0)$ with $(\bar\bff,0)$, and its analytic continuation to Lorentzian signature, in the $N=2$ case, i.e. Virasoro minimal model. The result was first derived in \cite{Jayaraman:1988ex}, and is a special case of our formulae for general $N$.

The formula in terms of summation over $t$ channel conformal blocks in this case is
\ie
\langle {\cal O}_{v_1}(z_1,\bar z_1){\cal O}_{v_2}(z_2,\bar z_2)\rangle_\tau 
%= {1\over N}\sum_{\lambda\in\Delta_1,~\lambda'\in\Delta_2,~w\in W}  \big|{\cal G}^t_{w(\lambda+\lambda')}(z_1,z_2|\tau)\big|^2
%\\
&= {1\over 2}\sum_{r=1}^{p-1}\sum_{s=1}^{p'-1} \left[  \left|{\cal C}_r{\cal G}^t_{p'r-ps\over \sqrt{2pp'}}(z_1,z_2|\tau)\right|^2 + \left|{\cal C}_{-r}{\cal G}^t_{-p'r+ps\over \sqrt{2pp'}}(z_1,z_2|\tau)\right|^2 \right]
\\
&= \sum_{r=1}^{p-1}\sum_{s=1}^{p'-1}  \left|{\cal C}_r{\cal G}^t_{p'r-ps\over \sqrt{2pp'}}(z_1,z_2|\tau)\right|^2 .
\fe
The subscript of the conformal block ${\cal G}_u^t$, $u={p'r-ps\over \sqrt{2pp'}}$, is the charge associated with the $(r,s)$ primary in the $t$-channel, normalized such that the fundamental weight is ${1\over \sqrt{2}}$. The normalization factor ${\cal C}_r$ is given by
\ie\label{crvir}
{\cal C}_r=  {1\over \Gamma(1-{p'\over p})} \left[ -\gamma({p'\over p})\gamma(2(1-{p'\over p})) {\sin (\pi{p'\over p}(r-1)) \over \sin(\pi {p'\over p}r)} \right]^{1\over 2} {\cal N}_r^{-1}.
\fe
We will also write ${\cal G}^t_u$ as ${\cal G}^t_{(r,s)}$. It is obtained from the free boson correlator by the contour integral
\ie
& {\cal G}^{t}_{(r,s)}(z_1,z_2|\tau) = \int_{L(0,z_1)} dt \, {\cal G}^{bos}_{(r,s)}(z_1,z_2,t|\tau),
\\
& {\cal G}^{bos}_{(r,s)}(z_1,z_2,t|\tau) = G^{bos}_{(r,s)}(z_1,z_2,t|\tau) - G^{bos}_{(r,-s)}(z_1,z_2,t|\tau).
\fe
$G^{bos}$ is given explicitly by
\ie\label{minexaaa}
&G^{bos}_{(r,s)}(z_1,z_2,t|\tau) 
%= {1\over \eta(\tau)}  \left( {\theta_1(z_{12}|\tau) \over \partial_z\theta_1(0|\tau)} \right)^{p'\over 2p} \left( {\theta_1(z_1-t|\tau) \over \partial_z\theta_1(0|\tau)} \right)^{-{p'\over p}} \left( {\theta_1(z_2-t|\tau) \over \partial_z\theta_1(0|\tau)} \right)^{-{p'\over p}}
%\\
%&~~~\times
%\sum_{n=-\infty}^\infty q^{{1\over 2} (u+\sqrt{2pp'} n)^2} \exp\left[ 2\pi i \sqrt{p'\over 2p} (u+\sqrt{2pp'} n) (z_1+z_2-2t) \right].
%\fe
%For $u={p'r-ps\over \sqrt{2pp'}}$, we will also write it as
%\ie
%&G^{bos}_{(r,s)}(z_1,z_2,t|\tau) 
= {1\over \eta(\tau)}  \left( {\theta_1(z_{12}|\tau) \over \partial_z\theta_1(0|\tau)} \right)^{p'\over 2p} \left( {\theta_1(z_1-t|\tau) \over \partial_z\theta_1(0|\tau)} \right)^{-{p'\over p}} \left( {\theta_1(z_2-t|\tau) \over \partial_z\theta_1(0|\tau)} \right)^{-{p'\over p}}
\\
&~~~\times
\sum_{n=-\infty}^\infty q^{pp' ({p'r-ps\over 2pp'}+ n)^2} \exp\left[ 2\pi i ({p'r-ps\over 2p}+ p'n) (z_1+z_2-2t) \right].
\fe
In the explicit evaluation of the two-point function below, we will restrict to the special case $\tau=i\beta$, $z_1=0$, $z_2=1/2$, and compute
\ie
{\cal G}^t_{(r,s)}\big(0,{1\over 2}|i\beta\big).
\fe
%Then, we analytically continue ${\rm Im}z$, and write the Lorentzian two point function as
%\ie
%\langle {\cal T}{\cal O}_{v_2}(0,t){\cal O}_{v_2}({1\over 2},0)\rangle_\beta &= \langle {\cal O}_{v_2}(0,t-i\epsilon){\cal O}_{v_2}({1\over 2},0)\rangle_\beta
%\\
%&
%= \sum_{r=1}^{p-1}\sum_{s=1}^{p'-1}  {\cal G}^t_{(r,s)}(t-i\epsilon,{1\over 2}|\tau)\overline{\cal G}^t_{(r,s)}(-t+i\epsilon,{1\over 2}|\tau) .
%\fe
At positive integer values of time, $t=m>0$, we have
\ie
& \langle {\cal O}_{v_1}(0,m){\cal O}_{v_2}({1\over 2},0)\rangle_\beta
= \sum_{r=1}^{p-1}\sum_{s=1}^{p'-1} e^{2\pi i m {p'\over p}(-r+{1\over 2}) } \left|{\cal G}^t_{(r,s)}(0,{1\over 2}|i\beta)\right|^2.
\fe
The integral is evaluated numerically using the following contour, which is convenient when the fractional powers of $\theta_1(z|\tau)$ in (\ref{minexaaa}) is defined with a branch cut along the positive real $z$ axis.

\bigskip
\bigskip

\centerline{  \begin{fmffile}{numcontour}
        \begin{tabular}{c}
            \begin{fmfgraph*}(200,200)
                \fmfstraight
                \fmfleft{i1,i2,i3,i4,e,i5}
                \fmfright{o1,o2,o3,o4,f,o5}
                \fmf{dashes,tension=0}{i1,i2,i3,i4,i5,o5,o4,o3,o2,o1,i1}
                \fmf{phantom}{i2,a1,a2,a3,a4,o2}
                \fmf{phantom}{i4,a,z,b,c,o4}
                \fmffixed{(0.06w,-0.03h)}{a3,w1}
                \fmffixed{(0.03w,0)}{w1,w2}
                \fmffixed{(0,-0.03h)}{o2,k2}
                \fmffixed{(0,-0.03h)}{i2,k1}
                \fmffixed{(0.03w,0)}{a3,u}
                \fmffixed{(0.03w,0)}{b,b1}
                \fmffixed{(0,-0.03h)}{b,b2}
                \fmffixed{(0.03w,-0.03h)}{b,b3}
                \fmffixed{(0.06w,-0.03h)}{b,v1}
                \fmffixed{(0.09w,-0.03h)}{b,v2}
                \fmf{plain_arrow,tension=0}{w2,k2}
                \fmf{plain_arrow,tension=0}{k1,w1}
                \fmf{plain_arrow,tension=0}{a3,i2}
                \fmf{plain_arrow,tension=0}{o2,u}
                \fmf{plain_arrow,tension=0}{b2,a3}
                \fmf{plain_arrow,tension=0}{u,b3}
                \fmf{plain_arrow,tension=0}{w1,v1}
                \fmf{plain_arrow,tension=0}{v2,w2}
                \fmf{plain,tension=0}{b,v1}
                \fmf{plain,tension=0}{b1,v2}
                \fmf{plain_arrow,tension=0,right=30}{b,b2}
                \fmf{plain_arrow,tension=0,left=30}{b3,b1}
                \fmf{wiggly,tension=0}{z,c}
                \fmfv{decor.shape=cross, label=$z_1$}{z}
             \end{fmfgraph*}
        \end{tabular}
        \end{fmffile}
}

\bigskip
\bigskip

\noindent The results for minimal models up to $k=30$ are plotted in Figures 1 and 2. At large values of $k$, while the Poincare recurrence times is of order $k$, the two-point function is already  ``thermalized" at $t=1$.

\begin{figure}
\begin{center}
\includegraphics[width=150mm]{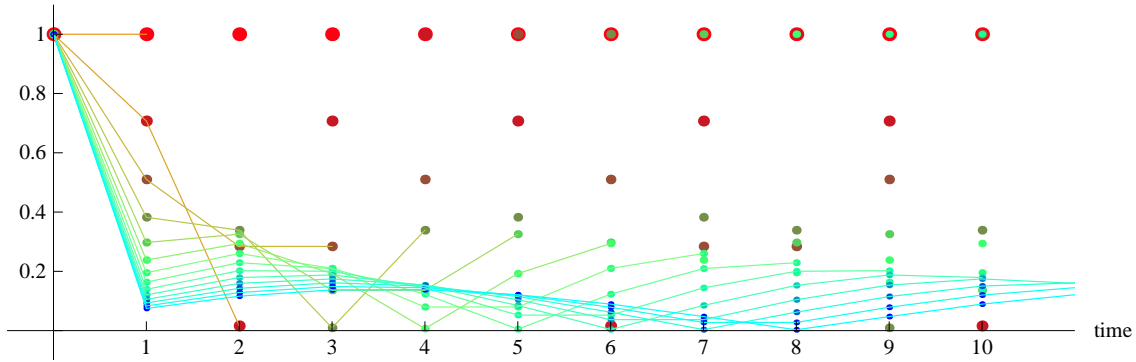}
\parbox{13cm}{
\caption{
The modulus of the two-point function $\langle {\cal O}(0,t) {\cal O}(0,0)\rangle_\beta$ (normalized to 1 at $t=0$) at inverse temperature $\beta=0.3$ is plotted at integer values of time $t=0,1,2,\cdots,10$. The results for Virasoro minimal models with $k=1,2,\cdots,14$ are shown in colors ranging from red to green and then to blue. For each $k$, the values of the modulus of the two-point function at integer times before Poincar\'e recurrence are connected with straight lines, for the purpose of illustration only.
\label{smallkplot}}}
\end{center}
\end{figure}

\begin{figure}
\begin{center}
\includegraphics[width=150mm]{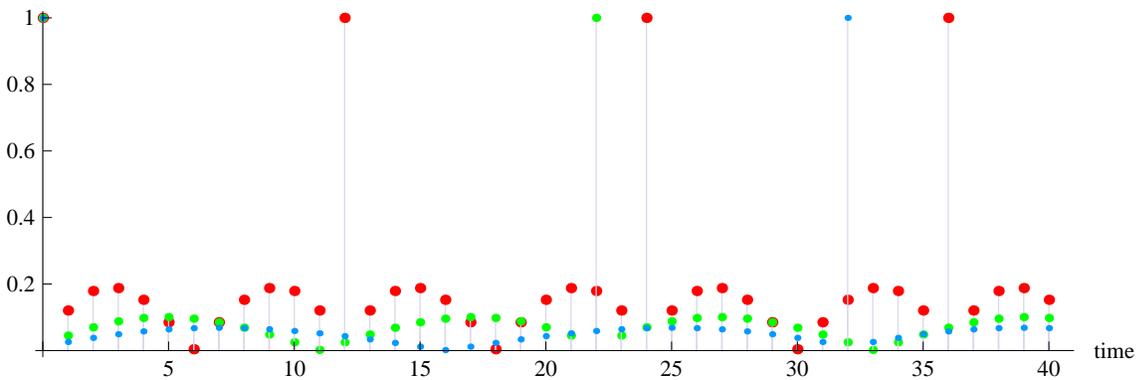}
\parbox{13cm}{
\caption{
The modulus of the two-point function $\langle {\cal O}(0,t) {\cal O}(0,0)\rangle_\beta$ (normalized to 1 at $t=0$) at inverse temperature $\beta=0.3$ is plotted at integer values of time $t=0,1,2,\cdots,40$, in Virasoro minimal models of $k=10,20,30$ (shown in red, green, and blue).
\label{smallkplot}}}
\end{center}
\end{figure}

We also plotted the two-point function at various temperatures, ranging from $0.05$ to $20$ (times the self-dual temperature), at integer times in the $k=4$ Virasoro minimal model, in Figure 3.

\begin{figure}
\begin{center}
\includegraphics[width=120mm]{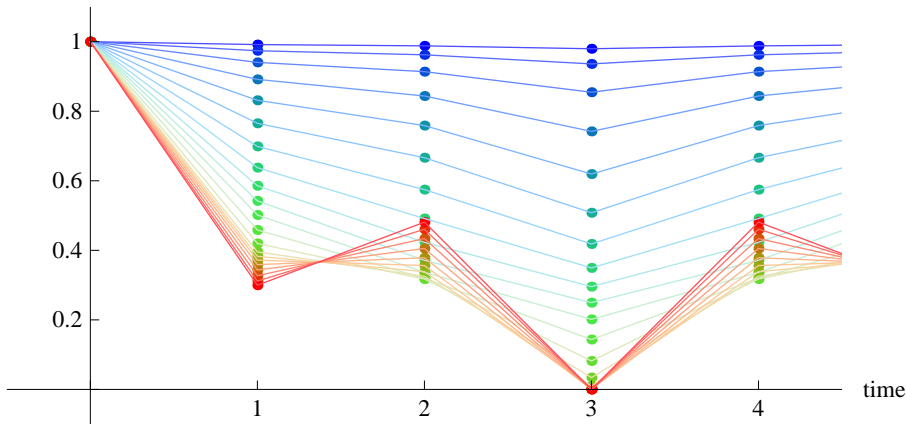}
\parbox{13cm}{
\caption{
Plots of the modulus of the two-point function $\langle {\cal O}(0,t) {\cal O}(0,0)\rangle_\beta$ (normalized to 1 at $t=0$) in the $k=4$ Virasoro minimal model, at integer values of time $t=0,1,\cdots,4$ (connected with fictitious straight lines for illustration only), at different values of the temperature $T=1/\beta$. $T$ ranges from $\sim 0.05$ to $20$ (depicted in colors ranging from blue to red), evenly spaced in logarithmic scale.
\label{smallkplot}}}
\end{center}
\end{figure}


\begin{thebibliography}{}

%\cite{Maldacena:1997re}
\bibitem{Maldacena:1997re}
  J.~M.~Maldacena,
  ``The large N limit of superconformal field theories and supergravity,''
  Adv.\ Theor.\ Math.\ Phys.\  {\bf 2}, 231 (1998)
  [Int.\ J.\ Theor.\ Phys.\  {\bf 38}, 1113 (1999)]
  [arXiv:hep-th/9711200];
  S.~S.~Gubser, I.~R.~Klebanov and A.~M.~Polyakov,
  ``Gauge theory correlators from non-critical string theory,''
  Phys.\ Lett.\  B {\bf 428}, 105 (1998)
  [arXiv:hep-th/9802109];
  E.~Witten,
  ``Anti-de Sitter space and holography,''
  Adv.\ Theor.\ Math.\ Phys.\  {\bf 2}, 253 (1998)
  [arXiv:hep-th/9802150].
  %%CITATION = 00203,2,253;%%

%\cite{Klebanov:2002ja}
\bibitem{Klebanov:2002ja}
  I.~R.~Klebanov and A.~M.~Polyakov,
  ``AdS dual of the critical O(N) vector model,''
  Phys.\ Lett.\  B {\bf 550}, 213 (2002)
  [arXiv:hep-th/0210114].
  %%CITATION = PHLTA,B550,213;%%
  
%\cite{Sezgin:2002rt}
\bibitem{Sezgin:2002rt}
  E.~Sezgin and P.~Sundell,
  ``Massless higher spins and holography,''
  Nucl.\ Phys.\  B {\bf 644}, 303 (2002)
  [Erratum-ibid.\  B {\bf 660}, 403 (2003)]
  [arXiv:hep-th/0205131].
  %%CITATION = NUPHA,B644,303;%%

%\cite{Giombi:2011kc}
\bibitem{Giombi:2011kc} 
  S.~Giombi, S.~Minwalla, S.~Prakash, S.~P.~Trivedi, S.~R.~Wadia and X.~Yin,
  ``Chern-Simons Theory with Vector Fermion Matter,''
  arXiv:1110.4386 [hep-th].
  %%CITATION = ARXIV:1110.4386;%%

%\cite{Vasiliev:1999ba}
\bibitem{Vasiliev:1999ba}
  M.~A.~Vasiliev,
  ``More On Equations Of Motion For Interacting Massless Fields Of All Spins In
  (3+1)-Dimensions,''
  Phys.\ Lett.\  B {\bf 285}, 225 (1992); M.~A.~Vasiliev,
  ``Higher-spin gauge theories in four, three and two dimensions,''
  Int.\ J.\ Mod.\ Phys.\  D {\bf 5}, 763 (1996)
  [arXiv:hep-th/9611024]; M.~A.~Vasiliev,
  ``Higher spin gauge theories: Star-product and AdS space,''
  arXiv:hep-th/9910096; M.~A.~Vasiliev,
  ``Nonlinear equations for symmetric massless higher spin fields in
  (A)dS(d),''
  Phys.\ Lett.\  B {\bf 567}, 139 (2003)
  [arXiv:hep-th/0304049].
  %%CITATION = HEP-TH/9910096;%%

%\bibitem{Vasiliev:1991}
%  M.~A.~Vasiliev,
%  ``Higher Spin Algebras and Quantization on the Sphere and Hyperboloid,''
%  Int.\ J.\ Mod.\ Phys.\ A {\bf 6}, 1115 (1991).


%\cite{Giombi:2009wh}
\bibitem{Giombi:2009wh}
  S.~Giombi and X.~Yin,
  ``Higher Spin Gauge Theory and Holography: The Three-Point Functions,''
  arXiv:0912.3462 [hep-th].
  %%CITATION = ARXIV:0912.3462;%

%\cite{Giombi:2010vg}
\bibitem{Giombi:2010vg}
  S.~Giombi and X.~Yin,
  ``Higher Spins in AdS and Twistorial Holography,''
  arXiv:1004.3736 [hep-th].
  %%CITATION = ARXIV:1004.3736;%%

%\cite{Koch:2010cy}
\bibitem{Koch:2010cy}
  R.~d.~M.~Koch, A.~Jevicki, K.~Jin, J.~P.~Rodrigues,
  ``$AdS_4/CFT_3$ Construction from Collective Fields,''
  Phys.\ Rev.\  {\bf D83}, 025006 (2011).
  [arXiv:1008.0633 [hep-th]].
  
%\cite{Douglas:2010rc}
\bibitem{Douglas:2010rc}
  M.~R.~Douglas, L.~Mazzucato, S.~S.~Razamat,
  ``Holographic dual of free field theory,''
  Phys.\ Rev.\  {\bf D83}, 071701 (2011).
  [arXiv:1011.4926 [hep-th]].


%\cite{Giombi:2011ya}
\bibitem{Giombi:2011ya} 
  S.~Giombi and X.~Yin,
  ``On Higher Spin Gauge Theory and the Critical O(N) Model,''
  arXiv:1105.4011 [hep-th].
  %%CITATION = ARXIV:1105.4011;%%


%\cite{Maldacena:2011jn}
\bibitem{Maldacena:2011jn} 
  J.~Maldacena and A.~Zhiboedov,
  ``Constraining conformal field theories with a higher spin symmetry,''
  arXiv:1112.1016 [hep-th].
  %%CITATION = ARXIV:1112.1016;%%

%\cite{Bouwknegt:1992wg}
\bibitem{Bouwknegt:1992wg}
  P.~Bouwknegt and K.~Schoutens,
  ``W symmetry in conformal field theory,''
  Phys.\ Rept.\  {\bf 223}, 183 (1993)
  [arXiv:hep-th/9210010].
  %%CITATION = PRPLC,223,183;%%


%\cite{Gaberdiel:2010ar}
\bibitem{Gaberdiel:2010ar}
  M.~R.~Gaberdiel, R.~Gopakumar, A.~Saha,
  ``Quantum $W$-symmetry in $AdS_3$,''
  JHEP {\bf 1102}, 004 (2011).
  [arXiv:1009.6087 [hep-th]].

%\cite{Gaberdiel:2010pz}
\bibitem{Gaberdiel:2010pz}
  M.~R.~Gaberdiel, R.~Gopakumar,
  ``An $AdS_3$ Dual for Minimal Model CFTs,''
  Phys.\ Rev.\  {\bf D83}, 066007 (2011).
  [arXiv:1011.2986 [hep-th]].


%\cite{Henneaux:2010xg}
\bibitem{Henneaux:2010xg}
  M.~Henneaux, S.~-J.~Rey,
  ``Nonlinear $W_{infinity}$ as Asymptotic Symmetry of Three-Dimensional Higher Spin Anti-de Sitter Gravity,''
  JHEP {\bf 1012}, 007 (2010).
  [arXiv:1008.4579 [hep-th]].

%\cite{Campoleoni:2010zq}
\bibitem{Campoleoni:2010zq}
  A.~Campoleoni, S.~Fredenhagen, S.~Pfenninger, S.~Theisen,
  ``Asymptotic symmetries of three-dimensional gravity coupled to higher-spin fields,''
  JHEP {\bf 1011}, 007 (2010).
  [arXiv:1008.4744 [hep-th]].
  
%\cite{Gaberdiel:2011wb}
\bibitem{Gaberdiel:2011wb}
  M.~R.~Gaberdiel, T.~Hartman,
  ``Symmetries of Holographic Minimal Models,''
  JHEP {\bf 1105}, 031 (2011).
  [arXiv:1101.2910 [hep-th]].

%\cite{Kiritsis:2010xc}
\bibitem{Kiritsis:2010xc}
  E.~Kiritsis, V.~Niarchos,
  ``Large-N limits of 2d CFTs, Quivers and $AdS_3$ duals,''
  JHEP {\bf 1104}, 113 (2011).
  [arXiv:1011.5900 [hep-th]].

%\cite{Castro:2010ce}
\bibitem{Castro:2010ce}
  A.~Castro, A.~Lepage-Jutier, A.~Maloney,
  ``Higher Spin Theories in $AdS_3$ and a Gravitational Exclusion Principle,''
  JHEP {\bf 1101}, 142 (2011).
  [arXiv:1012.0598 [hep-th]].


%\cite{Ahn:2011pv}
\bibitem{Ahn:2011pv}
  C.~Ahn,
  ``The Large N 't Hooft Limit of Coset Minimal Models,''
  [arXiv:1106.0351 [hep-th]].
  

%\cite{Gaberdiel:2011zw}
\bibitem{GGHR}
  M.~R.~Gaberdiel, R.~Gopakumar, T.~Hartman and S.~Raju,
  ``Partition Functions of Holographic Minimal Models,''
  arXiv:1106.1897 [hep-th].
  %%CITATION = ARXIV:1106.1897;%

  
  
%\cite{Chang:2011mz}
\bibitem{Chang:2011mz} 
  C.~-M.~Chang and X.~Yin,
  ``Higher Spin Gravity with Matter in $AdS_3$ and Its CFT Dual,''
  arXiv:1106.2580 [hep-th].
  %%CITATION = ARXIV:1106.2580;%%

%\cite{Papadodimas:2011pf}
\bibitem{Papadodimas:2011pf} 
  K.~Papadodimas and S.~Raju,
  ``Correlation Functions in Holographic Minimal Models,''
  arXiv:1108.3077 [hep-th].
  %%CITATION = ARXIV:1108.3077;%%

%\cite{Gutperle:2011kf}
\bibitem{Gutperle:2011kf} 
  M.~Gutperle and P.~Kraus,
  ``Higher Spin Black Holes,''
  JHEP {\bf 1105}, 022 (2011)
  [arXiv:1103.4304 [hep-th]].
  %%CITATION = ARXIV:1103.4304;%%

%\cite{Kraus:2011ds}
\bibitem{Kraus:2011ds} 
  P.~Kraus and E.~Perlmutter,
  ``Partition functions of higher spin black holes and their CFT duals,''
  JHEP {\bf 1111}, 061 (2011)
  [arXiv:1108.2567 [hep-th]].
  %%CITATION = ARXIV:1108.2567;%%

%\cite{Ahn:2011by}
\bibitem{Ahn:2011by} 
  C.~Ahn,
  ``The Coset Spin-4 Casimir Operator and Its Three-Point Functions with Scalars,''
  arXiv:1111.0091 [hep-th].
  %%CITATION = ARXIV:1111.0091;%%
  
%\cite{Castro:2011iw}
\bibitem{Castro:2011iw} 
  A.~Castro, R.~Gopakumar, M.~Gutperle and J.~Raeymaekers,
  ``Conical Defects in Higher Spin Theories,''
  arXiv:1111.3381 [hep-th].
  %%CITATION = ARXIV:1111.3381;%%

%\cite{Ammon:2011ua}
\bibitem{Ammon:2011ua} 
  M.~Ammon, P.~Kraus and E.~Perlmutter,
  ``Scalar fields and three-point functions in D=3 higher spin gravity,''
  arXiv:1111.3926 [hep-th].
  %%CITATION = ARXIV:1111.3926;%%


%\cite{Fateev:2007ab}
\bibitem{Fateev:2007ab} 
  V.~A.~Fateev and A.~V.~Litvinov,
  ``Correlation functions in conformal Toda field theory. I.,''
  JHEP {\bf 0711}, 002 (2007)
  [arXiv:0709.3806 [hep-th]].
  %%CITATION = ARXIV:0709.3806;%%

%\cite{Fateev:2001mj}
\bibitem{Fateev:2001mj} 
  V.~A.~Fateev,
  ``Normalization factors, reflection amplitudes and integrable systems,''
  hep-th/0103014.
  %%CITATION = HEP-TH/0103014;%%

%\cite{Jayaraman:1988ex}
\bibitem{Jayaraman:1988ex} 
  T.~Jayaraman and K.~S.~Narain,
  ``Correlation Functions For Minimal Models On The Torus,''
  Nucl.\ Phys.\ B {\bf 331}, 629 (1990).
  %%CITATION = NUPHA,B331,629;%%

%\cite{Maldacena:2001kr}
\bibitem{Maldacena:2001kr} 
  J.~M.~Maldacena,
  ``Eternal black holes in anti-de Sitter,''
  JHEP {\bf 0304}, 021 (2003)
  [hep-th/0106112].
  %%CITATION = HEP-TH/0106112;%%

%\cite{Maldacena:1996ds}
\bibitem{Maldacena:1996ds} 
  J.~M.~Maldacena and L.~Susskind,
  ``D-branes and fat black holes,''
  Nucl.\ Phys.\ B {\bf 475}, 679 (1996)
  [hep-th/9604042].
  %%CITATION = HEP-TH/9604042;%%

%\cite{Iizuka:2008hg}
\bibitem{Iizuka:2008hg} 
  N.~Iizuka and J.~Polchinski,
  ``A Matrix Model for Black Hole Thermalization,''
  JHEP {\bf 0810}, 028 (2008)
  [arXiv:0801.3657 [hep-th]].
  %%CITATION = ARXIV:0801.3657;%%

%\cite{Iizuka:2008eb}
\bibitem{Iizuka:2008eb} 
  N.~Iizuka, T.~Okuda and J.~Polchinski,
  ``Matrix Models for the Black Hole Information Paradox,''
  JHEP {\bf 1002}, 073 (2010)
  [arXiv:0808.0530 [hep-th]].
  %%CITATION = ARXIV:0808.0530;%%



    

%\cite{Bais:1987dc}
%\bibitem{Bais:1987dc}
%  F.~A.~Bais, P.~Bouwknegt, M.~Surridge, K.~Schoutens,
%  ``Extensions of the Virasoro Algebra Constructed from Kac-Moody Algebras Using Higher Order Casimir Invariants,''
%  Nucl.\ Phys.\  {\bf B304}, 348-370 (1988).
  

\end{thebibliography}
\end{document}